\documentclass[11pt,a4paper]{article}
\pdfoutput=1
\usepackage{jheppub}
\usepackage{amsthm,amsbsy,amsfonts,mathrsfs,enumerate,float,wrapfig,amsmath}
\usepackage[utf8]{inputenc}
\usepackage{subfigure}

\usepackage{tikz}
\usetikzlibrary{positioning}
\usetikzlibrary{arrows}

\usepackage{color}

\theoremstyle{definition}

\newcommand{\N}{\mathcal{N}}

\title{Thermodynamic limit of Nekrasov partition function for 5-brane web with $O5$-plane}

\author[a]{Xiaobin Li}
\author[a]{and Futoshi Yagi}
\affiliation[a]{School of Mathematics, Southwest Jiaotong University,\\
West zone, High-tech district, Chengdu, Sichuan 611756, China}
\emailAdd{lixiaobin@swjtu.edu.cn}
\emailAdd{futoshi\_yagi@swjtu.edu.cn}
\abstract{
In this paper, we study 5d $\mathcal{N}=1$ $Sp(N)$ gauge theory with $N_f ( \leq 2N + 3 )$ flavors based on 5-brane web diagram with $O5$-plane. On the one hand, we discuss Seiberg-Witten curve based on the dual graph of the 5-brane web with $O5$-plane. On the other hand, we compute the Nekrasov partition function based on the topological vertex formalism with $O5$-plane. Rewriting it in terms of profile functions, we obtain the saddle point equation for the profile function after taking thermodynamic limit.
By introducing the resolvent, we derive the Seiberg-Witten curve and its boundary conditions as well as its relation to the prepotential in terms of the cycle integrals. They coincide with those directly obtained from the dual graph of the 5-brane web with $O5$-plane.
This agreement gives further evidence for mirror symmetry which relates Nekrasov partition function with Seiberg-Witten curve in the case with orientifold plane and shed light on the non-toric Calabi-Yau 3-folds including D-type singularities.
}
\begin{document}

\maketitle

\section{Introduction and Main results}

Supersymmetric gauge theory has rich structures and applications in the study of non-perturbative quantum field theories. Especially, Seiberg-Witten solutions of four dimensional (4d) $\mathcal{N}=2$ gauge theory plays an important role in understanding analytic properties in SUSY gauge theory. The basic idea of Seiberg-Witten's work \cite{Seiberg:1994rs, Seiberg:1994aj} is that the low energy physics of SUSY gauge theory can be described by geometry, i.e. a Riemann surface (which is called Seiberg-Witten curve) and periods which can be obtained by integrating a meromorphic differential one form (i.e. Seiberg-Witten differential) along two conjugate homology cycles. Based on comparing infrared limit and ultraviolet limit of certain gauge theory, Nekrasov predicts a relation between SUSY $\mathcal{N}=2$ Yang-Mills instanton partition functions and Seiberg-Witten prepotential \cite{Nekrasov:2002qd}. This relation is also called Nekrasov conjecture which has been verified by Nekrasov-Okounkov \cite{Nekrasov:2003rj} for 4d gauge theories with/without matter content and five dimensional (5d) gauge theory compactified on a circle of circumference $\beta$ based on random partition technique, and proven by Nakajima-Yoshioka \cite{Nakajima:2003pg} and Braverman-Etingof \cite{braverman:2004} by using blow-up formula and Whittaker vectors technique respectively. Furthermore, Nekrasov conjecture has been also proven for instantons on toric surface which extends the case of instantons counting on $\mathbb{C}^2 \cong \mathbb{R}^4$ by equivariant localization \cite{Gasparim:2008ri}. Look at it from another angle, this relation can be also understood as mirror symmetry which relates Nekrasov partition function in A-model side with Seiberg-Witten prepotential in B-model side. It is natural to extend this correspondence to more general cases with different gauge groups and different matter contents. It is interesting to ask the following question: How far shall we go along this line? In this paper, we will explore 5d $\mathcal{N}=1$ $Sp(N)$ gauge theories with matter contents in the presence of orientifold 5 branes.

 Recall that there is a correspondence between $(p, q)$ 5-brane web diagrams and toric diagram underlying Calabi-Yau 3-folds which give the identical gauge theories \cite{Leung:1997tw}, so it is natural and convenient to use 5-brane web as a main tool to analyze and understand supersymmetric gauge theories. Moreover, a wide class of such geometric examples are constructed in this way and shown to be related by chain of string dualities \cite{Karch:1998yv}. Recent development indicates that this brane/toric geometry correspondence can be also generalized to some non-toric examples of Calabi-Yau 3-folds and the methods to construct non-toric Calabi-Yau 3-folds are not unique. For example, the first method to construct a class of non-toric Calabi-Yau 3-folds like $n$-th local del-Pezzo surface with $n=7,8$ can be realized in terms of 5-brane web with the inclusion of 7-branes at infinity \cite{Benini:2009gi}. Correspondingly, topological string partition functions for these examples can be computed by generalizing topological vertex formalism from 5-brane web diagrams to 5-brane web diagrams with 7-branes at infinity \cite{Hayashi:2013qwa, Hayashi:2015xla}. The second way to construct non-toric Calabi-Yau 3-folds is to consider 5-brane web diagram with orientifold 5-plane ($O5$-plane). Indeed, some of them are expected to correspond to the resolutions of D-type singularity \cite{Hanany:1999sj}, which is 
 not toric. However, unlike the toric cases,
 the systematic method to construct non-toric Calabi-Yau 3-fold from an arbitrary 5-brane web diagram with $O5$-plane is not yet known enough.

In this paper, we consider 5d $\mathcal{N}=1$ $Sp(N)$ gauge theories with $N_f$ flavors and discuss its relation with Seiberg-Witten theory based on 5-brane web diagram with $O5$-plane. On the one hand, Seiberg-Witten curve can be obtained from dual graph of 5-brane web with $O5$-plane. In \cite{Witten:1997sc}, it is discussed that M-theory uplift of the type IIA analogue of the Hanany-Witten brane setup \cite{Hanany:1996ie} produces the Seiberg-Witten curve \cite{Seiberg:1994rs, Seiberg:1994aj} of 4d $\mathcal{N}=2$ gauge theories with $SU$ gauge groups. Correspondingly, the Seiberg-Witten curve of the 4d $\mathcal{N}=2$ gauge theory with $SO$ or $Sp$ gauge group can be also constructed in \cite{Brandhuber:1997cc, Landsteiner:1997vd} by use of inclusion of an orientifold 4-plane. This construction can be generalized to 5d $\mathcal{N}=1$ supersymmetric gauge theory compactified on $S^1$ \cite{Brandhuber:1997cc}. It is noted that the more systematical way is to construct based on 5-brane web diagram in \cite{Aharony:1997bh}. In a similar fashion, the Seiberg-Witten curve for 5d $\mathcal{N}=1$ gauge theories with $Sp(1)$ gauge group can be constructed by inclusion of $O5$-plane to the 5-brane web \cite{Hayashi:2017btw}. On the other hand, since supersymmetric gauge theories can be realized and analysed by considering the string theory on Calabi-Yau 3-fold \cite{Katz:1996fh, Katz:1997eq}, then the Seiberg-Witten solution can be obtained from the Nekrasov's partition function by taking thermodynamic limit \cite{Nekrasov:2003rj, Nakajima:2003pg, Nakajima:2005fg}. Nekrasov's partition function is known to agree with Topological string partition function by geometric engineering \cite{Nekrasov:2002qd, Eguchi:2003sj, Iqbal:2003ix}. Especially, when a toric Calabi-Yau 3-fold is given, the topological string partition function can be computed by using topological vertex formalism \cite{Aganagic:2003db, Li:2004uf, Iqbal:2007ii, Awata:2008ed}. Recently, new generalized formalism for the topological vertex based on the 5-brane web with $O5$-plane was conjectured \cite{Kim:2017jqn}. According to this new formalism, the topological string partition function can be systematically computed for a given 5-brane web with $O5$-plane by using cut-reflect-join technique with the assumption that the given toric-like diagram corresponds to a certain Calabi-Yau 3-fold with involution. Similarly, the Nekrasov partition function for the 5d $\N=1$ pure $Sp(1)$ gauge theory can be also computed explicitly based on this method. Although the expression obtained in this way looks different from the known expression, it is checked to agree up to 10 instantons, which gives a support for the validity of new topological vertex formalism for 5-brane web with $O5$-plane.

The main results of this paper consist of two parts: In the first half, we compute 5d Seiberg-Witten curve directly from dual graph of 5-brane web with $O5$-plane \cite{Hayashi:2017btw}.
Especially, we discuss the boundary conditions on the Seiberg-Witten curve \eqref{eq:sec2SWcurve}, which is a significant characteristic induced by the $O5$-plane.
In the second half, we obtain 5d Seiberg-Witten curve by taking thermodynamic limit of Nekrasov partition function \cite{Nekrasov:2003rj} based on topological vertex formalism for 5-brane web with $O5$-plane \cite{Kim:2017jqn}
and derive its boundary conditions.
The comparison result shows that Nekrasov's conjecture relating Nekrosov partition function with Seiberg-Witten prepotential still holds for 5-brane web with $O5$-plane. 
%
As an effective check, we verify the agreement for the prepotentials for 5d $\mathcal{N}=1$ pure $Sp(1)=SU(2)$ gauge theory with discrete theta angle 0 based on 
Seiberg-Witten curves from 5-brane web diagram with and without $O5$-plane in Appendix \ref{App:SWwwoO5}.


The structure of this article is as follows. In section \ref{sec:SW}, we obtain Seiberg-Witten curve from 5-brane web with $O5$-plane; In section \ref{sec:Top}, we review topological vertex formalism for 5-brane web with $O5$-plane and compute partition function for 5d $\mathcal{N}=1$ $Sp(N)$ gauge theory with $N_f ( \leq 2N + 3)$ flavors by cut-reflect-join techniques; In section \ref{sec:4}, we rewrite partition function as profile function of random partition and obtain 5d Seiberg-Witten curve by taking thermodynamic limit; In section \ref{sec:concl}, we conclude the main results of this paper and give some perspective for future study.
In Appendix \ref{app:SW}, we describe explicit expressions for the Seiberg-Witten curves; In Appendix \ref{App:SWwwoO5}, we compare Seiberg-Witten prepotentials for $SU(2)=Sp(1)$ gauge theory with $O5$-plane and without $O5$-plane; In Appendix \ref{app:param}, we discuss the parametrization of the 5-brane web diagram for 5d $Sp(N)$ gauge theory with $N_f$ flavors. Especially, In Appendix \ref{app:Kahler}, we discuss the derivation of the parametrization and reproduce the relation between $a_I$ and $m_I$; In Appendix \ref{app:Sduality}, we place importance on parametrization of the pure $Sp(1)$ gauge theory and discuss the transformation of the parameters induced by $S$-duality; In Appendix \ref{app:IMS}, we obtain IMS prepotential from the tropical limit; In Appendix \ref{app:proofkey}, we give the proof of the key identity expressing $R_{\boldsymbol{\lambda} \boldsymbol{\mu}}$ in terms of profile functions; In Appendix \ref{app:X0}, 
we show that if $\sum_{\boldsymbol{\lambda} } C_{ \boldsymbol{\mu} \varnothing \boldsymbol{\lambda} } X_{\boldsymbol{\lambda}} = 0$ is satisfied for arbitrary $\boldsymbol{\mu}$, then $X_{\boldsymbol{\lambda}}$ vanishes for all Young diagram $\boldsymbol{\lambda}$.


\section{Seiberg-Witten curve from 5-brane web with $O5$-plane}\label{sec:SW}

\subsection{Seiberg-Witten curve}\label{sec:SWcurve}
\begin{figure}
\centering
\begin{minipage}{10cm}
\includegraphics[width=10cm]{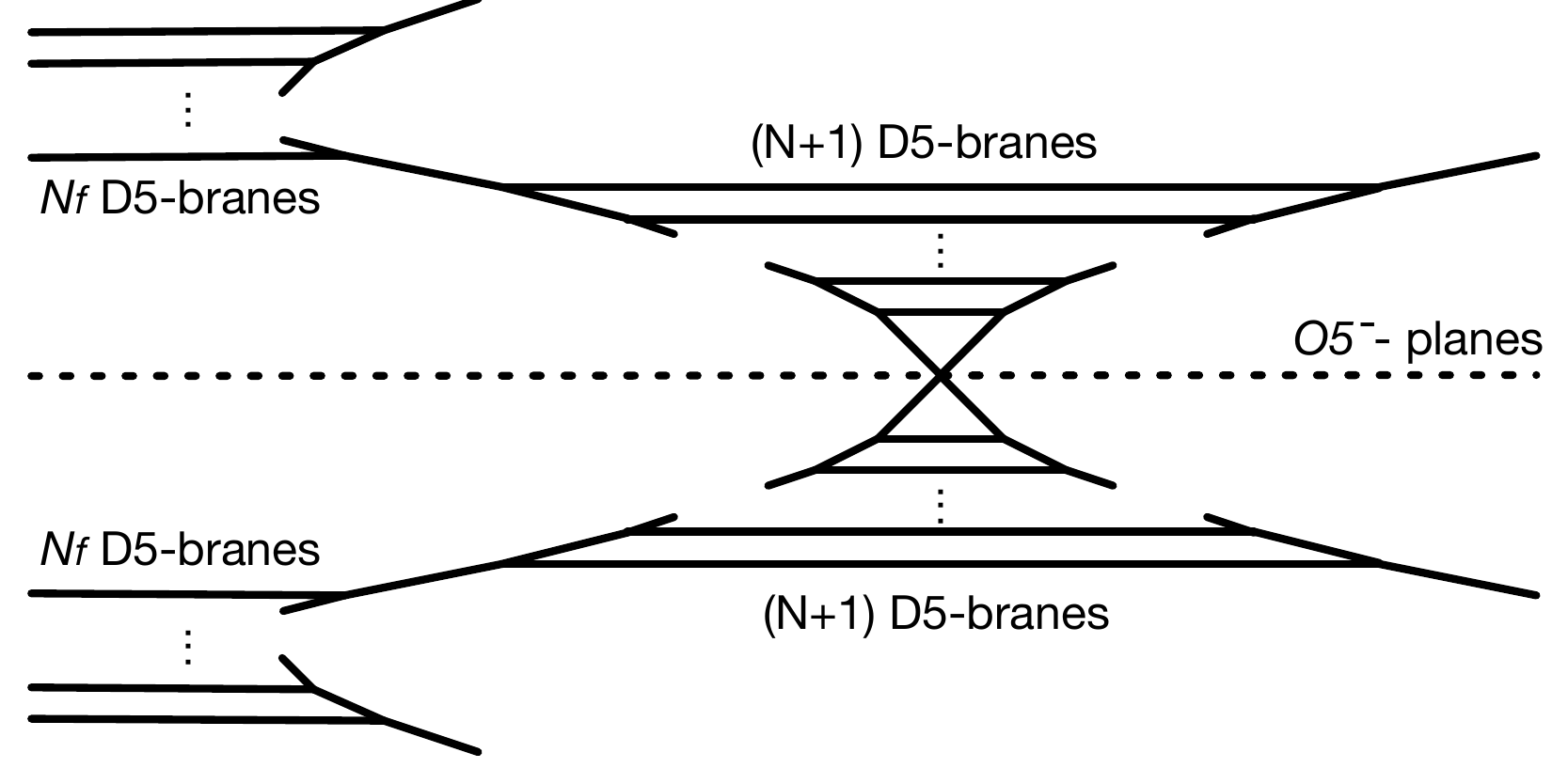}
\caption{The 5-brane web diagram with $O5$-plane including the mirror image. This corresponds to the 5d $\N=1$ $Sp(N)$ gauge theory with $N_f$ flavors.}
\label{Fig:SpN-Nf-web}
\end{minipage}
\hspace{5mm}
\begin{minipage}{4cm}
\centering
\includegraphics[width=2.3cm]{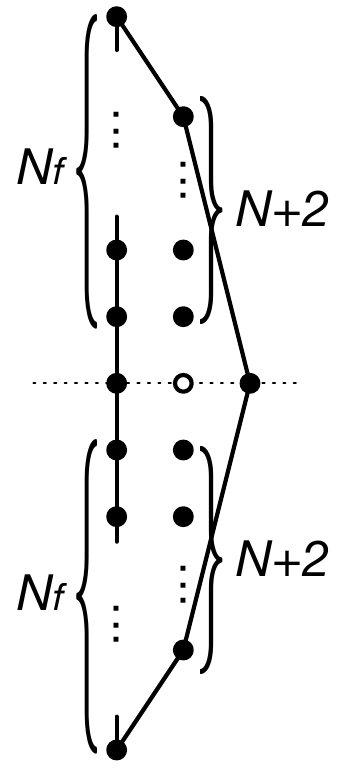}
\caption{Dual graph of the 5-brane web diagram with $O5$-plane.}
\label{Fig:SpN-Nf-dot}
\end{minipage}
\end{figure}

In this section, we construct the Seiberg-Witten curve for 5d $\N=1$ $Sp(N)$ gauge theory with $N_f$ flavors from the 5-brane web diagram with $O5$-plane by generalizing the computation in \cite{Hayashi:2017btw}. Although this class of theories are expected to have UV fixed point for $N_f \le 2N+6$, we consider $N_f \le 2N+3$ for simplicity in this paper. The 5-brane web diagram is depicted in Figure \ref{Fig:SpN-Nf-web}. This diagram includes $N+1$ internal D5-branes as well as $N_f$ half infinite D5-branes. Note that the mirror image is also included in this figure. Motivated by the ``dot diagram'' introduced in \cite{Benini:2009gi}, we introduced the white dot corresponding to the shrunken face. This diagram is interpreted as the brane construction for the 5d $\N=1$ $Sp(N)$ gauge theory with $N_f$ flavors up to the phase transition discussed in \cite{Hayashi:2017btw}.

The Seiberg-Witten curve should be constructed in such a way that its tropical limit reproduces the original 5-brane web diagram. Practically, it can be read off from its dual graph, which is depicted in Figure \ref{Fig:SpN-Nf-dot}. In this figure, we omitted the triangulation because it does not affect the Seiberg-Witten curve. Analogous to the method discussed in \cite{Aharony:1997bh}, each dot in the dual graph of the 5-brane web diagram corresponds to the monomial appearing in the Seiberg-Witten curve
\begin{align}\label{eq:SW-general}
\sum_{(m,n) \in \mathbb{Z}^2} C_{m,n} t^m w^n = 0
\end{align}
where $(m,n)$ represents the integer coordinates for the dots and the sum is over all the dots included in the dual graph. In our setup, Seiberg-Witten curve is given of the form
\begin{align}\label{eq:SW-start}
t^2 + P(w) t + Q(w) = 0
\end{align}
where
\begin{align}\label{eq:PQ}
P(w) = \sum_{n=-(N+2)}^{N+2} C_{1,n} w^n,
\qquad
Q(w) = \sum_{n=-N_f}^{N_f} C_{0,n} w^n.
\end{align}
Since the expression \eqref{eq:SW-general} has ambiguity of multiplying non-zero value to both sides of the equation, we can choose $C_{2,0} = 1$ without the loss of generality.

The effect of the existence of $O5$-plane is the following two aspects. First, the Seiberg-Witten curve is invariant under $w \to w^{-1}$. This indicates that $P(w)$ and $Q(w)$ satisfy the constraints
\begin{align}\label{eq:wtowinv}
P(w^{-1}) = P(w), \qquad Q(w^{-1}) = Q(w).
\end{align}
This gives the constraints on the coefficients to satisfy $C_{m,n} = C_{m,-n}$. Second, the Seiberg-Witten curve has double roots at $w=1$ and $w=-1$. These constraints can be written as\footnote{The Seiberg-Witten curve for 5d $Sp(N)$ gauge theory with $N_f$ flavors is also computed in \cite{Brandhuber:1997ua}. However, they impose different constraints $P(1)=P(-1)=0$. As a result, their curve does not agree with ours.}
\begin{align}\label{eq:Doubleroot}
P(1)^2 - 4 Q(1) = 0,
\qquad
P(-1)^2 - 4 Q(-1) = 0.
\end{align}
They are based on the following interpretation:
When we take T-duality and uplift to M-theory, the 5-brane web becomes a single M5-brane, whose configuration is identified as the Seiberg-Witten curve \cite{Witten:1997sc}, while the $O5$-plane becomes two OM5-planes at $w=1$ and $w=-1$.
The constraints \eqref{eq:Doubleroot} are understood as the boundary conditions of the M5-brane at the OM5-planes \cite{Landsteiner:1997vd, Hayashi:2017btw}.

In the following, we introduce some parameters which appear naturally in the region where $|t|$ and/or $|w|$ are large and/or small. They are known to be simply related to the gauge theory parameters. We first consider the region where $|t|$ is small while $|w|$ is finite. In the limit $t \to 0$, the Seiberg-Witten reduces to
\begin{align}
Q(w) = 0 .
\end{align}
From the constraint \eqref{eq:wtowinv}, the solution of this equation should be invariant under $w \to w^{-1}$. Taking this into account, we denote the $2N_f$ solutions of this equation to be
\begin{align}
w = e^{ \pm \beta m_i} \qquad (i=1,2,\cdots, N_f),
\end{align}
where $m_i$ is known to be identified as the mass parameter in the gauge theory \cite{Witten:1997sc, Brandhuber:1997ua, Aharony:1997bh}. This can be regarded as the condition that $Q(w)$ can be written as
\begin{align}\label{eq:def-M}
Q(w) = C \prod_{i=1}^{N_f} (w-e^{ - \beta m_i})(w^{-1}-e^{ - \beta m_i})
\end{align}
with $C$ being a constant. We assume $|e^{ - \beta m_i}| \le 1$ ($i=1,2,\cdots, N_f$) for later convenience.

We also consider the region where $|w|$ is small. If we solve the Seiberg-Witten curve \eqref{eq:SW-start} in terms of $t$, we have two solutions $t=t_+(w)$ and $t=t_-(w)$ which can be approximated as
\begin{align}\label{eq:tpm}
t_+(w) &=  \frac{1}{2} \left( - P(w) + \sqrt{P(w)^2 - 4 Q(w)} \right)  = - \frac{Q(w)}{P(w)} \left( 1 + \mathcal{O}(w^{2N+4-N_f}) \right)
\,\,\,\,\text{as} \,\,\,\, w \to 0,
\cr
t_-(w) &= \frac{1}{2} \left( - P(w) - \sqrt{P(w)^2 - 4 Q(w)} \right) = - P(w) \left( 1 + \mathcal{O}(w^{2N+4-N_f}) \right)
\quad \text{as} \quad w \to 0.
\end{align}
Without the loss of generality, we assume that Re$(\sqrt{P(w)^2 - 4 Q(w)})>0$ at the region $|w| \ll 1$, Re$(P(w))>0$ in order to fix the convention for the branches.
Note that $|t_+(w)| \ll |t_-(w)|$ if $|w|$ is small enough since we are considering the case $N_f \le 2N+3$. Here, we denote the coefficient of the leading order term of the ratio of these two solutions as $(-1)^{N_f} e^{- 2 \beta m_0}$. That is,
\begin{align}\label{eq:q}
\frac{t_+(w)}{t_-(w)}
&=
\frac{ Q(w)} {P(w){}^2}  \left( 1 + \mathcal{O}(w^{2N+4-N_f}) \right)
\cr
&= (-1)^{N_f} e^{- 2 \beta m_0} w^{2N+4-N_f} + \mathcal{O}(w^{2N+5-N_f})
\quad \text{as} \quad w \to 0.
\end{align}
The factor $e^{- 2 \beta m_0}$ is identified as the instanton factor in the $Sp(N)$ gauge theory and $m_0$ is identified as the mass of the instanton particle. The sign $(-1)^{N_f}$ is introduced to make the convention consistent with the past literatures including \cite{Hayashi:2017btw}.

We now see that the conditions above are enough to determine essentially all the coefficients of the Seiberg-Witten curve \eqref{eq:SW-start} with \eqref{eq:PQ} by counting degrees of freedom for its coefficients. Originally, there are totally $2N_f+2N+6$ coefficients in \eqref{eq:PQ}. The conditions \eqref{eq:wtowinv}, \eqref{eq:Doubleroot}, \eqref{eq:def-M} and \eqref{eq:q} reduce the degrees of freedom by $(N+2)+N_f$, $2$, $N_f$ and $1$, respectively. Thus, we now have $N+1$ independent coefficients. Finally, we see that one degree of freedom of the coefficients can be absorbed into the rescaling of $t$. The remaining $N$ coefficients correspond to the gauge invariant Coulomb moduli parameters, which appear in the Seiberg-Witten curve as it is. We discuss more concrete expressions for the Seiberg-Witten curve in Appendix \ref{app:SW} by writing down the coefficients $C_{m,n}$ more explicitly.

\subsection{Seiberg-Witten 1-form and cycle integrals}\label{sec:SW1form}

In the following, we discuss the Seiberg-Witten 1-form which is defined on the Seiberg-Witten curve. Based on the interpretation that the Seiberg-Witten curve is identified as the M5-brane configuration \cite{Witten:1997sc}, the Seiberg-Witten 1-form is derived as
\cite{Fayyazuddin:1997by, Henningson:1997hy, Mikhailov:1997jv}
\begin{align}\label{eq:lSW}
\lambda_{SW}= - \frac{1}{2 \pi i \beta} (\log w) \, d (\log t).
\end{align}
Since the Seiberg-Witten curve \eqref{eq:SW-start} can be understood as a double cover of the complex plane with the coordinate $w$, we introduce the 1-forms $\lambda_{SW}^+(w)$ and $\lambda_{SW}^-(w)$ defined on the complex plane (except on the branch cuts and on other singularities) so that they give the Seiberg-Witten 1-form $\lambda_{SW}$ defined on the Seiberg-Witten curve as a whole. They are obtained by substituting the solution $t=t_{+}(w)$ and $t=t_{-}(w)$ to \eqref{eq:lSW}, respectively:
\begin{align}\label{eq:r-explicit}
\lambda_{SW}^{\pm} (w)
:=&
- \frac{1}{2 \pi i \beta} (\log w) \, d (\log t_{\pm})
\cr
=&
\frac{\log w}{4 \pi i \beta Q(w)}
\left(
- Q'(w) \pm \frac{2 P'(w) Q(w) - P(w) Q'(w)}{\sqrt{ P(w)^2 - 4Q(w)} }
\right) dw.
\end{align}

Especially, at small $|w|$
\begin{align}\label{eq:SW1asym}
\lambda_{SW}^{+} (w) \sim \frac{\log w}{2 \pi i \beta} \left( \frac{P'(w) }{P(w)} -   \frac{Q'(w) }{Q(w)} \right) dw, \qquad
\lambda_{SW}^{-} (w) \sim - \frac{\log w}{2 \pi i \beta} \frac{P'(w) }{P(w)}dw,
\end{align}
 to be consistent with \eqref{eq:tpm}.

We see that the square root in the denominator of the second term in \eqref{eq:r-explicit} creates the branch cuts, where the two sheets are connected. Since we are considering the case $N_f \le 2N+3$, we expect $2N+4$ branch cuts at first sight. However, note that the conditions \eqref{eq:wtowinv} and \eqref{eq:Doubleroot} indicate that the function inside the square root is written in the form
\begin{align}\label{eq:def-deltatilde}
P(w)^2 - 4 Q(w) = C_{1,N+2}^2 w^{-2} (w-1)^2 (w+1)^2 \Delta (w),
\end{align}
where $w^{2N+2} \Delta(w)$ is a polynomial of degree $4N+4$ satisfying $ \Delta (w) = \Delta (w^{-1})$. Therefore, the number of the branch cuts are reduced by two from the naive expectation and we actually have $2N+2$ branch cuts.
We denote these $2N+2$ branch cuts as $C_I$ $(I=1,2, \cdots, 2N+2)$ and their associated branch points as $e^{- \beta \alpha^{+}_{I}}$ and $e^{- \beta \alpha^{-}_{I}}$, where we use this exponentiated expression to make the comparison easier in section \ref{sec:4}.
We can choose them in such a way that
\begin{align}\label{eq:bpt}
\alpha^{\pm}_{I}  = - \alpha^{\mp}_{2N+3 - I} \qquad (I=1,2, \cdots, 2N+2)
\end{align}
is satisfied, which is required due to the invariance under $w \to w^{-1}$ \eqref{eq:wtowinv}. For later convenience, we assume that $|e^{- \beta \alpha^{\pm}_{I}} | < 1$ for $I=1,2, \cdots, N+1$ and that all the $\alpha^{\pm}_{I}$ are different from each other.
With this notation, $\Delta(w)$ is explicitly written as
\begin{align}\label{eq:Deltafactor}
\Delta (w) =  w^{-2N-2} \prod_{I=1}^{N+1} (w - e^{- \beta \alpha^{+}_{I}})(w - e^{- \beta \alpha^{-}_{I}})(w - e^{\beta \alpha^{+}_{I}})(w - e^{\beta \alpha^{-}_{I}}).
\end{align}
We note that the branch cut structure of $\sqrt{P(w)^2-4Q(w)}$ discussed above indicates
\begin{align}\label{eq:wtowinv-tlambda}
t_{\pm} (w^{-1}) = t_{\mp}(w),
\qquad
\lambda_{SW}^{\pm} (w^{-1}) = \lambda_{SW}^{\mp} (w).
\end{align}
That is, $t_{+}(w)$ and $t_{-}(w)$ exchange with each other under the transformation $w \to w^{-1}$.

We also comment on the singularities of the Seiberg-Witten 1-form at $w=\pm 1$.  From \eqref{eq:def-deltatilde}, we can derive that the numerator of the second term in \eqref{eq:r-explicit} can be written as
\begin{align}\label{eq:PQK}
2 P'(w) Q(w) - P(w) Q'(w) = C_{1,N+2} w^{-2} (w-1) (w+1) \tilde{\Delta}(w)
\end{align}
with
\begin{align}\label{eq:deltatilde}
\tilde{\Delta}(w) = C_{1,N+2} \left[ \frac{1}{2} (w+w^{-1}) P(w) \Delta(w) - \frac{1}{4} (w^2-1) \left( 2 P'(w)  \Delta(w) -  P(w)  \Delta'(w) \right) \right],
\end{align}
where $\tilde{\Delta}(w^{-1}) = \tilde{\Delta}(w)$.
This means that the singularities at $w = \pm 1$ in the second term in \eqref{eq:r-explicit} are removable singularities because the factor $(w-1)(w+1)$ cancels with each other between the denominator and the numerator.
As a result, the Seiberg-Witten 1-form \eqref{eq:r-explicit} can be rewritten as
\begin{align}\label{eq:SW1cancelled}
\lambda_{SW}^{\pm} (w)
=&
- \frac{\log w}{4 \pi i \beta Q(w)}
\left(
w Q'(w) \pm \frac{\tilde{\Delta}(w)}{\sqrt{\Delta(w)} }
\right) \frac{dw}{w}.
\end{align}

\begin{figure}
\centering
\includegraphics[width=15cm]{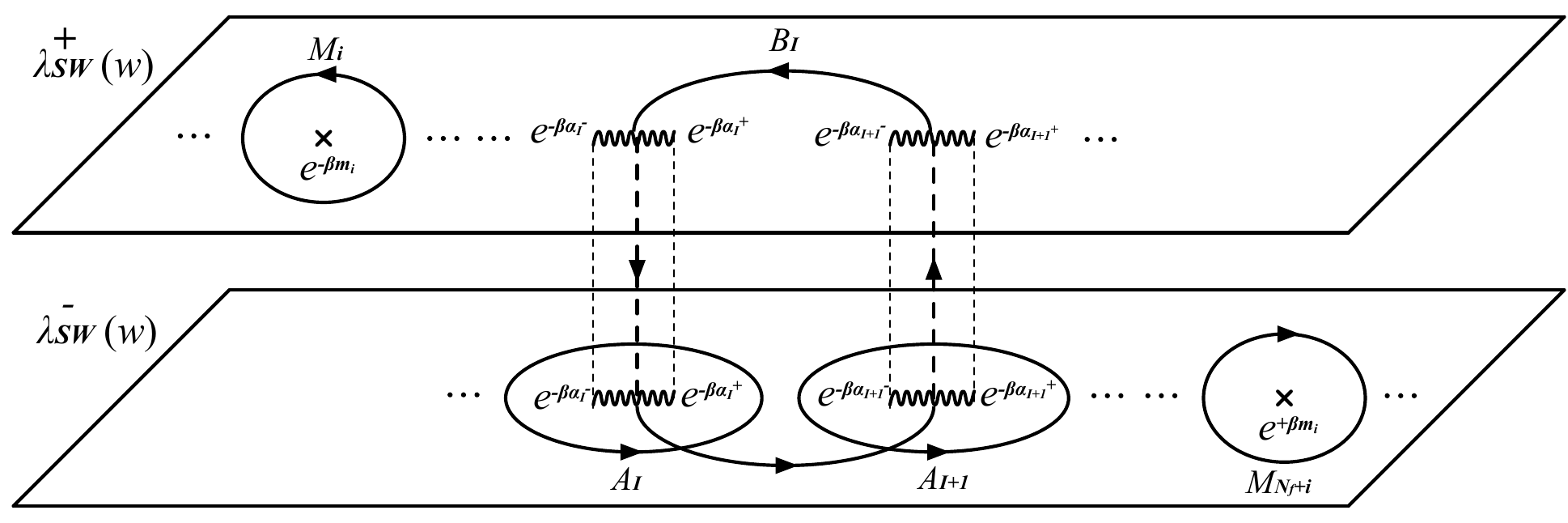}
\caption{
$A_I$, $B_I$, and $M_i$ -cycles depicted on the two copies of complex plane with coordinate $w$. $\lambda_{SW}^+(w)$ is defined on the upper complex plane while $\lambda_{SW}^-(w)$ is defined on the lower complex plane. The two complex planes are connected at the branch cuts and they give the Seiberg-Witten curve as a whole.}
\label{Fig:ABM}
\end{figure}

We introduce $A_I$-cycles ($I=1,2, \cdots 2N+2$) on the Seiberg-Witten curve as the contour
going around the branch cut $C_I$ counterclockwise on the complex plane on which $\lambda_{SW}^-$ is defined as depicted in Figure \ref{Fig:ABM}.
This means that the $A_I$-cycle integral of the Seiberg-Witten 1-form $\lambda_{SW}$ is given as the contour integral of $\lambda_{SW}^{\pm}(w)$ as
 \footnote{ As an abuse of notation, we use the same symbol $A_I$ both for the cycles on the Seiberg-Witten curve and the contour on the complex plane which goes around the branch cut $C_I$ counterclockwise. $A_I$ for the integral of $\lambda^+_{SW}$ in \eqref{eq:defAint} means the latter, which corresponds to the opposite of the $A_I$-cycle on the Seiberg-Witten curve. Due to this convention, the minus sign appears. }
\begin{align}\label{eq:defAint}
\oint_{A_I} \lambda_{SW}
=& \oint_{A_I} \lambda_{SW}^-(w) = - \oint_{A_I} \lambda_{SW}^+(w).
\end{align}
We also define $B_I$-cycles on the Seiberg-Witten curve so that the intersections of the cycles are given by%
\footnote{The choice of B-cycle in this paper is different from the standard non-compact cycle. We use this choice in order to avoid the divergence.}
\begin{align}\label{eq:ABintersect}
A_I \cdot B_J = \delta_{I,J} - \delta_{I,J+1}.
\end{align}
Then, $B_I$-cycle integral of the Seiberg-Witten 1-form is
\begin{align}
\oint_{B_I} \lambda_{SW}
=&  \int_{e^{- \beta \alpha^{+}_{I}} }^{e^{- \beta \alpha^{-}_{I+1}}} \lambda_{SW}^+(w) + \int^{e^{- \beta \alpha^{+}_{I}} }_{e^{- \beta \alpha^{-}_{I+1}}} \lambda_{SW}^-(w)
\end{align}
Finally, we denote the contour going around the pole at $w = e^{- \beta m_i}$ counterclockwise on the complex plane on which $\lambda_{SW}^{+}$ is defined as ``$M_i$-cycles'' ($i=1,2, \cdots, N_f$) in this paper.%
\footnote{Although $M_i$-cycles are not rigorously homological 1-cycles of the Riemann surface, we call so in this paper because they often appears as degenerating limit of the 1-cycles.
}

 The approximation \eqref{eq:SW1asym} is valid also around $w=e^{- \beta m_i}$, where $Q(w)$ vanishes due to \eqref{eq:def-M}. This indicates that $\lambda_{SW}^{+} $ has simple poles at $w=e^{-\beta m_i}$ while $\lambda_{SW}^{-} $ does not.
The relation \eqref{eq:wtowinv-tlambda} indicates that $\lambda_{SW}^{-} $ has simple poles at $w=e^{+\beta m_i}$ while $\lambda_{SW}^{+} $ does not. We denote the contour going around the pole at $w = e^{+ \beta m_i}$ clockwise on the complex plane on which $\lambda_{SW}^{-}$ is defined as ``$M_{N_f+i}$-cycles'' ($i=1,2, \cdots, N_f$).
This means that the $M_i$-cycle integral of the Seiberg-Witten 1-form $\lambda_{SW}$ is given as \footnote{$M_i$ in this equation denotes the contour going around $w=e^{-\beta m_i}$ conterclockwise rather than the $M_i$-cycle on the Seiberg-Witten curve, analogous to the convention for $A_I$.}
\begin{align}\label{eq:M-cycle}
&\oint_{M_i} \lambda_{SW}
= \oint_{M_i} \lambda_{SW}^+(w) = m_i,
\quad (i=1,2, \cdots N_f)
\cr
&\oint_{M_{i+N_f}} \lambda_{SW}
= \oint_{M_{i+N_f}} \lambda_{SW}^-(w) = - m_i, \quad (i=1,2, \cdots N_f).
\end{align}
We also note that $\lambda_{SW}^{\pm}$ do not have a pole at $w = e^{\mp \beta m_i}$ and thus,
\begin{align}
& \oint_{M_{i}} \lambda_{SW}^-(w) = \oint_{M_{i+N_f}} \lambda_{SW}^+(w) = 0 \quad (i=1,2, \cdots N_f).
\end{align}

\begin{figure}
\centering
\includegraphics[width=15cm]{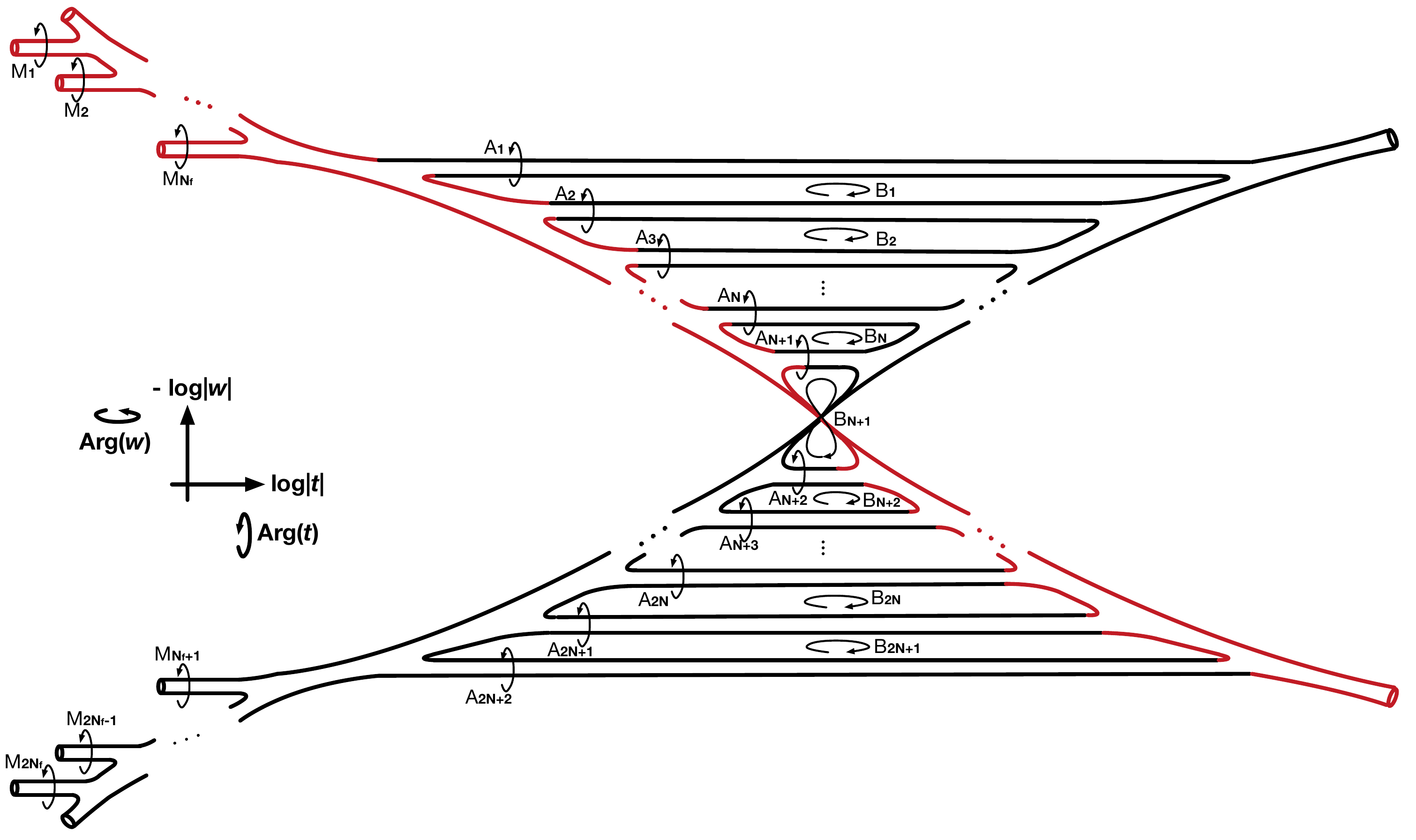}
\caption{Seiberg-Witten curve obtained by thickening the 5-brane web. $A_I$,$B_I$, and $M_i$ cycles are depicted. The red part is identified as one branch, on which $t^+(w)$ is defined.}
\label{fig:tropical}
\end{figure}

The cycles defined above can be also understood as depicted in Figure \ref{fig:tropical} due to the following interpretation. The Seiberg-Witten curve is constructed in such a way to reproduce the 5-brane web diagram in the tropical limit \cite{Aharony:1997bh}. In other word, the Seiberg-Witten curve is obtained by ``thickening'' the 5-brane web diagram. Analogous to the case in \cite{Witten:1997sc}, the $(-p,1)$ 5-brane at the left and the $(p,1)$ 5-brane at the right correspond to the two $w$-planes, whose positions are given by \eqref{eq:tpm}, respectively. The color D5 branes correspond to the cuts connecting them, and thus, the $A_I$-cycle goes around the tube corresponding to the $I$-th color brane. Analogously, the flavor D5 brane corresponds to the poles on the $w$-planes, and the $M_i$-cycle goes around the tube corresponding to the $i$-th flavor brane. The $B_I$-cycle goes to left along the $I$-th color brane, and goes back to right along the $I+1$-th color brane so that it intersects with the $A_I$-cycle and the $A_{I+1}$-cycle as in \eqref{eq:ABintersect}.

By using the $A_I$-cycles and $B_I$-cycles introduced above, the prepotential is given by the following Seiberg-Witten solution:
\begin{align}\label{eq:AB-integral}
& \oint_{A_I} \lambda_{SW} = a_I, \qquad (I=1,2, \cdots, N)
\cr
& \oint_{B_I} \lambda_{SW} = \frac{1}{2 \pi i \beta} \left( \frac{\partial F}{\partial a_I} - \frac{\partial F}{\partial a_{I+1}} \right),
\qquad (I=1,2, \cdots, N-1),
\cr
& \oint_{B_N}  \lambda_{SW} = \frac{1}{2 \pi i \beta} \frac{\partial F}{\partial a_N}.
\end{align}

\begin{figure}
\centering
\includegraphics[width=8cm]{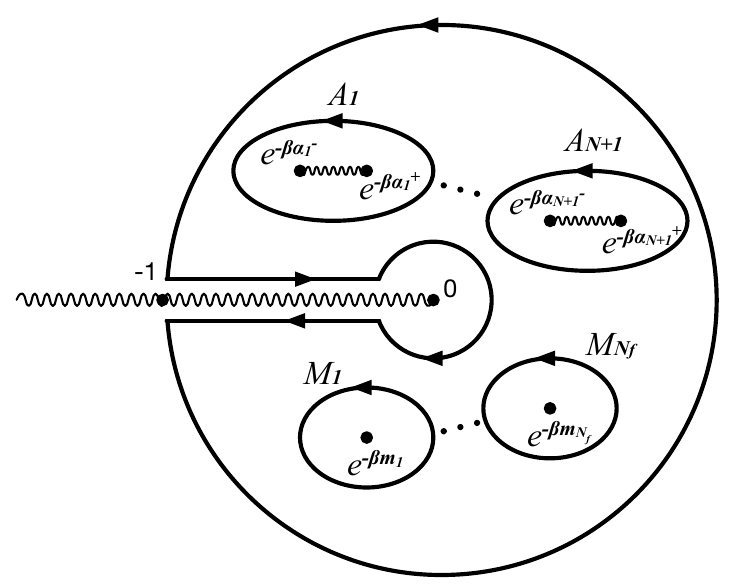}
\caption{Contour of the integral.}
\label{fig:contour}
\end{figure}

In the following, we compute the integrals over other cycles by using the parameters introduced above.
The integrals over $A_{N+1}$ cycle can be computed by considering the integral over the contour depicted in Figure \ref{fig:contour}. Since we have assumed $|e^{-\beta m_i}| \le 1$ ($i=1,2,\cdots, N_f$) and $|
e^{- \beta \alpha^{\pm}_{I}}| \le 1$ ($I=1,2,\cdots, N+1$), $A_I$-cycles and $M_i$-cycles are inside the unit circle $|w|=1$. Thus, we find the following identity
\begin{align}\label{eq:contourint}
\lim_{\epsilon \to 0+ }\left( \oint_{ |w|=\epsilon} + \int_{\epsilon e^{-\pi i}}^{e^{-\pi i}} +  \oint_{|w|=1} +  \int^{\epsilon e^{\pi i}}_{e^{\pi i}}  \right) \tilde{\lambda}_{SW}
= \sum_{I=1}^{N+1} \oint_{A_I} \tilde{\lambda}_{SW}
+ \sum_{i=1}^{N_f} \oint_{M_i} \tilde{\lambda}_{SW},
\end{align}
where we introduced the following 1-form defined on the complex plane
\begin{align}\label{eq:SW1shift}
\tilde{\lambda}_{SW}(w) := \lambda_{SW}^+(w) - \lambda_{SW}^- (w)
+  \frac{\log w}{2 \pi i \beta} (2N+4-N_f)
\, \frac{dw}{w}.
\end{align}
We have added the last term in \eqref{eq:SW1shift} so that the first integral of the left hand side in \eqref{eq:contourint} vanishes.
 The third integral of the left hand side in \eqref{eq:contourint} also vanishes because when we parametrize the path as $w=e^{i \theta}$ ($ -\pi < \theta < \pi$), the integrand is an odd function of $\theta$ due to \eqref{eq:wtowinv-tlambda}.
Here, note that the logarithmic branch cut of $\log w$, which appears in \eqref{eq:SW1shift} and \eqref{eq:SW1cancelled}, exists on the negative real axis of $w$ in our convention. Thus, $\log w$ included in the integrand in the second and the fourth terms at the left hand side of \eqref{eq:contourint} are different by $-2 \pi i$, and thus,
\begin{align}
\lim_{\epsilon \to 0+} \left(  \int_{\epsilon e^{-\pi i}}^{e^{-\pi i}} + \int^{\epsilon e^{\pi i}}_{e^{\pi i} }  \right) \tilde{\lambda}_{SW}
&= \frac{1}{\beta} \int_{w=0}^{w=-1} \left( d \log t_+ -  d \log t_-  - (2N+4-N_f) d \log w \right)
\cr
& = \left. \frac{1}{\beta} \log \left( \frac{ t_+(w)}{ w^{2N+4-N_f}  t_-(w)} \right) \right|^{-1}_0
 = 2 m_0.
\end{align}
Here, we used the expression \eqref{eq:lSW} and \eqref{eq:SW1shift} at the first equality while we used \eqref{eq:q} at the last equality.
The right hand side in \eqref{eq:contourint} can be obtained directly from \eqref{eq:M-cycle} and \eqref{eq:AB-integral} apart from the $A_{N+1}$-cycle integral. Therefore, $A_{N+1}$-cycle can be obtained from \eqref{eq:contourint} as
\begin{align}\label{eq:AN1}
a_{N+1}: = \oint_{A_{N+1}} \lambda_{SW} = - \sum_{I=1}^N a_I - m_0 + \frac{1}{2} \sum_{j=1}^{N_f} m_j.
\end{align}
Note that we could have imposed \eqref{eq:AN1} instead of the condition \eqref{eq:q}.

From the invariance under $w \to w^{-1}$ \eqref{eq:wtowinv-tlambda} with the convention \eqref{eq:bpt}, we have
\begin{align}
& \oint_{A_{2N+3-I}} \lambda_{SW} =  a_{I}=:  a_{2N+3-I} \qquad (I=1,2, \cdots, N, N+1),
\cr
&\oint_{B_{2N+2-I}} \lambda_{SW} = \frac{1}{2\pi i \beta} \left( \frac{\partial F}{\partial a_{I}} - \frac{\partial F}{\partial a_{I+1}} \right) \qquad (I=1,2, \cdots, N-1),
\cr
& \oint_{B_{N+2}} \lambda_{SW} = \frac{1}{2\pi i \beta} \frac{\partial F}{\partial a_{N}}
\end{align}
The integrals over $B_{N+1}$ is less obvious and we do not discuss in this paper since it is not necessary to determine the prepotential.

Finally, we comment on the difference between the Seiberg-Witten curves for 4d and 5d $Sp(N)$ gauge theories. 
The Seiberg-Witten curve for 4d $Sp(N)$ gauge theory is obtained from the Seiberg-Witten curve for 4d $SU(2N+2)$ gauge theory and by tuning the parameters in such a way that $a_{N+1}=0$ and $a_{2N+3-i} = - a_i $ $(i=1,2,…, N+1)$ are satisfied as mentioned, for example, in \cite{Nekrasov:2004vw}. 
The Seiberg-Witten curve for 5d $Sp(N)$ gauge theory is obtained from the Seiberg-Witten curve for 5d $SU(2N+4)$ gauge theory and by tuning the parameters in such a way that $a_{N+1}$ given in \eqref{eq:AN1}, $a_{N+2}=0$ and $a_{2N+5-i} = - a_i $ $(i=1,2,…, N+2)$ are satisfied. 
The appearance of the $A_{N+1}$-cycle, over which integral gives non-zero $a_{N+1}$ given in terms of the other parameters, is a remarkable feature for the 5d case. This phenomena is related to the duality between 5d $Sp(N)$ gauge theory and 5d $SU(N+1)$ gauge theory proposed in \cite{Gaiotto:2015una}, which duality does not hold for the 4d gauge theories.

\subsection{Seiberg-Witten solution from 5-brane web with $O5$-plane}\label{sec:summary}

We summarize the necessary information to reproduce the prepotential by collecting the necessary information discussed in the previous subsections. The prepotential $F(a,m)$ for the 5d $\mathcal{N}=1$ $Sp(N)$ gauge theory with $N_f(\le 2N+3)$ flavors is essentially uniquely determined\footnote{For $N_f=0$, due to $\pi_4 (Sp(N)) \cong \mathbb{Z}_2$, there is still an ambiguity of the choice of the discrete theta angle $\theta=0$ or $\theta=\pi$, which will be addressed in Appendix \ref{app:SW}.}  up to an integration constant by the following set of equations: The Seiberg-Witten curve
\begin{equation}\label{eq:sec2SWcurve}
\begin{split}
& t^2 + P(w) t + Q(w) = 0, \\
& P(w) = \sum_{n=1}^{N+2} C_{1,n} (w^n+w^{-n}) + C_{1,0},
\qquad
Q(w) = C \prod_{i=1}^{N_f} (w-e^{ - \beta m_i}) (w^{-1} - e^{ - \beta m_i}), \\
& P(1)^2 - 4 Q(1) = 0, \qquad P(-1)^2 - 4 Q(-1) = 0, \\
\end{split}
\end{equation}
with the cycle integrals of the Seiberg-Witten 1-form
\begin{equation}\label{eq:sec2SW1}
\begin{split}
& \lambda_{SW} := - \frac{1}{2 \pi i \beta} \log w \, d (\log t), \\
& \oint_{A_I} \lambda_{SW} = a_I, \quad (I=1,2, \cdots, N,N+1),
\qquad \sum_{I=1}^{N+1} a_I =  - m_0 + \frac{1}{2} \sum_{j=1}^{N_f} m_j,
\\
& \oint_{B_I} \lambda_{SW} =\frac{1}{2\pi i \beta} \left( \frac{\partial F}{\partial a_I} - \frac{\partial F}{\partial a_{I+1}} \right)
\quad (I=1,2, \cdots, N-1),
\qquad   \oint_{B_N}  \lambda_{SW} = \frac{1}{2\pi i \beta} \frac{\partial F}{\partial a_N}.
\end{split}
\end{equation}
We will reproduce these equations from the thermodynamic limit of the partition function in later section.


\section{Partition function via topological vertex formalism 
with $O5$-plane}\label{sec:Top}

\subsection{Topological vertex formalism for 5-brane web with $O5$-plane}\label{sec:top-rule}
In this subsection, we review the ``topological vertex formalism with $O5$-plane'' \cite{Kim:2017jqn, Hayashi:2020hhb}.
The (unrefined) topological vertex formalism \cite{Aganagic:2003db}
is a systematic algorithm to obtain the topological string partition function for a given toric-web diagram, which specifies the toric Calabi-Yau 3-fold. According to the conjecture in \cite{Leung:1997tw}, the toric web diagram to specify the toric Calabi-Yau 3-folds can be identified with the 5-brane web diagram of the same shape. This correspondence indicates that the topological vertex formalism can be reinterpreted as the method to compute partition function for a given 5-brane web diagram. This reinterpretation is useful when we generalize the formalism to the case where the 5-brane web diagram includes $O5$-plane \cite{Kim:2017jqn}.

Given a 5-brane web with $O5$-plane, for vertices and edges that are not attached to the $O5$-plane, the rules are the same as the case of 5-brane web without $O5$-plane. The topological string partition function can be computed based on $(p,q)$ 5-brane web. There are two basic rules about contributions from edges and vertices:
\begin{itemize}
\item
For each edge, we assign a partition $\boldsymbol{\lambda} = (\lambda_1 \geq \lambda_2 \geq \ldots \geq 0)$ which is a set of monotonically decreasing non-negative integers such that $\lambda_i = 0$ for $i\gg 0$.
The size of $ \boldsymbol{\lambda} $ is defined to be the sum of all nonnegative integers $| \boldsymbol{\lambda} | = \sum \lambda_i$.
The partition is  understood as a Young diagram, which is a collection of boxes with $\lambda_i$ boxes at the $i$-th column.%
\footnote{In this paper, we use bold fonts for Young diagrams. When we consider multiple Young diagrams, we often distinguish them by putting lower index as $\boldsymbol{\lambda}_1, \boldsymbol{\lambda}_2, \cdots$ in this paper. The Young diagram $\boldsymbol{\lambda}_i$ should not be confused with the $\lambda_i$.}
We introduce its transpose $\boldsymbol{\lambda}^T$ by exchanging the rows and the columns of $ \boldsymbol{\lambda}$.
We also define $|| \boldsymbol{\lambda} ||^2$ or $|| \boldsymbol{\lambda}^T ||^2$ as
\begin{align}\label{eq:lambdasquare}
& || \boldsymbol{\lambda}  ||^2 = \sum_{(i, j)\in \lambda} \lambda_i = \sum_{i=1}^{\lambda_1^T} \lambda_i^2 = \sum_{j=1}^{\lambda_1}\sum_{i=1}^{\lambda_j^T} \lambda_i
\\
&|| \boldsymbol{\lambda} ^T ||^2 = \sum_{(i, j)\in \lambda^T} \lambda_i^T = \sum_{i=1}^{\lambda_1} (\lambda_i^T)^2 = \sum_{j=1}^{\lambda_1^T}\sum_{i=1}^{\lambda_j} \lambda_i^T.
\end{align}

Based on these notations, we define the corresponding ``edge factors'' as
\begin{align}\label{eq:edgefactor}
E_{\boldsymbol{\lambda}}(Q;n) = Q^{| \boldsymbol{\lambda} |} (-1)^{(n+1) | \boldsymbol{\lambda} |} g^{\frac{n}{2}(\| \boldsymbol{\lambda}^T \| ^2 - \| \boldsymbol{\lambda} \|^2)}.
\end{align}
Here, the parameter $Q$ corresponding to an edge is given as
\begin{align}\label{eq:Q-length}
Q = \exp \left( - \frac{\beta (\text{``Length''} )}{\sqrt{p^2+q^2}} \right),
\end{align}
where (``Length'') denotes the length of the edge written in terms of the masses and the Coulomb branch parameters while $p$ and $q$ are the RR charge and the NSNS charge of the corresponding $(p,q)$ 5-brane.
This parameter $Q$ corresponds to the K\"ahler parameter of the corresponding 2-cycle in the geometry side.
The parameter $g=e^{-\beta \hbar}$ is given by the omega deformation parameters in terms of the following relation $\hbar = \epsilon_1 = - \epsilon_2$.
Finally, the power $n=p_1 q_2 - p_2 q_1 $ is determined by the charges $(p_1, q_1)$ and $(p_2,q_2)$ of the two 5-branes attached to the considered edge that are chosen diagonally.
\item For each trivalent vertex, where the three edges with Young diagrams $\boldsymbol{\lambda},  \boldsymbol{\mu} ,  \boldsymbol{\nu} $ meet,
we can introduce the following topological vertex $C_{\boldsymbol{\lambda} \boldsymbol{\mu}  \boldsymbol{\nu} }$
\begin{align}
C_{ \boldsymbol{\lambda} \boldsymbol{\mu} \boldsymbol{\nu} } = g^{\frac{-\| \boldsymbol{\mu}^T \| ^2+\| \boldsymbol{\mu} \| ^2 +\| \boldsymbol{\nu} \| ^2}{2}}\,
\widetilde{Z}_{\boldsymbol{\nu} }(g)\,\sum_{\boldsymbol{\eta}}
s_{\boldsymbol{\lambda}^T/ \boldsymbol{\eta} }(g^{- \rho - \boldsymbol{\nu} })\,s_{ \boldsymbol{\mu} / \boldsymbol{\eta}}(g^{- \rho - \boldsymbol{\nu} ^T}),
\end{align}
where
\begin{align}\label{eq:Ztilde}
\widetilde{Z}_{ \boldsymbol{\nu} }(g)
=\widetilde{Z}_{\boldsymbol{\nu}^T}(g) =
\prod_{i=1}^{{\ell}( \boldsymbol{\nu} )}\prod_{j=1}^{\nu_i}\frac1{1-g^{\nu_i+\nu^T_j-i-j+1}} ,
\end{align}
with
$g^{- \rho - \boldsymbol{\nu} }= (g^{\frac12-\nu_1}, g^{\frac32-\nu_2},\cdots)$ and $s_{ \boldsymbol{\lambda} / \boldsymbol{\eta}}(x)$ is a
skew-Schur function.
This topological vertex is known to satisfy the cyclic symmetry $C_{ \boldsymbol{\lambda}  \boldsymbol{\mu}  \boldsymbol{\nu} } = C_{ \boldsymbol{\mu}  \boldsymbol{\nu}  \boldsymbol{\lambda} } = C_{ \boldsymbol{\nu}  \boldsymbol{\lambda}  \boldsymbol{\mu} }$.
\end{itemize}
The topological string partition function is given by multiplying all the edge factors, the vertex factors and by summing them over all the Young diagrams
\begin{align}
Z = \sum_{\text{Young} \atop \text{diagrams}} \prod_{\text{edges}} E_{*}(*;*) \prod_{\text{vertices}} C_{***} .
\end{align}

\begin{figure}
        \centering
        \includegraphics[width=6cm]{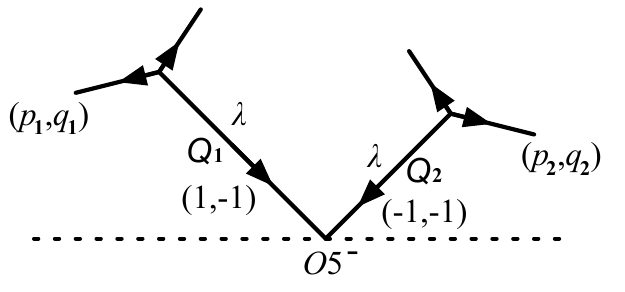}
        \caption{The two different 5-branes intersecting with each other on the $O5$-plane. The $O5$-plane factor  is introduced corresponding to such edge.}
        \label{fig:O5factor}
\end{figure}

If the 5-brane web digram includes edges depicted in Figure \ref{fig:O5factor}, then we need to introduce an ``$O5$-plane factor'' associated with the edges attached to the $O5$-plane.
\begin{itemize}
\item For two different 5-branes intersecting with each other on the $O5$-plane, we can introduce the following additional $O5$-plane factor by assigning identical Young diagrams to these two 5-branes
\begin{align}\label{eq:O5factor}
I_{\boldsymbol{\lambda}}(Q_1, Q_2; n') = (Q_1 Q_2)^{|\boldsymbol{\lambda}|} (-1)^{n' |\boldsymbol{\lambda}|} g^{\frac{n'}{2}(||\boldsymbol{\lambda}^T||^2 - ||\boldsymbol{\lambda}||^2)},
\end{align}
where the parameters $Q_1, Q_2$ are exponentials of the minus of the length of the two 5-branes divided by the factor $\sqrt{p_i^2+q_i^2}$
respectively, the index $n' = p_1 q_2 + p_2 q_1 + 1$. {This rule is reformulated in terms of ``O-vertex'' in \cite{Hayashi:2020hhb}. } 
\end{itemize}

The (main part of the) topological string partition function is given by multiplying all the edge factors, the vertex factors and the additional $O5$-factors and by summing them over all possible Young diagrams
\begin{align}
Z = \sum_{\text{Young} \atop \text{diagrams}} \prod_{\text{edges}} E_{*}(*;*) \prod_{\text{vertices}} C_{***} \prod_{\text{intersection} \atop \text{with }O5-\text{plane}}  I_{*}(*,*;*).
\end{align}

Although this formalism is based on the 5-brane web, we expect that this gives the Gromov-Witten invariants for the corresponding Calabi-Yau 3-fold $X$:
\begin{align}
Z = \exp \left[ \sum_{g=0}^{\infty} \sum_{\mathbf{d}} \hbar^{2g-2} N_{\mathbf{d}}^g e^{ - \mathbf{d} \cdot \mathbf{t}} \right],
\end{align}
where $\hbar$ is the string coupling constant, $\mathbf{t}$ is the collection of the K\"ahler parameters, and $N_{\mathbf{d}}^g$ is the genus $g$ Gromov-Witten invariants of the two cycles $\mathbf{d}$.
As discussed, for example, in \cite{Bershadsky:1993cx, Gopakumar:1998ii, Gopakumar:1998jq, Iqbal:2007ii, Dedushenko:2014nya, Codesido:2015dia},
in order to obtain the full topological string partition function,
we need to multiply the part which cannot be obtained from the computation based on the topological vertex:
\begin{align}\label{eq:Zfull}
Z_{\text{full}} = Z  \exp \left[
- \frac{1}{6\hbar^2} \sum_{i,j,k} a_{i,j,k} t_i t_j t_k
- \frac{1}{24} \sum_i b_i t_i
+ \sum_{g=2}^{\infty} \hbar^{2g-2} c_g
\right]
\end{align}
where $a_{ijk}$ and $b_i$ are related to the topological intersection number in $X$,
and $c_g$ is the constant called constant map contribution.
Especially, the first term is known to be identified as IMS prepotential \cite{Intriligator:1997pq} of the corresponding 5d $\mathcal{N}=1$ gauge theory.

\subsection{Comments on the $O5$-plane factor}
In this subsection, we focus on the $O5$-plane factor to give justification of the rule \eqref{eq:O5factor}. 

\begin{figure}
\centering
\begin{minipage}{6cm}
\centering
\includegraphics[width=5.5cm]{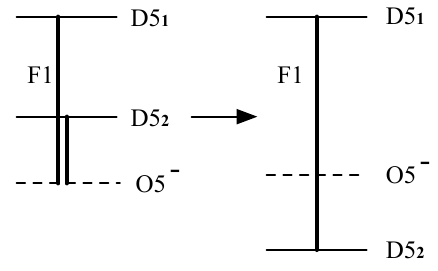}
\caption{Fundamental strings in the brane setup with D5 branes and an $O5$-plane. }
\label{fig:D5O5F1}
\end{minipage}
\hspace{1cm}
\begin{minipage}{7cm}
\centering
\includegraphics[width=6cm]{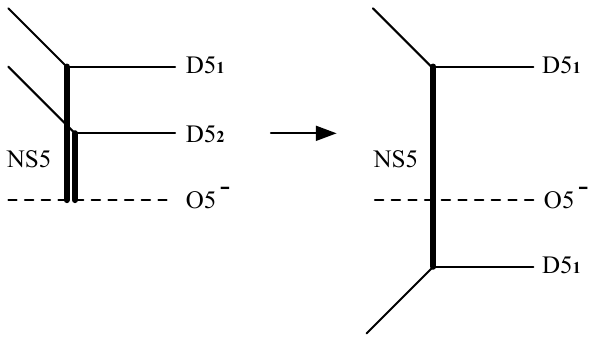}
\caption{5-brane web with an $O5$-plane, which is obtained by the combination of S-dualities and T-dualities.}
\label{fig:D5O5NS5}
\end{minipage}
\end{figure}

We start with a fundamental string in the brane setup with D5 branes and an $O5$-plane as in Figure \ref{fig:D5O5F1}. 
This fundamental strings can be understood as a single fundamental string stretched between the first D5-brane (D5$_1$) and the mirror image of the second D5-brane (D5$_2$). Especially, in the context of gauge theory, such fundamental string corresponds to one of the components of the W-boson, which should be treated in an equal footing with the other components of the W-boson corresponding to fundamental strings stretched between D5-branes without attaching to the $O5$-plane. 

As discussed in \cite{Hayashi:2015vhy}, certain combination of S-dualities and T-dualities leads this brane setup to a 5-brane web with an $O5$-plane in Figure \ref{fig:D5O5NS5}. Here, the fundamental string is mapped to an NS5-brane. Therefore, this NS5-brane can be also understood naturally as a single NS5-brane by reflecting part of the 5-brane web diagram and should be treated in an equal footing with other 5-branes which are not attached to the $O5$-plane. 

\begin{figure}
\centering
\centering
\includegraphics[width=8cm]{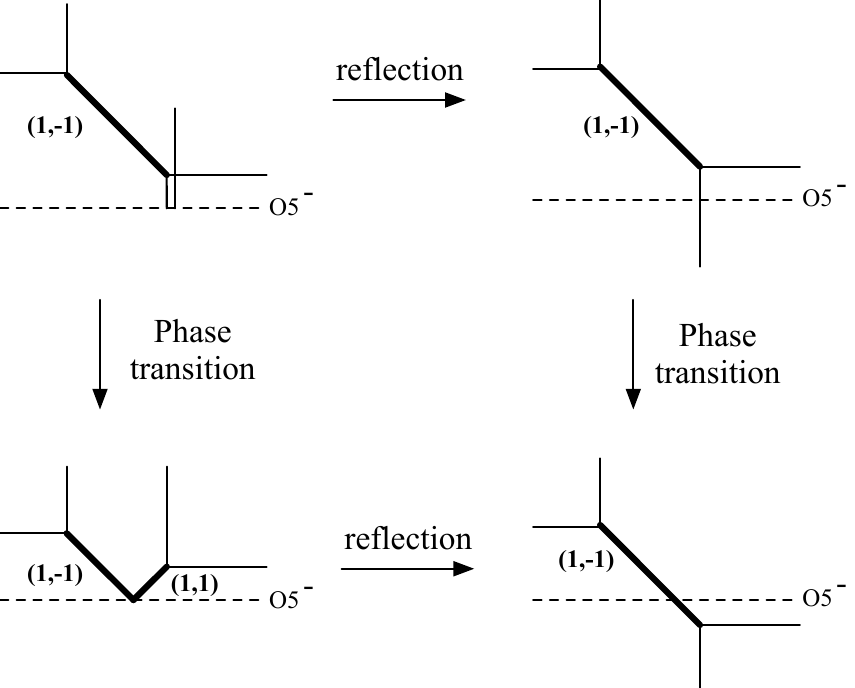}
\caption{The phase transition found in \cite{Hayashi:2017btw}. It can be understood naturally by partially reflecting the 5-brane web diagram. The $(1,-1)$ 5-brane given by the thick line in the upper left diagram is mapped to the thick lines in the other diagrams by the reflection or by the phase transition.}
\label{fig:Genflop}
\end{figure}

In order to generalize this observation to $(\pm 1, \pm 1)$ 5-branes, we consider the phase transition given in Figure \ref{fig:Genflop}. 
This phase transition was originally found in the process of considering the tropical limit of the Seiberg-Witten curve for rank 1 $E_2$ theory and was called ``generalized flop transition'' \cite{Hayashi:2017btw}. 
It may look non-trivial at first sight if we consider only the left half of Figure \ref{fig:Genflop}. However, if we reflect part of the web diagram, this can be understood more naturally since it is just moving the strip diagram until the $(1,-1)$ 5-brane intersect with the $O5$-plane as given in the right half of Figure \ref{fig:Genflop}. 
Since these two different phases are realized in different parameter regions of an identical gauge theory, the partition function should be invariant under this phase transition, 
analogous to the flop invariance of the topological string partition function \cite{Konishi:2006ev, Taki:2008hb}. 
Especially, the contribution from the $(1,-1)$ 5-brane at the upper left diagram of Figure \ref{fig:Genflop} should be identical to the contribution from the $(1,-1)$ 5-brane and the $(1,1)$ 5-brane intersecting at the $O5$-plane at the lower left of Figure \ref{fig:Genflop}.
This claim indicates the equivalence of the lower left diagram and the lower right diagram in Figure \ref{fig:Genflop}, which are related by the partial reflection of the 5-brane web diagram, 
because the shape of the $(1,-1)$ 5-brane is identical in the upper left diagram and in the lower right diagram in Figure \ref{fig:Genflop}. This observation motivates us to treat ``the two different 5-branes intersecting with each other on the $O5$-plane'' in Figure \ref{fig:O5factor} as a single $(1,-1)$ 5-brane or a single $(1,1)$ 5-brane by reflecting part of the 5-brane web and to treat it in an equal footing with other 5-branes which are not attached to the $O5$-plane. 

\begin{figure}
\centering
\includegraphics[width=15cm]{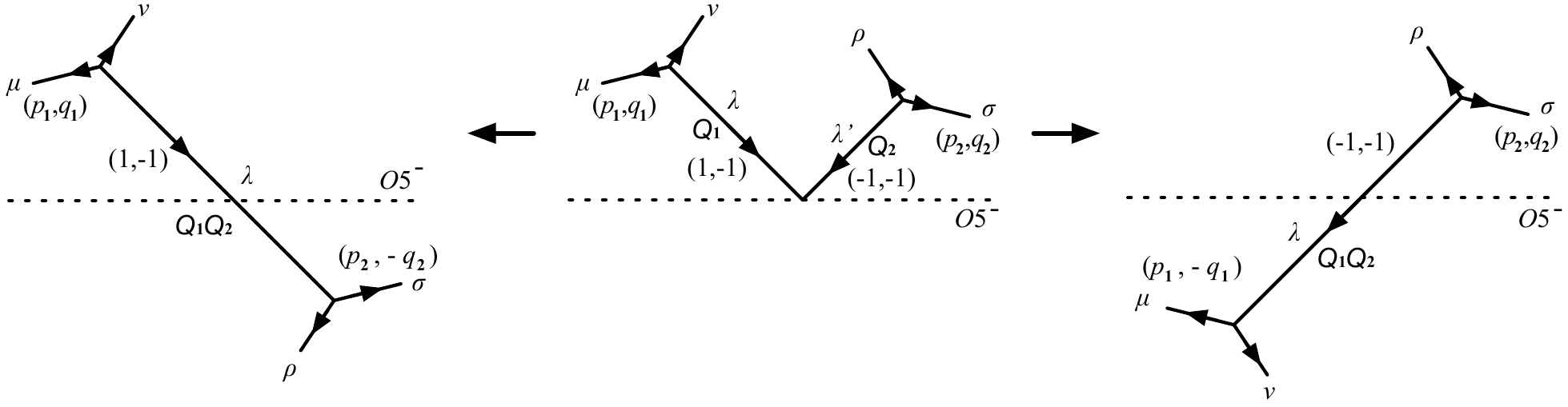}
\caption{The web diagram to which the $O5$-plane factor  is assigned and its partial reflection.}
\label{fig:reflection}
\end{figure}

The $O5$-plane factor \eqref{eq:O5factor} is basically determined from this idea. 
The 5-brane web diagram at the center in Figure \ref{fig:reflection} includes the configuration to which we should assign the $O5$-plane factor. 
The left diagram in Figure \ref{fig:reflection} is a strip diagram obtained by reflecting the right half of the 5-brane web while the right diagram in Figure \ref{fig:reflection} is obtained by reflecting the left half. 
Naively, these three should be basically equivalent. However, we need to be careful about the framing factor when we assign Young diagrams also to the external lines. 

In order to see this point, we first compare the partition function for the left strip and the right strip in Figure \ref{fig:reflection}. The partition function for the left strip is given by
\begin{align}\label{eq:Zleft}
Z_{\text{left}} = \sum_{ \boldsymbol{\lambda} } C_{  \boldsymbol{\lambda} \boldsymbol{\mu} \boldsymbol{\nu} } C_{  \boldsymbol{\lambda}^T \boldsymbol{\rho}  \boldsymbol{\sigma} } E_{ \boldsymbol{\lambda}} (Q_1 Q_2, n),
\end{align}
where $n=p_1 q_2 + p_2 q_1$. 
Here, we apply the identity 
\begin{align}\label{eq:Ctr}
C_{\mathbf{X} \mathbf{Y} \mathbf{Z}}
= g^{- \frac{1}{2} ( \| \mathbf{X}^T \|^2 - \| \mathbf{X} \|^2 ) - \frac{1}{2} ( \| \mathbf{Y}^T \|^2 - \| \mathbf{Y} \|^2 ) - \frac{1}{2} ( \| \mathbf{Z}^T \|^2 - \| \mathbf{Z} \|^2 ) } C_{\mathbf{Y}^T \mathbf{X}^T \mathbf{Z}^T}
\end{align}
to the vertex factors in \eqref{eq:Zleft} to obtain
\begin{align}
Z_{\text{left}} = &g^{
- \frac{1}{2} ( \|  \boldsymbol{\mu}^T \|^2 - \|  \boldsymbol{\mu} \|^2 ) 
- \frac{1}{2} ( \| \boldsymbol{\nu} ^T \|^2 - \| \boldsymbol{\nu}  \|^2 ) 
- \frac{1}{2} ( \| \boldsymbol{\rho}^T \|^2 - \| \boldsymbol{\rho} \|^2 ) 
- \frac{1}{2} ( \| \boldsymbol{\sigma}^T \|^2 - \| \boldsymbol{\sigma} \|^2 ) 
} 
\cr
& \sum_{ \boldsymbol{\lambda} } C_{  \boldsymbol{\mu}^T \boldsymbol{\lambda}^T \boldsymbol{\nu}^T } C_{  \boldsymbol{\rho}^T \boldsymbol{\lambda}  \boldsymbol{\sigma}^T } E_{ \boldsymbol{\lambda}} (Q_1 Q_2, n).
\end{align}
Since the partition function for the right strip is 
\begin{align}
Z_{\text{right}} = \sum_{ \boldsymbol{\lambda} } C_{  \boldsymbol{\mu}^T \boldsymbol{\lambda}^T \boldsymbol{\nu}^T } C_{  \boldsymbol{\rho}^T \boldsymbol{\lambda}  \boldsymbol{\sigma}^T } E_{ \boldsymbol{\lambda}} (Q_1 Q_2, n),
\end{align}
we find the relation
\begin{align}
Z_{\text{left}} = g^{
- \frac{1}{2} ( \|  \boldsymbol{\mu}^T \|^2 - \|  \boldsymbol{\mu} \|^2 ) 
- \frac{1}{2} ( \| \boldsymbol{\nu} ^T \|^2 - \| \boldsymbol{\nu}  \|^2 ) 
- \frac{1}{2} ( \| \boldsymbol{\rho}^T \|^2 - \| \boldsymbol{\rho} \|^2 ) 
- \frac{1}{2} ( \| \boldsymbol{\sigma}^T \|^2 - \| \boldsymbol{\sigma} \|^2 ) 
} 
Z_{\text{right}}. 
\end{align}
This indicates that the framing factors are multiplied when the corresponding external lines are reflected.
Therefore, we should impose that the partition function for the central web diagram in Figure \ref{fig:reflection} is related to the left one and the right one as
\begin{align}\label{eq:lcr}
Z_{\text{center}} =  g^{
 \frac{1}{2} ( \| \boldsymbol{\rho}^T \|^2 - \| \boldsymbol{\rho} \|^2 ) 
+ \frac{1}{2} ( \| \boldsymbol{\sigma}^T \|^2 - \| \boldsymbol{\sigma} \|^2 ) 
}  Z_{\text{left}} 
= g^{
- \frac{1}{2} ( \|  \boldsymbol{\mu}^T \|^2 - \|  \boldsymbol{\mu} \|^2 ) 
- \frac{1}{2} ( \| \boldsymbol{\nu} ^T \|^2 - \| \boldsymbol{\nu}  \|^2 ) 
} Z_{\text{right}} 
\end{align}

Here, we consider the partition function for the web diagram at the center in Figure \ref{fig:reflection}.
Suppose that we do not know the rule \eqref{eq:O5factor} in advance, we introduce a factor corresponding to the $(1,-1)$ 5-brane and $(-1,-1)$ 5-brane intersecting with each other on the $O5$-plane as 
$I_{\boldsymbol{\lambda}, \boldsymbol{\lambda}'} (Q_1 ,Q_2)$, which is unknown at this stage. 
Here, we assign different Young diagrams $\boldsymbol{\lambda}$ $\boldsymbol{\lambda}'$ to the $(1,-1)$ 5-brane and $(-1,-1)$ 5-brane, respectively, in order to keep the generality.
The idea of topological vertex formalism is that the whole partition function is obtained by gluing the contributions from the local geometries. 
Therefore, we impose that this factor depends only on the local structure and is not affected by the Young diagrams which are away from the $O5$-plane.
That is, $I_{\boldsymbol{\lambda}, \boldsymbol{\lambda}'} (Q_1 ,Q_2)$ does not depend on $ \boldsymbol{\mu}, \boldsymbol{\nu},  \boldsymbol{\rho},  \boldsymbol{\sigma}$. 
By using this factor the partition function is given in the form
\begin{align}
Z_{\text{center}} =  \sum_{\boldsymbol{\lambda}, \boldsymbol{\lambda}'} C_{  \boldsymbol{\lambda} \boldsymbol{\mu} \boldsymbol{\nu} }  C_{  \boldsymbol{\rho}^T \boldsymbol{\lambda}  \boldsymbol{\sigma}^T } I_{\boldsymbol{\lambda}, \boldsymbol{\lambda}'} (Q_1 ,Q_2).
\end{align}
Or, if we use the identity \eqref{eq:Ctr} to the second vertex factor $ C_{  \boldsymbol{\rho}^T \boldsymbol{\lambda}  \boldsymbol{\sigma}^T } $,
\begin{align}\label{eq:Zcenter}
Z_{\text{center}} =  g^{
 \frac{1}{2} ( \| \boldsymbol{\rho}^T \|^2 - \| \boldsymbol{\rho} \|^2 ) 
+ \frac{1}{2} ( \| \boldsymbol{\sigma}^T \|^2 - \| \boldsymbol{\sigma} \|^2 ) 
}  \sum_{\boldsymbol{\lambda}, \boldsymbol{\lambda}'} C_{  \boldsymbol{\lambda} \boldsymbol{\mu} \boldsymbol{\nu} }  C_{  \boldsymbol{\lambda}^T \boldsymbol{\rho}  \boldsymbol{\sigma} }  
g^{
 \frac{1}{2} ( \| \boldsymbol{\lambda}^T \|^2 - \| \boldsymbol{\lambda} \|^2 ) 
} I_{\boldsymbol{\lambda}, \boldsymbol{\lambda}'} (Q_1 ,Q_2),
\end{align}
where we also used the cyclic symmetry of the topological vertex.

Substituting \eqref{eq:Zleft} and \eqref{eq:Zcenter} into \eqref{eq:lcr}, we obtain the condition
\begin{align}\label{eq:lc-comp}
 \sum_{\boldsymbol{\lambda}, \boldsymbol{\lambda}'} C_{  \boldsymbol{\lambda} \boldsymbol{\mu} \boldsymbol{\nu} }  C_{  \boldsymbol{\rho}^T \boldsymbol{\lambda}  \boldsymbol{\sigma}^T } 
 \left( \delta_{\boldsymbol{\lambda}, \boldsymbol{\lambda}'} E_{ \boldsymbol{\lambda}} (Q_1 Q_2, n) - g^{
 \frac{1}{2} ( \| \boldsymbol{\lambda}^T \|^2 - \| \boldsymbol{\lambda} \|^2 ) 
} I_{\boldsymbol{\lambda}, \boldsymbol{\lambda}'} (Q_1 ,Q_2) \right) = 0.
\end{align}
The unknown factor $I_{\boldsymbol{\lambda}, \boldsymbol{\lambda}'} (Q_1 ,Q_2)$ is determined by imposing that this is satisfied for arbitrary external Young diagrams $ \boldsymbol{\mu}, \boldsymbol{\nu},  \boldsymbol{\rho},  \boldsymbol{\sigma}$. 

Here, we note that if 
\begin{align}\label{eq:CX=0}
\sum_{\boldsymbol{\lambda}} C_{ \boldsymbol{\lambda} \boldsymbol{\mu} \boldsymbol{\nu}} X_{\boldsymbol{\lambda}} = 0
\end{align}
is satisfied for arbitrary $\boldsymbol{\mu}$ and $\boldsymbol{\nu}$, where $X_{\boldsymbol{\lambda}} $ is an arbitrary factor which depends on $\boldsymbol{\lambda}$ 
but does not depend on $\boldsymbol{\mu}$ and $\boldsymbol{\nu}$, 
then 
\begin{align}\label{eq:Xis0}
X_{\boldsymbol{\lambda}} = 0
\end{align}
is satisfied for all $\boldsymbol{\lambda}$. 
By applying this to \eqref{eq:lc-comp} repeatedly, we conclude that 
\begin{align}
\delta_{\boldsymbol{\lambda}, \boldsymbol{\lambda}'} E_{ \boldsymbol{\lambda}} (Q_1 Q_2, n) - g^{
 \frac{1}{2} ( \| \boldsymbol{\lambda}^T \|^2 - \| \boldsymbol{\lambda} \|^2 ) 
} I_{\boldsymbol{\lambda}, \boldsymbol{\lambda}'} (Q_1 ,Q_2) = 0
\end{align}
have to be satisfied. 
Taking into account that the edge factor is given in \eqref{eq:edgefactor}, 
we find that the $O5$-plane factor  is given as
\begin{align}
I_{\boldsymbol{\lambda}, \boldsymbol{\lambda}'} (Q_1 ,Q_2) 
= \delta_{\boldsymbol{\lambda}, \boldsymbol{\lambda}'} (Q_1 Q_2)^{|\boldsymbol{\lambda}|} (-1)^{n' |\boldsymbol{\lambda}|} g^{\frac{n'}{2}(||\boldsymbol{\lambda}^T||^2 - ||\boldsymbol{\lambda}||^2)},
\end{align}
where $n'=n+1=p_1 q_2 + p_2 q_1 +1$.
This is exactly the rule for the $O5$-plane factor  given in \eqref{eq:O5factor}, 
where $\boldsymbol{\lambda} = \boldsymbol{\lambda}'$ was imposed from the beginning. 



\subsection{Partition function for 5d $\mathcal{N}=1$ $Sp(N)$ gauge theory with $N_f$ flavors}

\begin{figure}
\centering
\includegraphics[width=15cm]{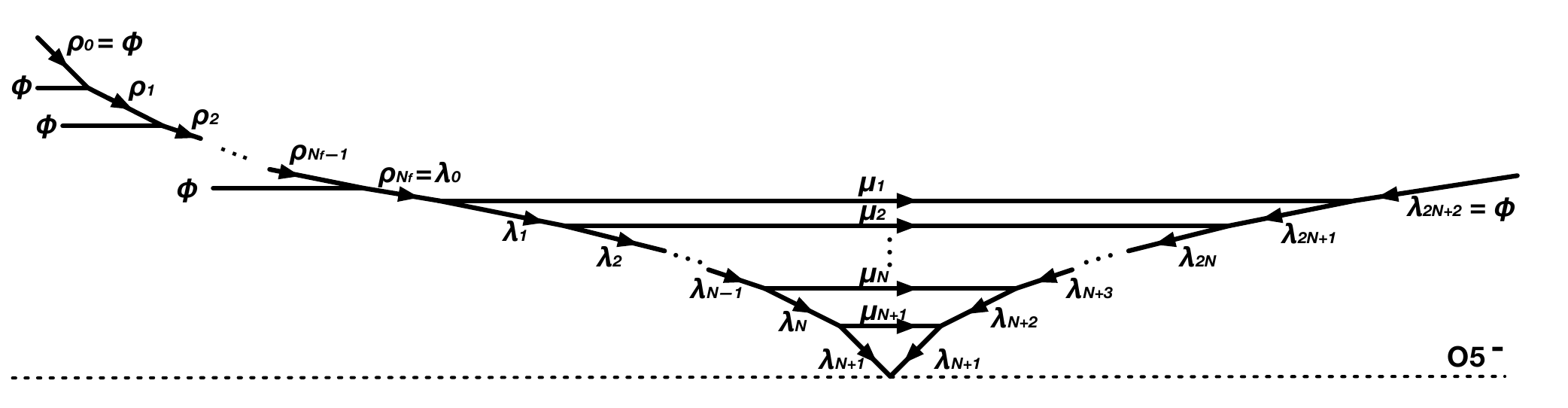}
\caption{Assingment of Young diagrams}
\label{fig:assign}
\end{figure}

In this subsection, we apply the previous rules to the 5-brane web diagram
corresponding to the 5d $\mathcal{N}=1$ $Sp(N)$ gauge theory with $N_f$ flavors.

We assign Young diagrams $ \boldsymbol{\rho} _i$, $ \boldsymbol{\lambda} _I$ and $ \boldsymbol{\mu} _I$ as depicted in Figure \ref{fig:assign}.
We define
\begin{align}
 \boldsymbol{\rho} _0 = \varnothing, \qquad
 \boldsymbol{\rho} _{N_f} =  \boldsymbol{\lambda} _0, \qquad
 \boldsymbol{\lambda} _{2N+2} = \varnothing, \qquad
\end{align}
for later convenience.
We denote the K\"ahler parameters $Q_{M,i}$ ($i=1,\cdots, N_f$) , $Q_{F,I}$ ($I=1,\cdots, 2N+2$) and $Q_{B,I}$ ($I=1,\cdots, N+1$)
to the edges where Young diagrams $ \boldsymbol{\rho} _i$, $ \boldsymbol{\lambda} _I$ and $ \boldsymbol{\mu} _I$ are assigned, respectively.
Explicitly in terms of gauge theory parameters,
\begin{align}\label{eq:Kahlerparameter}
&Q_{M,i} = e^{ - \beta (m_i - m_{i+1})} \quad (i=1,2,\cdots, N_f-1),
\cr
&Q_{M,N_f} = Q_{F,0} = e^{ - \beta (m_{N_f} - a_{1})}
\cr
& Q_{F, I} = Q_{F, 2N+2-I} =e^{ - \beta (a_{I} - a_{I+1})} \quad (I=1,2,\cdots, N),
\cr
& Q_{F, N+1}  =e^{ - \beta a_{N+1}}
\cr
& Q_{B, I} = e^{ - 2 \beta \left( (N-I + 2) a_I - \sum_{J=I+1}^{N+1} a_J \right)} \quad (I=1,2,\cdots, N+1)
\end{align}
as discussed in Appendix \ref{app:Kahler}. We note that the relation
\begin{align}\label{eq:AN1A}
\sum_{I=1}^{N+1} a_I + m_0 - \frac{1}{2} \sum_{i=1}^{N_f} m_j = 0
\end{align}
is also derived in Appendix \ref{app:Kahler}, which reproduces \eqref{eq:AN1}.

By applying the rules discussed in section \ref{sec:top-rule} to the web diagram in Figure \ref{fig:assign}, we obtain
\begin{align}\label{eq:Zfirst}
Z_0 =
&\sum_{\{  \boldsymbol{\rho}  \} \{  \boldsymbol{\lambda} \} \{ \boldsymbol{\mu} \} }
\left( \prod_{i=1}^{N_f} C_{ \boldsymbol{\rho} _{i-1}^T  \boldsymbol{\rho} _{i} \varnothing} \right)
\left( \prod_{I=1}^{N+1} C_{  \boldsymbol{\lambda} _{I}  \boldsymbol{\lambda} _{I-1}^T  \boldsymbol{\mu} _I} \right)
\left( \prod_{I=N+2}^{2N+2} C_{  \boldsymbol{\lambda} _{I}^T  \boldsymbol{\lambda} _{I-1}  \boldsymbol{\mu} _{2N+3-I}^T} \right)
\cr
& \quad \times
\left( \prod_{i=1}^{N_f-1} E_{ \boldsymbol{\rho} _i} (Q_{M, i};1) \right)
\left( E_{ \boldsymbol{\rho} _{N_f}} (Q_{M, N_f};0)  \right)
\left( \prod_{I=1}^{N} E_{ \boldsymbol{\lambda} _I} (Q_{F, I}; -1) \right)
\cr
& \quad \times
\left( \prod_{I=N+2}^{2N+1} E_{ \boldsymbol{\lambda} _I}(Q_F; 1) \right)
\left( \prod_{I=1}^{N+1} E_{ \boldsymbol{\mu} _I}(Q_B; 2N -2I +5 )  \right)
\cr
& \quad \times
\left( I_{ \boldsymbol{\lambda} _{N+1}} ( Q_{F, N+1}, Q_{F, N+1}; 1) \right).
\end{align}
The first line is the vertex factors,
the second and the third lines are the edge factors,
and the last line is the $O5$-plane factor.

\begin{figure}
\centering
\includegraphics[width=13cm]{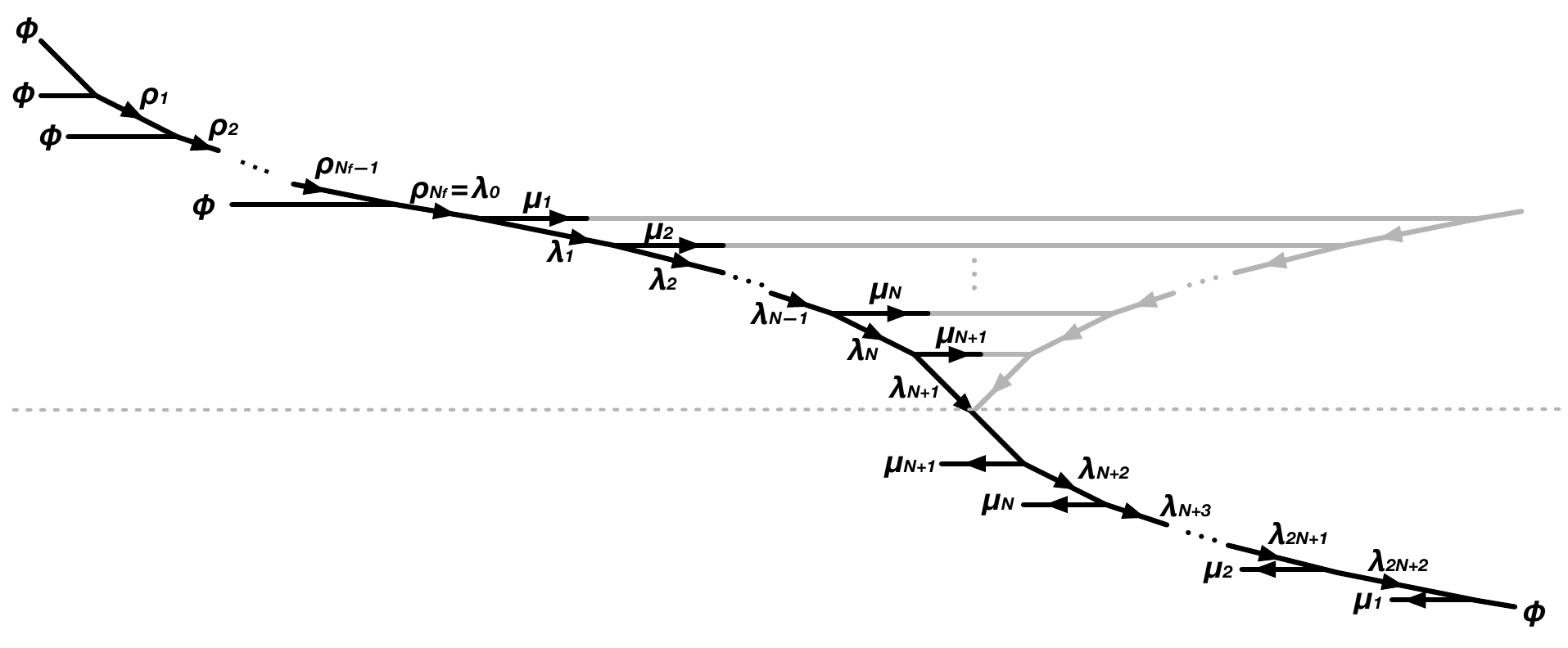}
\caption{Strip diagram obtained by reflecting the right half of the 5-brane web diagram.}
\label{fig:strip}
\end{figure}

Here, we apply 
the identity \eqref{eq:Ctr}
to the third factor in the first line
by identifying $\mathbf{X} =  \boldsymbol{\lambda} _i^T$, $\mathbf{Y}= \boldsymbol{\lambda} _{i-1}$, $\mathbf{Z}= \boldsymbol{\mu} _{2N+3-i}^T$ .
We also rewrite the $O5$-plane factors by using the edge factor function defined in \eqref{eq:edgefactor} as
\begin{align}\label{eq:IE}
I_{\boldsymbol{\lambda}}(Q, Q'; n') = g^{\frac{1}{2}(||\boldsymbol{\lambda}^T||^2 - ||\boldsymbol{\lambda}||^2)} E_{\boldsymbol{\lambda}} ( Q Q'; n'-1).
\end{align}
Then, \eqref{eq:Zfirst} can be rewritten as
\begin{align}\label{eq:Zfromstrip}
Z_0 =
&\sum_{ \{ \boldsymbol{\mu} \} }
\left( \prod_{I=1}^{N+1} E_{ \boldsymbol{\mu} _I} (Q_{B_I}, 2N -2I + 6) \right)
Z_{\text{strip}} (\{  \boldsymbol{\mu}  \})
\end{align}
where
\begin{align}\label{eq:strip-mu}
Z_{\text{strip}} (\{  \boldsymbol{\mu}  \}) =
&\sum_{\{  \boldsymbol{\rho}  \} \{  \boldsymbol{\lambda} \} } \left( \prod_{i=1}^{N_f} C_{ \boldsymbol{\rho} _{i-1}^T  \boldsymbol{\rho} _{i} \varnothing} \right)
\left( \prod_{I=1}^{N+1} C_{  \boldsymbol{\lambda} _{i}  \boldsymbol{\lambda} _{i-1}^T  \boldsymbol{\mu} _i} \right)
\left( \prod_{I=N+2}^{2N+2} C_{  \boldsymbol{\lambda} _{i}^T  \boldsymbol{\lambda} _{i-1}  \boldsymbol{\mu} _{2N+3-i}^T} \right)
\cr
& \quad \times
\left( \prod_{i=1}^{N_f-1} E_{ \boldsymbol{\rho} _i} (Q_{M, i};1) \right)
\left( E_{ \boldsymbol{\rho} _{N_f}} (Q_{M, N_f};0)  \right)
\left( \prod_{I=1}^{N} E_{ \boldsymbol{\lambda} _I} (Q_{F, I}; -1) \right)
\cr
& \quad \times
\left( \prod_{I=N+2}^{2N+1} E_{ \boldsymbol{\lambda} _I}(Q_{F,I}; 1) \right)
\left( E_{ \boldsymbol{\lambda} _{N+1}} ( Q_{F, N+1}{}^2; 0) \right).
\end{align}
This can be identified as strip diagram depicted in Figure \ref{fig:strip}.
Indeed, if we apply the rules discussed in section \ref{sec:top-rule} to the diagram in Figure \ref{fig:strip},
we reproduce \eqref{eq:strip-mu}.
Therefore, we can interpret that using the identity \eqref{eq:Ctr} with \eqref{eq:IE}
corresponds to the reflection the right half of the 5-brane web diagram.

\begin{figure}
\centering
\includegraphics[width=8cm]{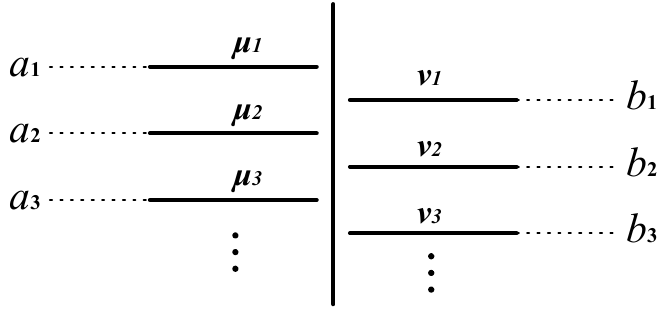}
\caption{A generic strip diagram. The central $(p,1)$ 5-brane is simply depicted as a vertical line for simplicity. $N_L$ D5-branes and $N_R$ D5-branes are attached to this
$(p,1)$ 5-brane from the left and from the right, respectively.}
\label{fig:generalstrip}
\end{figure}

The topological string amplitude for any strip diagram can be computed by following \cite{Iqbal:2004ne}.
Here, we consider a generic strip diagram depicted in Figure \ref{fig:generalstrip}.
In this diagram, $N_L$ D5-branes and $N_R$ D5-branes are attached to the central $(p,1)$ 5-brane from the left and from the right, respectively.
We denote the height of the left D5-branes as $a_i$ ($i=1,2,\cdots, N_L$) while the height of the right D5-branes as $b_j$ ($j=1,2,\cdots, N_R$).
We impose $a_1 > a_2 > a_3 > \cdots > a_{N_L}$ and $b_1 > b_2 > b_3 > \cdots > b_{N_R}$ while we do not impose any inequality relation between $a_i$ and $b_j$.
The Young diagrams $ \boldsymbol{\mu} _i$ are assigned to the left D5-branes while $ \boldsymbol{\nu} _j$ are assigned to the right D5-branes.
In order to proceed, we introduce some necessary building blocks for this expression:
\begin{align}\label{eq:defofR}
R_{ \boldsymbol{\lambda}  \boldsymbol{\mu} } (Q)
:= \prod_{i=1}^{\infty} \prod_{j=1}^{\infty} \left(1 - Q g^{i+j - \lambda_i - \mu_j - 1} \right), \quad \text{where} \quad g=e^{-\beta \hbar}.
\end{align}
In particular, $Q=e^{-\beta x}$ for some $x$ is the exponentiated length of some 5-brane corresponding to the K\"ahler parameter of toric Calabi-Yau geometry.
It is easy to observe that $R_{ \boldsymbol{\lambda}  \boldsymbol{\mu} } (Q) = R_{ \boldsymbol{\mu}  \boldsymbol{\lambda} } (Q)$ is satisfied.
Then, the amplitude for the strip diagram given in Figure \ref{fig:generalstrip} is given as follows:
\begin{align}
Z_{\text{strip}} =
& \prod_{i} g^{\frac{\|  \boldsymbol{\mu} _i^T \|^2}{2}} \tilde{Z}_{ \boldsymbol{\mu} _i}
\prod_{j} g^{\frac{\|  \boldsymbol{\nu} _j \|^2}{2}} \tilde{Z}_{ \boldsymbol{\nu} _j}
\cr
&\prod_{i<j} \left( R_{ \boldsymbol{\mu} _i  \boldsymbol{\mu} _j^T} ( e^{- \beta (a_i - a_j)} ) \right)^{-1}
\prod_{i<j} \left( R_{ \boldsymbol{\nu} _i  \boldsymbol{\nu} _j^T}  (e^{- \beta (b_i - b_j)} ) \right) {}^{-1}
\prod_{i, j} X_{ \boldsymbol{\mu} _i,  \boldsymbol{\nu} _j}  \left( a_i,  b_j \right)
\end{align}
with
\begin{align}
X_{ \boldsymbol{\mu} ,  \boldsymbol{\nu} } \left( a,  b  \right)
:= \left\{
\begin{array}{ll}
R_{ \boldsymbol{\mu}   \boldsymbol{\nu} ^T}\left( e^{- \beta (a - b)} \right) & \text{if }  a > b  \\
R_{ \boldsymbol{\nu}   \boldsymbol{\mu} {}^T} \left( e^{- \beta (b - a)} \right) & \text{if }  a < b \\
R_{ \boldsymbol{\mu}   \boldsymbol{\mu} {}^T} ( 1 ) \delta_{ \boldsymbol{\mu}   \boldsymbol{\nu} } & \text{if }  a = b \
\end{array}
\right. .
\end{align}
This can be shown by explicit computation for fixed ordering of the parameters, for example,
$a_1 > b_1 > a_2 > b_2 > \cdots$, and by using the flop invariance of the topological vertex \cite{Konishi:2006ev, Taki:2008hb}.

Applying this formula to the strip diagram given in Figure \ref{fig:strip}, we find that \eqref{eq:strip-mu} can be explicitly computed to give
\begin{align}\label{eq:strip-amp-O5}
Z_{\text{strip}} (\{  \boldsymbol{\mu}  \})
= & \prod_{I=1}^{N+1} \left( g^{ \|  \boldsymbol{\mu} _I \|^2} \tilde{Z}_{ \boldsymbol{\mu} _I}{}^2 \right)
\prod_{1 \le i<j \le N_f} \left( R_{\varnothing \varnothing} ( e^{- \beta (m_i - m_j)} ) \right)^{-1}
\cr
&
\prod_{1 \le I < J \le N+1} \left( R_{ \boldsymbol{\mu} _I  \boldsymbol{\mu} _J^T}  (e^{- \beta (a_I - a_J)} ) \right) {}^{-2}
\prod_{I=1}^{N+1} \prod_{J=1}^{N+1}  R_{ \boldsymbol{\mu} _I  \boldsymbol{\mu} _J} ( e^{- \beta (a_I + a_J)} )
\cr
&
\prod_{i=1}^{N_f} \prod_{I=1}^{N+1} \left( R_{\varnothing  \boldsymbol{\mu} _I^T} ( e^{- \beta (m_i - a_I)} ) \right)
\left( R_{\varnothing  \boldsymbol{\mu} _I} ( e^{- \beta (m_i + a_I)} ) \right)^{-1}.
\end{align}
It has been discussed in various contexts \cite{Konishi:2006ya, Bergman:2013ala, Bergman:2013aca, Bao:2013pwa, Hayashi:2013qwa, Hwang:2014uwa, Kim:2015jba} that the ``extra factor'' $Z_{\text{extra}}$ should be removed from the naive partition function $Z_0$,
which is \eqref{eq:Zfromstrip} in our case, in order to obtain the correct partition function
\begin{align}\label{eq:removeextra}
Z = \frac{Z_0}{Z_{\text{extra}}}.
\end{align}
The extra factor is the part which does not depend on the Coulomb moduli $a_I$.
Therefore, even though it is not always straightforward to obtain the extra factor exactly,
this subtlety does not affect our computation as long as we consider the Coulomb moduli dependent part
\begin{align}\label{eq:CBdependent}
\frac{\partial}{\partial a_I} \log Z = \frac{\partial}{\partial a_I} \log Z_0.
\end{align}
In the following, we concentrate only on such part and
treat the partition function up to the factor independent of the Coulomb moduli.
We consider the following partition function obtained from \eqref{eq:Zfromstrip} with \eqref{eq:strip-amp-O5} as
\begin{align}\label{eq:ZC}
Z
=
\, C \, & \sum_{ \{ \boldsymbol{\mu} \} }
\left( \prod_{I=1}^{N+1} E_{ \boldsymbol{\mu} _I} (Q_{B_I}, 2N -2I + 6) g^{ \|  \boldsymbol{\mu} _I \|^2} \tilde{Z}_{ \boldsymbol{\mu} _I}{}^2 \right)
\cr
&
\qquad \left( \prod_{1 \le I < J \le N+1} \left( R_{ \boldsymbol{\mu} _I  \boldsymbol{\mu} _J^T}  (e^{- \beta (a_I - a_J)} ) \right) {}^{-2} \right)
\left( \prod_{I=1}^{N+1} \prod_{J=1}^{N+1}  R_{ \boldsymbol{\mu} _I  \boldsymbol{\mu} _J} ( e^{- \beta (a_I + a_J)} ) \right)
\cr
&
\qquad \left( \prod_{i=1}^{N_f} \prod_{I=1}^{N+1} \left( R_{\varnothing  \boldsymbol{\mu} _I^T} ( e^{- \beta (m_i - a_I)} ) \right)
\left( R_{\varnothing  \boldsymbol{\mu} _I} ( e^{- \beta (m_i + a_I)} ) \right)^{-1} \right).
\end{align}
where $C$ is a factor which does not depend on the Coulomb moduli.
The factor $\left( R_{\varnothing \varnothing} ( e^{- \beta (m_i - m_j)} ) \right)^{-1}$
in the first line in \eqref{eq:strip-amp-O5} is also included in this $C$.

In order to further simplify this expression, we introduce the following notations:
First, we define $a_I$ with $N+2 \le I \le 2N+2$ as
\begin{align}\label{eq:aI2N}
a_{I}:= - a_{2N+3-I}.
\end{align}
Also, we define the Young diagram $ \boldsymbol{\mu} _I$ with $N+2 \le I \le 2N+2$ as
\begin{align}\label{eq:mu2N}
 \boldsymbol{\mu} _I :=  \boldsymbol{\mu} _{2N+3-I}^T.
\end{align}
Finally, we introduce
\begin{align}\label{eq:cI2N}
c_I = \left\{
\begin{array}{ll}
1 & (1 \le I \le N+1) \\
-1 & (N+2 \le I \le 2N+2) .
\end{array}
\right.
\end{align}
In terms of these notations, the partition function \eqref{eq:ZC} simplifies as
\begin{align}\label{eq:Zsimp}
Z
= &
\, C \, \sum_{ \{ \boldsymbol{\mu} \} }
\left( \prod_{I=1}^{2N+2} E_{ \boldsymbol{\mu} _I} (Q_{B_I}, 2N -2I + 6) g^{ \|  \boldsymbol{\mu} _I \|^2} \tilde{Z}_{ \boldsymbol{\mu} _I}{}^2  \right)
\cr
&
\left( \prod_{1 \le I < J \le 2N+2} \left( R_{ \boldsymbol{\mu} _I  \boldsymbol{\mu} _J^T}  (e^{- \beta (a_I - a_J)} ) \right)^{-c_I c_J} \right)
\left( \prod_{i=1}^{N_f} \prod_{I=1}^{2N+2} \left( R_{\varnothing  \boldsymbol{\mu} _I^T} ( e^{- \beta (m_i - a_I)} ) \right)^{c_I} \right).
\end{align}

For later convenience, we further symmetrize this expression by using the following identity
\begin{align}
\frac{ R_{ \boldsymbol{\lambda}   \boldsymbol{\mu} ^T} (Q) }{ R_{\varnothing \varnothing} (Q)}
=  E_{ \boldsymbol{\lambda} }(-Q,1) E_{ \boldsymbol{\mu} ^T}(-Q,1) \frac{ R_{ \boldsymbol{\mu}   \boldsymbol{\lambda} ^T} (Q^{-1}) }{ R_{\varnothing \varnothing} (Q^{-1})},
\end{align}
which plays a key role in discussing the flop invariance \cite{Konishi:2006ev, Taki:2008hb}.
We use this identity to ``half'' of the factor $R_{ \boldsymbol{\lambda}   \boldsymbol{\mu} ^T} (Q)$, or equivalently, use the following identity
\begin{align}
R_{ \boldsymbol{\lambda}   \boldsymbol{\mu} ^T} (Q)
=  \left( E_{ \boldsymbol{\lambda} }(-Q,1) E_{ \boldsymbol{\mu} ^T}(-Q,1)
\frac{R_{\varnothing \varnothing} (Q) }{R_{\varnothing \varnothing} (Q^{-1}) } R_{ \boldsymbol{\lambda}   \boldsymbol{\mu} ^T} (Q) R_{ \boldsymbol{\mu}   \boldsymbol{\lambda} ^T} (Q^{-1})
\right)^{\frac{1}{2}}.
\end{align}
We also rewrite the factor $\tilde{Z}_{ \boldsymbol{\lambda} }$ by using the identity
\begin{align}
\tilde{Z}_{ \boldsymbol{\lambda} }{}^2 = (-1)^{| \boldsymbol{\lambda} |} e^{ \frac{1}{2} \beta \hbar ( \|  \boldsymbol{\lambda} ^T \|^2 + \|  \boldsymbol{\lambda}  \|^2)} R_{\varnothing\varnothing} (1) R_{ \boldsymbol{\lambda} ^T  \boldsymbol{\lambda} } (1){}^{-1}.
\end{align}
After a straightforward computation, we find that the partition function can be rewritten as
\begin{align}\label{eq:Zsym}
Z
= &
\, C' \, \prod_{1 \le I<J\le2N+2}
\left( \frac{R_{\varnothing \varnothing} (e^{-\beta (a_I - a_J)}) }{R_{\varnothing \varnothing} (e^{-\beta (- a_I + a_J)}) } \right)^{-\frac{1}{2} c_I c_J}
\cr
& \sum_{ \{ \boldsymbol{\mu} \} }
\left( \prod_{I=1}^{2N+2} e^{ - \beta (N+2) c_I \left(  a_I | \boldsymbol{\mu} _I| - \hbar \|  \boldsymbol{\mu} _I \|^2  \right) } \right)
\left( \prod_{I=1}^{2N+2} \prod_{J=1}^{2N+2} \left( R_{ \boldsymbol{\mu} _I  \boldsymbol{\mu} _J^T}  (e^{- \beta (a_I - a_J)} ) \right)^{- \frac{1}{2}c_I c_J} \right)
\cr
&
\qquad \left( \prod_{i=1}^{N_f} \prod_{I=1}^{2N+2} \left( R_{\varnothing  \boldsymbol{\mu} _I^T} ( e^{- \beta (m_i - a_I)} ) \right)^{c_I} \right),
\end{align}
where $C'=C R_{\varnothing \varnothing} (1){}^{N+1}$ is the prefactor independent of the Coulomb moduli.


\section{Deriving Seiberg-Witten prepotentials from
partition function}\label{sec:4}

The goal of this section is to take thermodynamic limit of 5d $\mathcal{N}=1$ gauge theory partition function and evaluate the Seiberg-Witten prepotential in the presence of $O5$-plane. There are four steps to achieve this goal: the first step is to rewrite the gauge theory partition function for 5d $\mathcal N=1$ $Sp(N)$ gauge theory with $N_f$ flavors in terms of profile function $f_{ \boldsymbol{\lambda} }(x)$ of random partition $ \boldsymbol{\lambda} $; the second step is to derive the saddle point equation which profile function should satisfy in the thermodynamic limit; the third step is to introduce resolvent and to evaluate the integrals of resolvent over non-trivial cycles, which is related to the prepotential by Legendre transformation;
finally, we derive the Seiberg-Witten curve and its boundary conditions.
We find that the results obtained in this section coincide with with the results in section \ref{sec:SW}.
The technique used in this section is based on \cite{Nekrasov:2003rj} and also motivated by related works \cite{Hollowood:2003cv, Nekrasov:2004vw, Shadchin:2004yx, Klemm:2008yu, Nekrasov:2012xe, Ishii:2013nba, Haghighat:2016jjf, Zhang:2019msw}.

\subsection{Profile function of partition diagram}

\begin{figure}
\centering
\includegraphics[width=8cm]{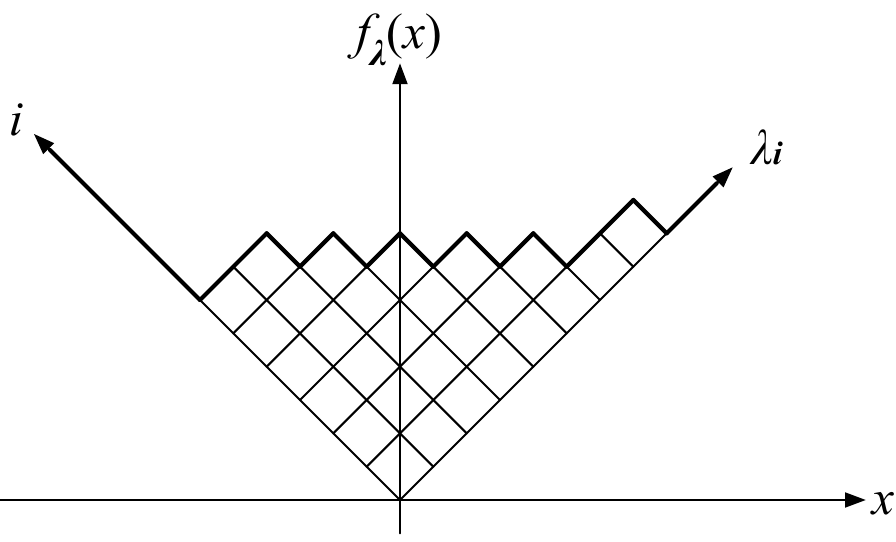}
\caption{Young diagram $\lambda$ and its corresponding profile function $f_{ \boldsymbol{\lambda}}(x)$.}
\label{fig:young diagram}
\end{figure}

Suppose a Young diagram $ \boldsymbol{\lambda}$ is given, which we depict in Russian style as Figure \ref{fig:young diagram}.
The profile function $f_{ \boldsymbol{\lambda} }(x)$ is a piecewise function with Lipschitz constant 1 corresponding to the upper boundary of the partition diagram $ \boldsymbol{\lambda} $. The precise expression for profile function is
$$
f_{ \boldsymbol{\lambda} }(x)= |x| + \sum_{i=1}^{\infty}\biggl[|x-\lambda_i + i - 1| - |x - \lambda_i + i| + |x + i| - |x + i -1|\biggr]
\quad (-\infty < x < \infty)
$$

The profile function $f_{  \boldsymbol{\lambda}  }(x)$ can be easily generalized to $f_{ \boldsymbol{\lambda} }(x|\epsilon_1, \epsilon_2)$ by adding two parameters $\epsilon_1$ and $\epsilon_2$ ($\epsilon_2 < 0 < \epsilon_1$) where $\epsilon_1$ and $-\epsilon_2$ are scaling constants for two axes respectively. By setting $\hbar=\epsilon_1 = -\epsilon_2$, the profile function $f_{ \boldsymbol{\lambda} }(x|\hbar): = f_{ \boldsymbol{\lambda} }(x|\hbar, -\hbar)$ can be simplified as
\begin{align}
\!\!\!\!\!\! f_{ \boldsymbol{\lambda} } (x | \hbar)
:= & |x|
+ \sum_{i=1}^{\infty} \biggl[ |x - \hbar (  i - 1 - \lambda_i)  |
- |x - \hbar(  i  -  \lambda_i ) |
- |x - \hbar (i - 1) |
+ |x - \hbar i |
\biggr]
\cr
= & |x|
+ \sum_{j=1}^{\infty} \biggl[
|x + \hbar ( j - 1 - \lambda^T_j )  |
- |x + \hbar ( j - \lambda^T_j ) |
- |x + \hbar (j - 1) |
+ |x + \hbar  j |
\biggr] \! .
\end{align}

The second derivative $f_{ \boldsymbol{\lambda}}'' (x|\hbar)$ of the profile function (i.e. a compactly supported distribution on a real line which behaves like a density function for large size partition) can be obtained as
\begin{align}\label{eq:profile-sec}
f_{ \boldsymbol{\lambda} }'' (x|\hbar)
&= 2 \delta (x)
+ 2 \sum_{i=1}^{\infty} \biggl[
\delta (x - \hbar (  i - 1 - \lambda_i)  )
- \delta ( x - \hbar(  i  -  \lambda_i ) )
\cr
& \qquad \qquad \qquad \qquad  \qquad \qquad   \qquad \qquad
- \delta ( x - \hbar (i - 1) )
+ \delta ( x - \hbar i )
\biggr]
\cr
&= 2 \delta (x)
+ 2 \sum_{j=1}^{\infty} \biggl[
\delta (x + \hbar ( j - 1 - \lambda^T_j )  )
- \delta (x + \hbar ( j - \lambda^T_j ) )
\cr
& \qquad \qquad  \qquad \qquad  \qquad \qquad  \qquad \qquad
- \delta ( x + \hbar (j - 1))
+ \delta ( x + \hbar  j )
\biggr].
\end{align}
It is also not hard to check that the second derivative $f_{ \boldsymbol{\lambda}}'' (x| \hbar)$ of profile function  satisfies the following identities
\begin{align}
&\int^{\infty}_{-\infty} f_{ \boldsymbol{\lambda} }'' (x| \hbar) dx = 2,
\label{eq:intprofile}
\\
&\int^{\infty}_{-\infty} x f_{ \boldsymbol{\lambda} }'' (x| \hbar) dx = 0,
\label{eq:intxprofile}
\\
&\int^{\infty}_{-\infty} x^2 f_{ \boldsymbol{\lambda} }'' (x| \hbar) dx = 4 \hbar^2 | \boldsymbol{\lambda} |,
\\
&\int^{\infty}_{-\infty} x^3 f_{ \boldsymbol{\lambda} }'' (x| \hbar) dx = 6 \hbar^3
\left(  || \boldsymbol{\lambda} ^T ||^2 - || \boldsymbol{\lambda}  ||^2 \right).
\end{align}
Here $| \boldsymbol{\lambda}|$ is the size of partition $ \boldsymbol{\lambda}$, and $|| \boldsymbol{\lambda} ||^2$ (or $|| \boldsymbol{\lambda}^T ||^2$ ) are defined in \eqref{eq:lambdasquare}.

The technique playing an important role in our paper is to rewrite the partition function \eqref{eq:Zsym} in terms of the profile function $f_{\lambda}(x|\hbar)$ with the help of the following key identity:
\begin{align}\label{eq:keyid}
R_{ \boldsymbol{\lambda}   \boldsymbol{\mu} } \bigl( e^{ -\beta (a-b)} \bigr)
= \exp \left [
\frac{1}{4} \int^{\infty}_{-\infty}\int^{\infty}_{-\infty}
f_{ \boldsymbol{\lambda} }''(x-a | \hbar) f_{ \boldsymbol{\mu} ^T}''(y-b | \hbar ) \gamma_{\hbar} (x-y) dx dy
\right]
\quad (a \neq b).
\end{align}
The function $\gamma_{\hbar} (x-y)$ is given in terms of Barnes double gamma function:
\begin{align}
\gamma_{\hbar} (x)
:= \sum_{n=1}^{\infty} \frac{1}{n} \frac{e^{-\beta n x}}{(1-e^{-\beta n\hbar})(1-e^{\beta n\hbar})}.
\end{align}
Interested readers can refer to Appendix \ref{app:proofkey} for details about proof of this key identity.
In order to rewrite the partition function in a concise way, we introduce
\begin{align}\label{eq:deffvec}
f_{\vec{ \boldsymbol{\mu} }} (x|\hbar) &:= \sum_{I=1}^{N+1} f_{ \boldsymbol{\mu}_I }(x-a_I|\hbar)
\cr
\tilde{f}_{\vec{ \boldsymbol{\mu} }} (x|\hbar) &:= f_{\vec{ \boldsymbol{\mu} }} (x|\hbar) - f_{\vec{ \boldsymbol{\mu} }} (-x|\hbar)
= \sum_{I=1}^{2N+2} c_I f_{ \boldsymbol{\mu}_I }(x-a_I|\hbar)
\end{align}
where we use the notation given in \eqref{eq:aI2N}, \eqref{eq:mu2N} and \eqref{eq:cI2N}.
Here, we tune $a_I$ to be real value.
Then, the full Nekrasov partition function 
for 5d $\N =1$ $Sp(N)$ gauge theory with $N_f (N_f \leq 2N + 3)$ flavors can be rewritten in terms of profile functions as
\begin{align}\label{eq:Z5dNek}
Z
&=
\exp \left [ \tilde{C}
 - \frac{1}{2} \sum_{1 \le I < J \le 2N+2} \sum_{s = \pm 1} c_I c_J s \gamma_{\hbar} \left( s(a_I-a_J) \right)
+ \frac{\beta (N+2)}{3 \hbar^2} \sum_{I=1}^{N+1} a_I{}^3
\right]
\cr
& \qquad \times \sum_{\vec{  \boldsymbol{\mu}  } }
\exp \left[
- \frac{1}{8} \int^{\infty}_{-\infty}
\tilde{f}_{\vec{ \boldsymbol{\mu} }}''(x|\hbar) \tilde{f}_{\vec{ \boldsymbol{\mu} }}''(y|\hbar) \gamma_{\hbar} (x-y) dx dy
\right.
\cr
& \left.
\qquad \qquad
- \frac{1}{4} \sum_{i=1}^{N_f} \int^{\infty}_{-\infty}
\tilde{f}_{\vec{ \boldsymbol{\mu} }}''(x) \gamma_{\hbar} (m_i-x|\hbar) dx
- \frac{\beta (N+2)}{12 \hbar^2} \int^{\infty}_{-\infty} x^3 \tilde{f}''_{\vec{ \boldsymbol{\mu} }} (x|\hbar) dx
\right],
\end{align}
where $\tilde{C}:=\log C'$.

\subsection{Thermodynamic limit and saddle point equation}

In this section, we take the thermodynamic limit of partition function and derive its Seiberg-Witten geometry (curve, differential and prepotential).
In the thermodynamic limit $\hbar = \epsilon_1 = - \epsilon_2\rightarrow 0$, the Seiberg-Witten prepotential $F$ can be extracted as the leading order contribution of the logarithm of the Nekrasov partition function. That is,
\begin{align}
F
:= \beta^2 \lim_{\hbar= \epsilon_1 = - \epsilon_2 \to 0} \epsilon_1 \epsilon_2 \log Z
\end{align}
or equivalently
\begin{align}\label{eq:prepdef2}
Z
= \exp \left[ - \frac{1}{\beta^2 \hbar^2} F
+ \mathcal{O} (\hbar^{-1}) \right].
\end{align}

In order to obtain the prepotential, we first expand the exponents of the expression in \eqref{eq:Z5dNek}
in terms of $\hbar$ in the following form:
\begin{align}\label{eq:summu}
Z
=
\sum_{\vec{ \boldsymbol{\mu} }}
\exp \left[ - \frac{1}{\beta^2 \hbar^2}
\left( \mathcal{E}[ f^{''}_{\vec{ \boldsymbol{\mu} }} (x) ]  +  {C}_0 \right)
+ \mathcal{O} (\hbar^{-1}) \right].
\end{align}
Here, $\mathcal{E}[ f^{''}_{\vec{ \boldsymbol{\mu} }} (x) ]$ is a functional of the second derivative of profile function $f^{''}_{\vec{ \boldsymbol{\mu} }} (x)$
\begin{align}
\mathcal{E}[ f^{''}_{\vec{ \boldsymbol{\mu} }} (x) ]
:=
&
- \frac{1}{8}
\int^{\infty}_{-\infty}
\int^{\infty}_{-\infty}
\tilde{f}''_{\vec{ \boldsymbol{\mu} }}(x)
\tilde{f}''_{\vec{ \boldsymbol{\mu} }}(y)
\mathrm{Li}_3 \left( e^{ -\beta (x - y)} \right)
dx dy
\cr
&
- \frac{1}{4} \sum_{i=1}^{N_f} \int^{\infty}_{-\infty}
\tilde{f}''_{\vec{ \boldsymbol{\mu} }}(x)
\mathrm{Li}_3 \left( e^{ -\beta (m_i - x)} \right) dx
+ \frac{\beta^3 (N+2)}{12} \int^{\infty}_{-\infty}
x^3 \tilde{f}''_{\vec{ \boldsymbol{\mu} }}(x) dx
\end{align}
and the constant $C_0$, which does not depend on $ f^{''}_{\vec{ \boldsymbol{\mu} }} (x) $, is
\begin{align}\label{eq:C0explicit}
{C}_0
&= \tilde{C}
- \beta^3 F_{\text{IMS}}
+ 2 \pi i \beta^2 \sum_{1 \le I < J \le N+1 } a_I a_J  +  \pi i \beta^2 \sum_{I=1}^{N+1} a_I{}^2
- \frac{\pi^2}{3} \beta \sum_{I=1}^{N+1} (2I-1)a_I,
\cr
\end{align}
where $F_{\text{IMS}}$ is the IMS prepotential for 5d $\mathcal{N}=1$ $Sp(N)$ gauge theory with $N_f$ flavors
\begin{align}
F_{\text{IMS}} = & m_0\sum_{I=1}^N a_I{}^2 + \frac{1}{6} \left( \sum_{1\le I<J \le N} ( |a_I-a_J|^3 + |a_I+a_J|^3 )+ \sum_{I=1}^N |2 a_I|^3 \right)
\cr
& \qquad - \frac{1}{6} \sum_{I=1}^N \sum_{i=1}^{N_f} \left( |a_I + m_i |^3 + | - a_I + m_i |^3 \right)
\end{align}
at the region $m_1 > m_2 > \cdots > m_{N_f} > a_1 > a_2 > \cdots > a_{N} > 0$.
We have used the expansion
\begin{align}
\gamma_{\hbar} (x) = - \frac{1}{\beta^2 \hbar^2} \mathrm{Li}_3 (e^{- \beta x}) + \mathcal{O}(\hbar^{-1})
\end{align}
as well as the identity
\begin{align}
\mathrm{Li}_3 (- z) + \mathrm{Li}_3 (- z^{-1}) =  - \frac{1}{6} ( \log z )^3 - \frac{1}{6} \pi^2 \log z.
\end{align}

Here, if we consider the full partition function,
the IMS prepotential included in $C_0$ in \eqref{eq:C0explicit} is cancelled by the cubic term in \eqref{eq:Zfull}.
Although the interpretation of the remaining terms in $C_0$ is not very clear,
we omit $C_0$ in the following discussion.
Since $C_0$ does not depend on the profile function, this omission does not affect the following discussion in any case.

%
%

When $\hbar$ is small enough, the profile function is approximated by a continuous function
\begin{align}\label{eq:defcontf}
\tilde{f}(x) := \lim_{\hbar \to 0} \tilde{f}_{\vec{ \boldsymbol{\mu} }} (x|\hbar),
\qquad
f(x) := \lim_{\hbar \to 0} f_{\vec{ \boldsymbol{\mu} }} (x|\hbar),
\end{align}
where we have removed the index $\vec{ \boldsymbol{\mu} }$ of the profile function.
Since the summand in \eqref{eq:summu} depends on Young diagrams $\boldsymbol{\mu}_I$ only through the profile functions $\tilde{f}_{\vec{\boldsymbol{\mu}}}''(x|\hbar)$,
the partition function, which is a statistical sum over partitions, can be approximated by path integral over the space of continuous functions $f''$.
That is
\begin{align}\label{eq:pathint}
Z
 \simeq \int D f'' \,\, d^{N+1} \xi \,\, d^{N+1} \zeta  \exp\left[ -\frac{1}{\beta^2 \hbar^2} \mathcal{S}[f''] (\xi,\zeta) + \mathcal{O}(\hbar^{-1}) \right].
\end{align}
In order to take into account the constraints on the profile function induced from \eqref{eq:intprofile} and \eqref{eq:intxprofile}, we
introduce the following auxiliary functional
\begin{align}
\mathcal{S}[f''] (\xi,\zeta)
: = \mathcal{E} [f''] + \sum_{I=1}^{N+1} \xi_I \left( \int_{\mathcal{C}_I} x f''(x) dx - 2 a_I \right)
+ \sum_{I=1}^{N+1} \zeta_I \left( \int_{\mathcal{C}_I} f''(x) dx - 2 \right) .
\end{align}
Here, $\mathcal{S}$ is a functional of $f''$ and a function of Lagrangian multipliers $\xi_I$ and $\zeta_I$ ($I=1, 2, \cdots, N+1$).

Now, it would be reasonable to discuss the issues on the measures on profile function.
When $\hbar$ is small enough, the dominant contributions in \eqref{eq:summu} come from the Young diagrams with sizes of partitions $|\boldsymbol{\mu}_I |$ are large in such a way that $\hbar^2 |\boldsymbol{\mu}_I |$ is of order 1.
This indicates that the dominant contribution in the path integral \eqref{eq:pathint} comes from the functions with finite and non-trivial configuration in the finite region around $x=a_I$.
Especially, in the thermodynamic limit $\hbar \to 0$, the measures on the profile function $f_{ \vec{\boldsymbol{\mu}} }(x)$ weakly converges to the Delta measure on $f$.
In other word, the profile function $f_{ \vec{\boldsymbol{\mu}} }(x)$ has a limit shape $f_{*}(x)$,
where $f_{*}(x)$ gives the critical point of the exponent in \eqref{eq:pathint}.

For each fixed constants $\{ a_I \}$, we denote the solution of the following (functional) equations to be $f_*(x)$, $\xi_{I *}$ and $\zeta_{I *}$:
\begin{align}
\frac{\delta \mathcal{S} }{\delta f''} = 0,
\quad
\frac{\partial \mathcal{S} }{\partial \xi_I} =0,
\quad
\frac{\partial \mathcal{S} }{\partial \zeta_I} =0.
\end{align}
Moreover, we denote $\tilde{f}_*(x):=f_*(x) - f_*(-x)$. With this solution, the leading order contribution of the partition function is simply given as
\begin{align}
Z
= \exp \left[ -\frac{1}{\beta^2 \hbar^2} \mathcal{S}[f_*''] (\xi_*, \zeta_*)  + \mathcal{O}(\hbar^{-1}) \right],
\end{align}
from which we read off the Seiberg-Witten prepotential defined in \eqref{eq:prepdef2} as
\begin{align}\label{eq:prepS}
F = \mathcal{S}[f_*''] (\xi_*, \zeta_*).
\end{align}

We denote the union of all distinct local compact supports of $\tilde{f}_* ''(x)$ as $\mathcal{C} = \bigcup_{I=1}^{2N+2} \mathcal{C}_I$, where each $\mathcal{C}_I$ ($I=1,2,\cdots, 2N+2$) is the closure of $\{x | \tilde{f}''_{*}(x) \neq 0 \}$. For convenience, we can use the endpoints $\alpha_I^{-}$ and $\alpha_I^{+}$ of $\mathcal{C}_I$ to represent the region $\mathcal{C}_I$ as $\mathcal{C}_I: \alpha_I^- \le  x \le \alpha_I^+$.
On the other hand, it is easy to observe that
\begin{align}
\tilde{f}_* ''(x) = 0 \quad \text{ for } \alpha_I^+ < x < \alpha_{I+1}^-.
\qquad (I=1,2,\cdots, 2N+1)
\end{align}
Here for $I=1, 2, \ldots, 2N+2$, $\alpha_I^+ =  \alpha_{2N+3-I}^-, \quad \alpha_I^- =  \alpha_{2N+3-I}^+$.

To evaluate the extremum $f''_*(x)$ of $\mathcal{S}[f''](\xi,\zeta)$, which is the limit shape for $f''_{\vec{\boldsymbol{\mu}}}(x|\hbar)$, we need to take functional derivatives (i.e. variations) of $\mathcal{S} [f''] (\xi,\zeta)$ with respect to $f''(z)$ for $z \in \mathcal{C}_I$ as well as $\xi_I$ and $\zeta_I$ ( $I=1, 2, \ldots 2N+2$ ) respectively:
\begin{align}
& - \frac{1}{4}
\sum_{s = \pm 1}  s
\int^{\infty}_{-\infty} \tilde{f}_* ''(x) \mathrm{Li}_3 \left( e^{ -\beta (x + s z)} \right) dx
- \frac{1}{2} \sum_{i=1}^{N_f} \sum_{s = \pm 1} s \mathrm{Li}_3 \left( e^{ -\beta (m_i + s z)} \right)
\cr
& \qquad \qquad \qquad \qquad \qquad \qquad \qquad \qquad \qquad \qquad
- \frac{\beta^3 (N+2)}{6} z^3
+ \xi_{I*} z + \zeta_{I*} = 0 \label{eq:crit}
\\
& \int_{\mathcal{C}_I} x \tilde{f}_*''(x) dx = 2 a_I
\label{eq:constf1}
\\
& \int_{\mathcal{C}_I} \tilde{f}_* ''(x) dx =2
\label{eq:constf2}
\end{align}

Differentiating \eqref{eq:crit} with respect to $z$ repeatedly, it reduces to
\begin{align}
& \frac{1}{4}
\sum_{s = \pm 1}
\int^{\infty}_{-\infty} \tilde{f}_* ''(x)
\mathrm{Li}_2 \left( e^{ -\beta (x + s z)} \right) dx
- \frac{1}{2} \sum_{i=1}^{N_f} \sum_{s = \pm 1} \mathrm{Li}_2 \left( e^{ -\beta (m_i + s z)} \right)
\cr
& \qquad \qquad \qquad \qquad \qquad \qquad \qquad \qquad \qquad \qquad
- \frac{(N+2) \beta^2}{2} z^2
+ \xi_{I*} = 0, \label{eq:seqderiv1}
\\
&
\frac{1}{4}
\sum_{s = \pm 1}  s
\int^{\infty}_{-\infty} \tilde{f}_* ''(x)
\log \left( 1 - e^{ -\beta (x + s z)} \right) dx
- \frac{1}{2} \sum_{i=1}^{N_f} \sum_{s = \pm 1} s \log \left( 1 - e^{ -\beta (m_i + s z)} \right)
\cr
& \qquad \qquad \qquad \qquad \qquad \qquad \qquad \qquad \qquad \qquad
- (N+2) \beta z = 0, \label{eq:seqderiv2}
\\
&
\frac{1}{4}
\sum_{s = \pm 1}
\int^{\infty}_{-\infty} \tilde{f}_* ''(x)
\frac{e^{ -\beta (x + s z)}}{1 - e^{ -\beta (x + s z)}} dx
- \frac{1}{2} \sum_{i=1}^{N_f} \sum_{s = \pm 1} \frac{e^{ -\beta (m_i + s z)}}{1 - e^{ -\beta (m_i + s z)}}
- (N+2) = 0.
\label{eq:fff}
\end{align}
Since
\begin{align}\label{eq:simpleodd}
\sum_{s = \pm 1}
\int^{\infty}_{-\infty} \tilde{f}_* ''(x)
\frac{e^{ -\beta (x + s z)}}{1 - e^{ -\beta (x + s z)}} dx
= 2 \int^{\infty}_{- \infty}
\frac{\tilde{f}_* ''(x) }{1 - e^{ -\beta ( x - z)}} d x
\end{align}
due to $\tilde{f}(-x) = \tilde{f}(x)$,
the last equation \eqref{eq:fff} can be simplified as
\begin{align}
\int^{\infty}_{- \infty}
\frac{\tilde{f}_* ''(x) }{1 - e^{ -\beta ( x - z)}} d x
= \sum_{i=1}^{N_f}\sum_{s=\pm 1} \frac{e^{-\beta(m_i+sz)}}{1- e^{-\beta(m_i+sz)} } + 2(N+2), \qquad z \in \mathcal{C}
\label{eq:criteq3}
\end{align}
where the integral should be understood as the principal value integral.


\subsection{Resolvent and its integrals over cycles}\label{sec:resolvent}

In this section, we will introduce a complex holomorphic
function $R^+(z)$ \footnote{The upper label $+$ is put because it corresponds to $\lambda_{SW}^+(w)$ defined in section \ref{sec:SW1form} as we see later. }
over $\mathbb C \backslash \mathcal{C}$ which is called ``Resolvent''. 
The resolvent $R^+(z)$ is defined as follows:
\begin{align}\label{eq:defresolvent}
R^+(z): = \int^{\infty}_{- \infty}
\frac{\tilde{f}_* ''(x) }{1 - e^{ -\beta ( x - z)}} d x
\qquad
z \in \mathbb{C}\backslash \mathcal{C}
\end{align}

The Resolvent $R^+(z)$ has the following properties:
\begin{itemize}
\item $R^+(z)$ is an even function satisfying $R^+(-z) = R^+(z)$ since $\tilde{f}''_*(x)$ is an odd function;
\item $R^+(z)$ is periodic in pure imaginary direction: $R^+ \left( z + \frac{2 \pi i}{\beta} \right) = R(z)$;
\item $R^+(z)$ satisfies  boundary conditions: $\displaystyle \lim_{z \to \infty} R^+(z) = \displaystyle \lim_{z \to - \infty} R^+(z) = 0$.
\item $R^+(z)$ is regular at $\mathbb C \backslash \mathcal{C}$ while discontinuous at $\mathcal{C}$.
\end{itemize}

In order to understand the last property more in detail, we consider the points just above or below $\mathcal{C}$, that is, $z \pm i\epsilon$ with $z \in \mathcal{C}$ and $\epsilon>0$.
We find that the resolvent $R^+(z)$ satisfies the following relations:
\begin{align}\label{eq:Rplusminus}
& \lim_{\epsilon\rightarrow 0+} R^+ (z \pm i \epsilon) = \sum_{i=1}^{N_f} \sum_{s=\pm1} \frac{e^{-\beta(m_i+sz)}}{1- e^{-\beta(m_i+sz)} } + 2(N+2)
 \mp \frac{\pi i \tilde{f}_*''(z)}{\beta}.
\end{align}
Here, we have used \eqref{eq:criteq3}, \eqref{eq:defresolvent} and the formula 
\begin{align}\label{eq:SokhotskiPlemelj}
\lim_{\epsilon \to 0+ } \int^{\infty}_{-\infty} \frac{\varphi(x)}{x \pm i \epsilon} dx = P.V. \int^{\infty}_{-\infty} \frac{\varphi(x)}{x} \mp \pi i \varphi(0),
\end{align}
where $P.V.$ denotes the principal value integral, which we omit whenever it is clear from the context.

Based on this relation, it would be natural to
decompose the resolvent $R^+(z)$ as the sum of regular part and singular part:
\begin{align}
R^+(z) = R_{\text{reg}} (z) + R_{\text{sing}} (z). 
\end{align}
Here, the regular part $R_{\text{reg}} (z)$ is the first two terms in \eqref{eq:Rplusminus}, which is continuous at $z \in \mathcal{C}$.
That is
\begin{align}\label{eq:defRreg}
R_{\text{reg}} (z) := \sum_{i=1}^{N_f} \sum_{s=\pm1} \frac{e^{-\beta(m_i+sz)}}{1- e^{-\beta(m_i+sz)} } + 2(N+2).
\end{align}
The singular part changes its sign at $z \in \mathcal{C}$ as
\begin{align}
\lim_{\epsilon\rightarrow 0+} R_{\text{sing}}(z \pm i \epsilon) = \mp \frac{\pi i \tilde{f}_*''(z)}{\beta}.
\end{align}
This indicates the square root branch cuts at $\mathcal{C}_I$.

\begin{figure}
\centering
\includegraphics[width=14cm]{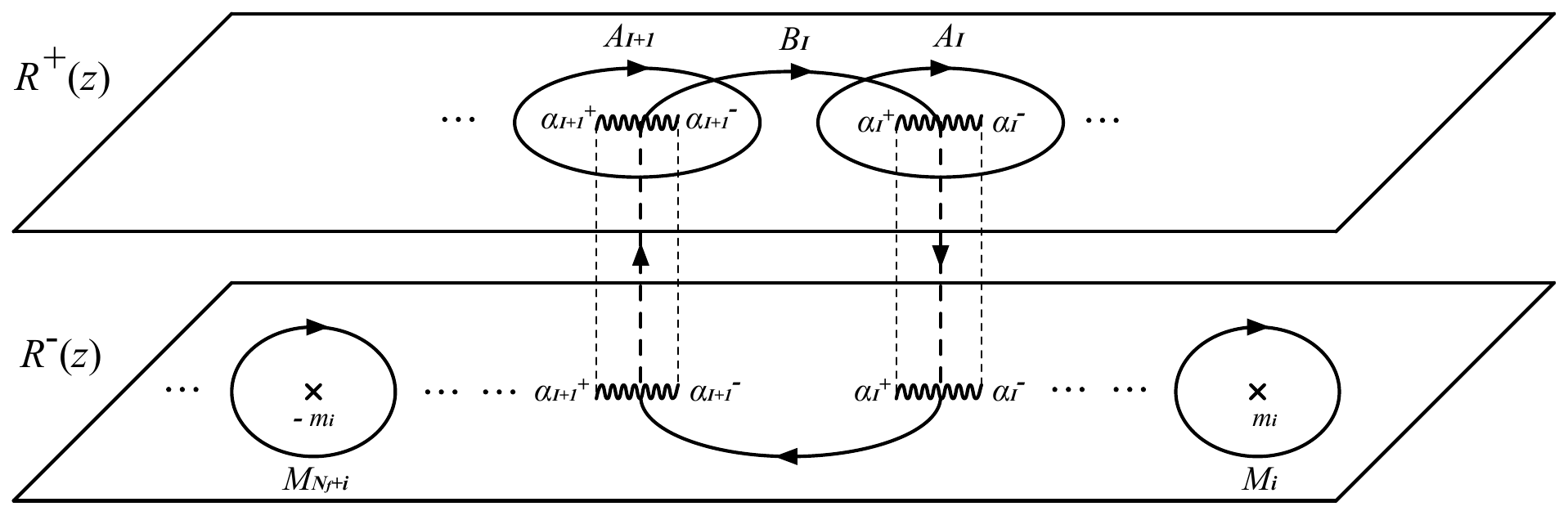}
\caption{$A_I$, $B_I$, and $M_i$-cycles depicted on the two copies of complex plane with coordinate $z$. $R^-(z)$ is defined on the upper complex plane while $R^+(z)$ is defined on the lower complex plane. The two complex planes are connected at the branch cuts and they give the Riemann surface as a whole. Note that the structure is analogous to Figure \ref{Fig:ABM}, apart from
the locations of $M_{i+N_f}$-cycles ($i=1,\cdots, N_f$).}
\label{fig:ABMcycleR}
\end{figure}

We consider analytic continuation $\hat{R}(z)$
of the function $R^+(z)$ from the complex plane 
to Riemann surface which is constructed by gluing the cuts in the upper sheet and lower sheet respectively as in Figure \ref{fig:ABMcycleR}.
The function $\hat{R}(z)$ can be regarded as a multi-valued function, which satisfies
\begin{align}\label{eq:upperR}
\hat{R}(z) = R^{+}(z) = R_{\text{reg}}(z) +  R_{\text{sing}}(z)
\end{align}
on the upper sheet, while
\begin{align}\label{eq:lowerR}
\hat{R}(z) = R^{-}(z) = R_{\text{reg}}(z) -  R_{\text{sing}}(z) = - R^+(z) + 2 R_{\text{reg}}(z)
\end{align}
on the lower sheet.

As depicted in Figure \ref{fig:ABMcycleR},
$A_I$-cycles, $B_I$-cycles, and $M_i$-cycles are defined in the $z$-plane as follows:
$A_I$-cycles go around $\mathcal{C}_I$ on the upper sheet.
$B_I$-cycles are closed loops starting from $\alpha_{I+1}\in (\alpha_{I+1}^{-}, \alpha_{I+1}^{+})$ to $\alpha_{I}\in (\alpha_{I}^{-}, \alpha_{I}^{+}) $ clockwise and they are determined by intersection condition $A_I \cdot B_J = \delta_{IJ}$.
$M_i$-cycles ($i=1,\cdots, N_f$) go around the point $z= m_i$ on the lower sheet
and $M_{N_f+i}$-cycles ($i=1,\cdots, N_f$) go around the point $z= - m_i$.
In section \ref{sec:SW}, the corresponding cycles are described in the $w$-plane by the same name.
In the following, we consider the integrals of $z^n \hat{R}(z)$ over $A_I$-cycles, $B_I$-cycles, and $M_i$-cycles, where the case $n=0$ and $n=1$ are especially our interest.

\subsubsection*{A-cycle integrals}
Since $\hat{R}(z) = R^+(z)$ on the upper sheet, we can compute the $A_I$-cycle integral by using the definition of the resolvent \eqref{eq:defresolvent}.
By exchanging the order of the integrals, we find
\begin{align}\label{eq:ointAzR}
\oint_{A_I} z^n \hat{R}(z)dz
&= \int_{-\infty}^{\infty} \tilde{f}^{''}_{*}(x) \left( \oint_{A_I} \frac{z^n}{1-e^{-\beta(x-z)}} dz \right) dx
\end{align}
Since
\begin{align}
\oint_{A_I} \frac{z^n}{1-e^{-\beta(x-z)}}dz
= \left\{
\begin{array}{ll}
2\pi i x^n \beta^{-1} & x \in \mathcal{C}_I \\
0 & x \notin \mathcal{C}_I
\end{array}
\right.
\end{align}
the $A_I$-cycle integrals are given by
\begin{align}
\oint_{A_I} z^n \hat{R}(z) dz = 2 \pi i \beta^{-1} \int_{\mathcal{C}_I} x^n \tilde{f}^{''}_*(x) dx = 4\pi i \beta^{-1} a_I^n \quad (n=0, 1)
\end{align}
due to the constraints \eqref{eq:constf1} and \eqref{eq:constf2}.

\subsubsection*{B-cycle integrals}
\begin{figure}
\centering
\includegraphics[width=8cm]{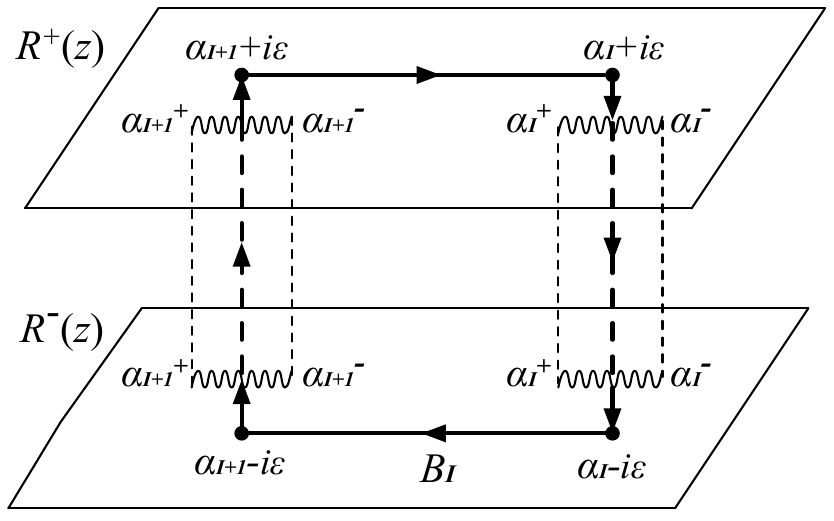}
\caption{Computation of B cycle integrals}
\label{fig:Bcycle}
\end{figure}

To compute the integrals of the Resolvent $\hat{R}(z)$ along $B_I (I=1, 2, \ldots, N+1)$ cycles, it is useful to consider contour integrals depicted in Figure \ref{fig:Bcycle}. Here the $B_I$ cycle is a closed path which starts and ends at $\alpha_{I+1} + i \epsilon $ and consists of 4 intervals: $\alpha_{I+1} + i \epsilon \rightarrow \alpha_{I} + i \epsilon$, $\alpha_{I} + i \epsilon \rightarrow \alpha_{I} - i \epsilon$, $\alpha_{I} - i \epsilon \rightarrow \alpha_{I+1} - i \epsilon$, $\alpha_{I+1} - i \epsilon \rightarrow \alpha_{I+1} + i \epsilon$.

\begin{align}\label{eq:Bint1}
\int_{B_I} z^n \hat{R}(z) dz = \lim_{\epsilon\rightarrow 0+} \Biggl[& \int_{(\alpha_{I+1} + i\epsilon)_{+}}^{(\alpha_I + i\epsilon)_{+}}z^n \hat{R}(z)dz + \int_{(\alpha_{I} + i\epsilon)_{+}}^{(\alpha_I - i\epsilon)_{-}}z^n \hat{R}(z)dz
\cr
&+ \int_{(\alpha_{I} - i\epsilon)_{-}}^{(\alpha_{I+1} - i\epsilon)_{-}}z^n \hat{R}(z)dz + \int_{(\alpha_{I+1} - i\epsilon)_{-}}^{(\alpha_I + i\epsilon)_{+}}z^n \hat{R}(z)dz\Biggr]
\end{align}
Here $(z)_{+}$ denotes the point on the upper sheet while $(z)_{-}$ denotes the point on the lower sheet. Assume that $\hat{R}(z)$ does not diverge at $z=\alpha_I$, then the second and the fourth terms will vanish when we take the limit $\epsilon\rightarrow 0$. Taking into account the expression of $\hat{R}(z)$ at the upper sheet and the lower sheet given in \eqref{eq:upperR} and \eqref{eq:lowerR} respectively, we find the $B_I$-cycle integral reduces to
\begin{align}
\oint _{B_I} z^n\hat{R}(z) dz
&=
\lim_{\epsilon \to 0+}  \left(
\int^{\alpha_I + i \epsilon}_{\alpha_{I+1} + i \epsilon} z^n R^+(z) dz
+  \int_{\alpha_I - i \epsilon}^{\alpha_{I+1} - i \epsilon}z^n ( - R^+(z) + 2 R_{\text{reg}} (z) )dz
\right)
\cr
&=
\lim_{\epsilon \to 0+} \left(
\int^{\alpha_I}_{\alpha_{I+1}} z^n \left( R^+(z - i \epsilon) + R^+(z + i\epsilon) dz
\right)  \right)
- 2 \int^{\alpha_I}_{\alpha_{I+1}} z^n R_{\text{reg}} (z)dz.
\end{align}
Here, we use again 
the formula \eqref{eq:SokhotskiPlemelj}
as well as the definitions of $R_+(z)$ in \eqref{eq:defresolvent} and $R_{\text{reg}}(z)$ in \eqref{eq:defRreg} to obtain\footnote{Note that we cannot simply use \eqref{eq:Rplusminus} here because $z \in \mathcal{C}_I$ is not satisfied in all the integral region. }

\begin{align}
\oint _{B_I} z^n\hat{R}(z) dz
=&
2 \int^{\alpha_{I}}_{\alpha_{I+1}  }
z^n\left( P.V. \int^{\infty}_{-\infty} \frac{\tilde{f}_* ''(x)}{1 - e^{ -\beta ( x - z )}} d x  \right) dz
\cr
& \qquad
-  2 \int^{\alpha_I }_{\alpha_{I+1} } z^n
\left(
\sum_{i=1}^{N_f} \sum_{s = \pm 1} \frac{e^{ -\beta (m_i + s z)}  }{1 - e^{ -\beta (m_i + s z)} } + 2 (N+2)
\right)
 dz
\cr
=&
\sum_{s=\pm 1}P.V.\int^{\infty}_{-\infty} \tilde{f}_* ''(x)
\left( \int^{\alpha_{I}}_{\alpha_{I+1}}z^n \frac{e^{ -\beta (x + s z)}}{1 - e^{ -\beta (x + s z)} }dz\right)dx
\cr
& \,\,
-  2 \sum_{i=1}^{N_f} \sum_{s = \pm 1}  \int^{\alpha_I }_{\alpha_{I+1} } z^n
\frac{e^{ -\beta (m_i + s z)}  }{1 - e^{ -\beta (m_i + s z)} }
 dz
- 4 (N+2) \int^{\alpha_I }_{\alpha_{I+1} }  z^n dz,
\end{align}
where we changed the order of integral and used \eqref{eq:simpleodd} at the last equality.

We first consider the case $n=0$. Since the integral is computed as
\begin{align}
& \int
\frac{e^{ -\beta (x + s z)}}{1 - e^{ -\beta (x + s z )}} dz
=
\frac{1}{\beta s} \log (1 - e^{- \beta ( x + s z ) }) + \text{Const},
\cr
& \int
\frac{e^{ -\beta (m_i + s z)}}{1 - e^{ -\beta (m_i + s z)}} dz
=
\frac{1}{\beta s} \log (1 - e^{- \beta ( m_i + s z) })  + \text{Const},
\end{align}
we find
\begin{align}
\oint _{B_{I}} \hat{R}(z) dz
=&
\sum_{s=\pm 1}  \int^{\infty}_{-\infty} \tilde{f}_* ''(x)
\left( \left. \frac{1}{\beta s} \log (1 - e^{- \beta ( x + s z ) }) \right|^{\alpha_{I}}_{\alpha_{I+1}  }  \right) dx
\cr
& \qquad
-  2 \sum_{i=1}^{N_f} \sum_{s = \pm 1}
\left. \frac{1}{\beta s} \log (1 - e^{- \beta ( m_i + s z) }) \right|^{\alpha_I }_{\alpha_{I+1} }
- \left. 4 (N+2) z \right|^{\alpha_I }_{\alpha_{I+1} }
\cr
= & 0,
\end{align}
where the last equality holds due to the derivative of the saddle point equation \eqref{eq:seqderiv2}.

Next, we consider the case $n=1$. By integrating by parts, we find
\begin{align}
& \int
z \frac{e^{ -\beta (x + s z)}}{1 - e^{ -\beta (x + s z )}} dz
=
\frac{z}{\beta s} \log (1 - e^{- \beta ( x + s z ) }) - \frac{1}{\beta^2}  \text{Li}_2 (e^{ -\beta (x + sz)})  + \text{Const}
\cr
& \int
z \frac{e^{ -\beta (m_i + s z)}}{1 - e^{ -\beta (m_i + s z)}} dz
=
\frac{z}{\beta s} \log (1 - e^{- \beta (m_i + s z ) }) - \frac{1}{\beta^2}  \text{Li}_2 (e^{ -\beta (m_i + sz)})  + \text{Const}
\end{align}
For convenience, we define $\xi(z) := \xi_{I*}$ for $z \in \mathcal{C}_I$. Then the formula for B-cycle integral is
\begin{align}\label{eq:Bprocess}
& \oint_{B_I} z \hat{R}(z) dz
\cr
=& \frac{z}{\beta} \left( \left.
\sum_{s = \pm 1} s \int^{\infty}_{- \infty} \tilde{f}_* ''(x)
\log \left( 1 - e^{ -\beta (x + s z)} \right) dx
- 2 \sum_{i=1}^{N_f} \sum_{s = \pm 1}
s \log \left( 1 - e^{ -\beta (m_i + s z)} \right)
\right) \right|_{\alpha_{I+1}}^{\alpha_I}
\cr
& \qquad
- \frac{1}{\beta^2} \left.
\left(
 \sum_{s = \pm 1} \int^{\infty}_{- \infty} \tilde{f}_* ''(x)
\mathrm{Li}_2\left(  e^{ -\beta (x + s z)} \right)  dx
- 2 \sum_{i=1}^{N_f} \sum_{s = \pm 1}
\mathrm{Li}_2\left(  e^{ -\beta (m_i + s z)} \right)
\right)
\right|_{\alpha_{I+1}}^{\alpha_I}
\cr
& \qquad - 2 (N+2) z^2 \biggr|_{\alpha_{I+1}}^{\alpha_I}
\cr
=& \frac{z}{\beta} \biggl(
4(N+2) \beta z
\biggr) \biggr|_{\alpha_{I+1}}^{\alpha_I}
- \frac{1}{\beta^2} \left.
\biggr(
2(N+2) \beta^2 z^2 - 4 \xi(z)
\biggr)
\right|_{\alpha_{I+1}}^{\alpha_I} - 2 (N+2) z^2 \biggr|_{\alpha_{I+1}}^{\alpha_I}
\cr
= & \frac{4}{\beta^2} ( \xi_{I*} - \xi_{I+1*} ),
\end{align}
where in the second equality, we used the first and second $z$-derivatives of the saddle point equation \eqref{eq:seqderiv1} and \eqref{eq:seqderiv2}.

Moreover, the Lagrangian multipliers $\xi_I (I=1, 2, \ldots, N+1)$ are related to the prepotential as
\begin{align}\label{eq:xiIN+1}
\xi_{I*} = \frac{1}{2} \frac{\partial}{\partial a_I} F(a_1, \cdots, a_N, a_{N+1}) \quad (I=1, 2, \ldots, N+1)
\end{align}
by Legendre transformation.
In order to compare with the result in section \ref{sec:SW}, we should erase $a_{N+1}$ by using \eqref{eq:AN1A}.
Defining the composite function
$$\bar{F}(a_1, \cdots, a_{N}, m_0 ) := F(a_1, \cdots, a_{N}, a_{N+1}(a_1,\cdots, a_N, m_0) ), $$
with
$$a_{N+1}(a_1,\cdots, a_N, m_0) := - m_0 - \sum_{I=1}^N a_I + \frac{1}{2} \sum_{i=1}^{N_f} m_i $$
as given in \eqref{eq:AN1A}, we find


\begin{align}
\frac{\partial}{\partial a_I} \bar{F}(a_1, \cdots, a_{N}, m_0 )
&= \frac{\partial}{\partial a_I} F(a_1, \cdots, a_{N}, a_{N+1} ) - \frac{\partial}{\partial a_{N+1}} F(a_1, \cdots, a_{N}, a_{N+1} ),
\cr
&= 2 (\xi_{I*} - \xi_{N+1*})
\quad (I=1, 2, \ldots, N).
\end{align}
Thus, we find
\begin{align}
\xi_{I*} - \xi_{I+1*}&= \frac{1}{2} \frac{\partial}{\partial a_I} \bar{F}(a_1, \cdots, a_{N}, m_0 ) - \frac{1}{2} \frac{\partial}{\partial a_{I+1}} \bar{F}(a_1, \cdots, a_{N}, m_0 ),
\quad (I=1, 2, \ldots, N-1)
\cr
\xi_{N*} - \xi_{N+1*} &=  \frac{1}{2}  \frac{\partial}{\partial a_N} \bar{F}(a_1, \cdots, a_{N}, m_0 ).
\end{align}
%
%

Thus, the $B_I$-cycle integrals for $n=1$ are given by
\begin{align}
\oint_{B_I} z \hat{R}(z) dz &= 2 \beta^{-2} \left( \frac{\partial F}{\partial a_I} - \frac{\partial F}{\partial a_{I+1}} \right)
\qquad (I=1,2,\cdots N-1)
\cr
\oint_{B_N} z \hat{R}(z) dz &= 2 \beta^{-2} \frac{\partial F}{\partial a_N}
\end{align}
where $F$ denotes $\bar{F}(a_1, \cdots, a_{N}, m_0 )$ for simplicity.

\subsubsection*{M-cycle integrals}
Since $M_i$-cycles are on the lower sheet, we find
%
\begin{align}
\oint_{M_i} z^n \hat{R}(z) dz
= \oint_{M_i} z^n \left( - R^+(z) dz + 2  \oint_{M_i} R_{\text{reg}}(z) \right) dz
= 2  \oint_{M_i} z^n R_{\text{reg}}(z) dz,
\end{align}
where we note that $R^+(z)$ is regular outside of the branch cuts.
With the explicit expression for $R_{\text{reg}} (z)$ in \eqref{eq:defRreg}, this can be further computed as
\begin{align}
2 \oint_{M_i} z^n R_{\text{reg}} (z)  dz
& =
2 \sum_{i'=1}^{N_f} \sum_{s = \pm 1} \oint_{M_i} z^n
\frac{e^{ -\beta (m_{i'} + s z)}  }{1 - e^{ -\beta (m_{i'} + s z)} } dz
+ 4(N+2) \oint_{M_i} z^n dz
\label{eq:aRreg}
\end{align}
Since $n \neq -1$, the $M_i$-cycle integral of $z^n$ vanishes: $\oint_{M_i} z^n= 0$.
Assuming that $M_i$ cycle is small enough to include only the pole at $z=m_i$,
then only the term for $i'=i$ $s=-1$ remains:
\begin{align}
2 \sum_{i'=1}^{N_f} \sum_{s = \pm 1} \oint_{M_i} z^n
\frac{e^{ -\beta (m_{i'} + s z)}  }{1 - e^{ -\beta (m_{i'} + s z)} } dz
= 2 \oint_{M_i} z^n \frac{e^{ -\beta (m_{i} - z)}  }{1 - e^{ -\beta (m_{i} - z)} } dz
= 4 \pi i \beta^{-1} m_i{}^n.
\end{align}
We also carry out the analogous computation for $M_{N_f+i}$-cycles. If we define $m_{N_f+i}:= - m_i$,
The expression is valid for $i=1,2,\cdots 2N_f$:
\begin{align}
\oint_{M_i} z^n \hat{R}(z) dz
= 4 \pi i \beta^{-1} m_i{}^n.
\end{align}



\subsubsection*{Results of $A_I,B_I,M_i$-cycle integrals}

In conclusion, the $A_I, B_I, M_i$-cycle integrals are summarized as follows:
\begin{align}\label{eq:intABMR}
\oint_{A_I} \hat{R}(z) = 4 \pi i \beta^{-1}, \quad
\oint_{B_I} \hat{R}(z) = 0, \quad
\oint_{M_i} \hat{R}(z) = 4 \pi i \beta^{-1},  \quad
\end{align}
and
\begin{align}\label{eq:intABMzRz}
&\oint_{A_I} z \hat{R}(z) = 4 \pi i \beta^{-1} a_I,
\cr
&\oint_{B_I} z \hat{R}(z) = 2 \beta^{-2} \left( \frac{\partial F}{\partial a_{I} }  - \frac{\partial F}{\partial a_{I+1} } \right) \quad (I=1,2,\cdots N-1 ),
\cr
&\oint_{B_N} z \hat{R}(z) = 2 \beta^{-2} \frac{\partial F}{\partial a_{N} },
\cr
&\oint_{M_i} z \hat{R}(z) = 4 \pi i \beta^{-1} m_i.
\end{align}



\subsection{Deriving the Seiberg-Witten curve}
In this section, we will derive the Seiberg-Witten curve in terms of the resolvent.

When we consider the integral of the resolvent
\begin{align}\label{eq:antiderivR}
\int^z_0 \hat{R}(z') dz',
\end{align}
we find that there are two kinds of ambiguity, which are related to the logarithmic branch points and the square root branch points, respectively.
One ambiguity is the dependence on the integral path as can be seen from \eqref{eq:intABMR},
which is to add $4 \pi i \beta^{-1} n$ $(n \in \mathbb{Z})$.
This can be resolved by considering the exponentiated value
\begin{align}\label{eq:expR}
\exp \left[ \frac{\beta}{2} \int^z_0 \hat{R}(z) dz \right]
\end{align}
since $e^{2 \pi i n} = 1$ $(n \in \mathbb{Z})$.
In other word, \eqref{eq:expR} does not have logarithmic branch points although \eqref{eq:antiderivR} does.
The other ambiguity is due to the fact that the resolvent $\hat{R}(z)= R^{\pm}(z)$ is multivalued on the complex plane, which is related to the square root branch points as discussed before.
This can be also resolved by considering the combination
\begin{align}\label{eq:defMz}
M(z) := \exp \left[ \frac{\beta}{2} \int^z_0 R^+(z') dz' \right] +  \exp \left[ \frac{\beta}{2} \int^z_0 R^- (z') dz' \right]
\end{align}
which is invariant under the exchange of $R^+(z)$ and $R^-(z)$.
This indicates that $M(z)$ does not have any branch points and is single valued on the complex plane.
Since $R^-(z)$ can be rewritten in terms of $R^+(z)$ and $R_{\text{reg}}(z)$ as in \eqref{eq:lowerR}, where $R_{\text{reg}}(z)$ is defined in \eqref{eq:defRreg},
we note that the second term in \eqref{eq:defMz} is rewritten as
\begin{align}\label{eq:Msecond}
\exp \left[ \frac{\beta}{2} \int^z_0 R^- (z') dz' \right]
=
\exp \left[ - \frac{\beta}{2} \int^z_0 R^+(z') dz' \right]
e^{2(N+2) \beta z} \left( \prod_{i=1}^{N_f}\frac{1-e^{-\beta(m_i + z)}}{1-e^{-\beta(m_i - z)}} \right)
\end{align}
by explicitly computing the integral of $R_{\text{reg}}(z)$.

Here, we show the periodicity $z \to z + \frac{2 \pi i }{\beta}$ of $M(z)$.
Since the periodicity of the last two factors in \eqref{eq:Msecond} is obvious, it is enough to consider the integral of $R^+(z)$.
Note that
\begin{align}\label{eq:intRpiibeta}
\int^{z+\frac{2 \pi i }{\beta}}_0 R^+(z') dz'
&= \int^{\frac{\pi i }{\beta}}_0 R^+(z') dz' + \int^{\frac{2 \pi i }{\beta}}_{\frac{\pi i }{\beta}} R^+(z') dz' + \int^{z+\frac{2 \pi i }{\beta}}_{\frac{2 \pi i }{\beta}} R^+(z') dz'
\cr
&= \int^{\frac{\pi i }{\beta}}_{- \frac{\pi i }{\beta}}  R^+(z') dz' + \int^{z}_{0} R^+(z') dz'
\end{align}
where we used the periodicity of the resolvent $R^{+}(z)$ in the second equality. The third equality holds since the resolvent is an even function.
\begin{figure}
\centering
\includegraphics[width=13cm]{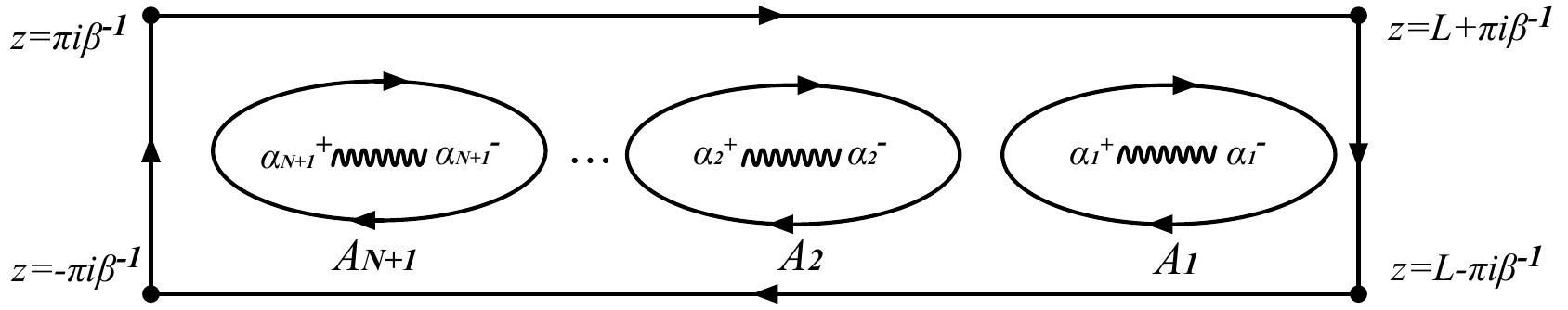}
\caption{Contour integral related to boundary condition}
\label{fig:contourbdy}
\end{figure}
In order to evaluate the first term in \eqref{eq:intRpiibeta}, we need to consider the following contour integral depicted in Figure \ref{fig:contourbdy}.
By Cauchy integral theorem, we have
\begin{align}
 \int_{-\frac{\pi i}{\beta}}^{\frac{\pi i}{\beta}} R^+(z)dz
 +  \int_{\frac{\pi i}{\beta}}^{L + \frac{\pi i}{\beta}} R^+(z)dz
 + \int_{L + \frac{\pi i}{\beta}}^{L-\frac{\pi i}{\beta}}R^+(z)dz
  + \int_{L-\frac{\pi i}{\beta}}^{-\frac{\pi i}{\beta}} R^+(z)dz
  = \sum_{I=1}^{N+1} \int_{A_I}R^+(z)dz ,
\end{align}
where we choose $L$ to be large enough.
If we take the limit $L\rightarrow \infty$, the third term at the left hand side vanishes.
Also, the second term and the fourth term cancel with each other due to the periodicity of $R^+(z)$.
The right hand side are given by the $A$-cycle integrals in \eqref{eq:intABMR}.
Thus, we can obtain
\begin{align}\label{eq:intpiibeta}
\int_{-\frac{\pi i}{\beta}}^{\frac{\pi i}{\beta}} R^+(z)dz = \frac{4\pi i (N+1)}{\beta}.
\end{align}
Combining \eqref{eq:defMz}, \eqref{eq:Msecond}, \eqref{eq:intRpiibeta} and \eqref{eq:intpiibeta} all together, we find the periodicity of $M(z)$
\begin{align}\label{eq:periodcityM}
M \left( z+\frac{2 \pi i }{\beta} \right)  = M(z).
\end{align}
This periodicity implies that $M(z)$ is rewritten more naturally as a function of $e^{ -\beta z}$.

Now, we discuss the singularities of $M(z)$. Due to the periodicity, it would be enough to consider the region $- \pi i \beta^{-1} \le \text{Im} (z) \le \pi i \beta^{-1}$.
As mentioned in section \ref{sec:resolvent}, $R^{+}(z)$ is regular outside of the branch cuts. Moreover,
its $B_I$-cycle integral in \eqref{eq:intABMR} does not depend on the choice of $\alpha_I$ and $\alpha_{I+1}$ in \eqref{eq:Bint1},
which indicates that the integral of $R^+(z)$ does not have singularities also on the branch cuts.
Therefore, the first term in \eqref{eq:defMz} has no singularities except at $z \to \pm \infty$.
The singularities of $M(z)$ at finite region are originated from the denominator of the last factor in \eqref{eq:Msecond}. That is, $M(z)$ has simple poles at $z= m_i$ ($i=1,2, \cdots, N_f$).

In order to remove the simple poles at $z=m_i$, we define
\begin{align}\label{defPexp}
P( e^{- \beta z})
:=& - M(z) e^{-(N+2)\beta z} \prod_{i=1}^{N_f} \left( 1-e^{-\beta(m_i - z)} \right)
\cr
=&
- \exp \left[ \frac{\beta}{2}\int_{0}^z R^+(z')dz' \right] e^{-(N+2)\beta z} \prod_{i=1}^{N_f} \left( 1-e^{-\beta(m_i - z)} \right)
\cr
& - \exp\left[ -\frac{\beta}{2}\int_{0}^z R^+(z')dz' \right] e^{(N+2)\beta z} \prod_{i=1}^{N_f} \left( 1-e^{-\beta(m_i + z)} \right),
\end{align}
where we have defined it as a function of $e^{ -\beta z}$ since it is invariant under $z \to z+\frac{2 \pi i }{\beta}$ due to \eqref{eq:periodcityM}.
Denoting
\begin{align}\label{eq:defwz}
w:= e^{ -\beta z},
\end{align}
$P(w)$ is a single valued function on the $w$-plane, which has isolated singularities only at $w=0$ and $w = \infty$.
Moreover, note that
\begin{align}\label{eq:symPth}
P(w^{-1}) = P(w),
\end{align}
which can be shown from \eqref{defPexp} by taking into account that $R^{+}(z)$ is an even function.

Since $R^+(z) \to 0$ as $z \to +\infty$ as discussed in section \ref{sec:resolvent}, we find
\begin{align}\label{eq:Romicron}
\exp \left[ \pm \frac{\beta}{2}\int_{0}^z R^+(z')dz' \right] = o ( e^{ \beta z} ) \qquad \text{ as } \quad z \to + \infty.
\end{align}
Since we are considering $N_f \le 2N+3$, we find from \eqref{defPexp} and \eqref{eq:Romicron} that
\begin{align}
P( w ) = o (w^{-N-3}) \qquad \text{ as } \quad  w \to 0,
\end{align}
which indicates that $w=0$ is a pole of (at most) order $N+2$.
Combined with \eqref{eq:symPth}, we find that $P(w)$ must be a Laurent polynomial of the form
\begin{align}
P(w) = \sum_{n=1}^{N+2} C_{1,n} (w^n+w^{-n}) + C_{1,0}.
\end{align}

If we define
\begin{align}\label{eq:tdef}
t_{+} := \exp \left[ \frac{\beta}{2}\int_{0}^z R^+(z')dz' \right] e^{-(N+2)\beta z} \prod_{i=1}^{N_f} \left( 1-e^{-\beta(m_i - z)} \right),
\end{align}
$P( w ) $ can be rewritten in terms of $t$ as
\begin{align}\label{eq:Pexpre}
P( w)
=&
- t_{+} - t_{+}^{-1} \prod_{i=1}^{N_f} \left( 1-e^{-\beta(m_i - z)} \right)  \left( 1-e^{-\beta(m_i + z)} \right).
\end{align}
Thus, $t_+$ defined above can be understood as one of the solutions of the following equation:
\begin{align}\label{eq:SWrederiv}
t^2 + P(w) t +  Q(w)  = 0,
\end{align}
where we defined
\begin{align}\label{eq:defQth}
Q(w) := \prod_{i=1}^{N_f} \left( w-e^{-\beta m_i } \right)  \left( w^{-1} - e^{-\beta m_i } \right).
\end{align}
This is identified as the Seiberg-Witten curve in \eqref{eq:sec2SWcurve} in section \ref{sec:SW}.

Also, from the definition of $w$ in \eqref{eq:defwz} and $t_+$ in \eqref{eq:tdef}, we find
\begin{align}
- \frac{1}{2 \pi i \beta} (\log w) d (\log t_+) =
\frac{\beta z}{4 \pi i } \left( R^+(z) - \sum_{i=1}^{N_f} \frac{2 e^{- \beta (m_i + z) }}{1 - e^{- \beta (m_i + z) }} - 4(N+2) \right) dz
\end{align}
This is identified as the Seiberg-Witten one form $\lambda^+_{SW}$.
Indeed, if we identify
\begin{align}
\lambda_{SW}^{\pm} =
\frac{\beta z}{4 \pi i } \left( R^{\pm}(z) - \sum_{i=1}^{N_f} \frac{2 e^{- \beta (m_i + z) }}{1 - e^{- \beta (m_i + z) }} - 4(N+2) \right) dz,
\end{align}
the cycle integrals in \eqref{eq:intABMzRz} reproduce the cycle integrals in \eqref{eq:sec2SW1}.

\subsection{Deriving the Boundary conditions}
In this section, we will derive the boundary conditions satisfied by
the Laurent polynomials $P(w)$ and $Q(w)$.

\subsubsection*{Boundary condition at $w=1$:}

Starting from \eqref{defPexp},
we set $z=0$, we find
\begin{align}
P(1)= -2 \prod_{i=1}^{N_f} (1-e^{-\beta m_i}).
\end{align}
From \eqref{eq:defQth}, we set $w=1$ and find
\begin{align}
Q(1)= \prod_{i=1}^{N_f} (1-e^{-\beta m_i})^2.
\end{align}
Thus, we derive the boundary condition at $w=1$:
\begin{align}\label{eq:BC1}
P(1)^2 - 4 Q(1)= 0,
\end{align}
which means that the curve \eqref{eq:SWrederiv} has a double root at $w=1$.

%

\subsubsection*{Boundary condition at $w=-1$:}

Starting from \eqref{defPexp}, we set $z=\frac{\pi i}{\beta}$, then we find
\begin{align}
P(-1)= (-1)^{N+1} \prod_{i=1}^{N_f} (1 + e^{-\beta m_i})\times [e^{\frac{\beta}{2}\int_{0}^{\frac{\pi i}{\beta}} R(z')dz'} + e^{-\frac{\beta}{2}\int_{0}^{\frac{\pi i}{\beta}} R(z')dz'} ].
\end{align}
Since the resolvent $R^+(z)$ is an even function, we find from \eqref{eq:intpiibeta} that
\begin{align}
\int_{0}^{\frac{\pi i}{\beta}} R^+(z)dz = \frac{1}{2}  \int_{-\frac{\pi i}{\beta}}^{\frac{\pi i}{\beta}} R^+(z)dz = 2\pi i (N+1)
\end{align}
and thus,
\begin{align}
P(-1)= 2 \prod_{i=1}^{N_f} (1 + e^{-\beta m_i}).
\end{align}
From \eqref{eq:defQth}, we find
\begin{align}
Q(-1)= \prod_{i=1}^{N_f} (1+e^{-\beta m_i})^2.
\end{align}
Thus, we derive the boundary condition at $w=-1$:
\begin{align}\label{eq:BC2}
P(-1)^2 - 4 Q(-1) = 0,
\end{align}
which means that the curve \eqref{eq:SWrederiv} has a double root also at $w=-1$.

It is remarkable that the boundary conditions \eqref{eq:BC1} and \eqref{eq:BC2},
which were originally claimed from the intuition that the M5-brane is attaching to the OM5-plane at one point \cite{Landsteiner:1997vd, Hayashi:2017btw},
are now derived independently from the partition function based on the topological vertex formalism with $O5$-plane.

In  this section, we have derived all the information in \eqref{eq:sec2SWcurve} and \eqref{eq:sec2SW1} from the thermodynamic limit of the partition function.
As is indeed explicitly found in Appendix \ref{app:SW}, they are enough to determine the Seiberg-Witten curve.
Thus, the prepotential defined as the leading order of the partition function as in \eqref{eq:prepdef2} agrees with the prepotential computed from the Seiberg-Witten curve in section \ref{sec:SW} up to a constant term which does not depend on the Coulomb moduli $a_I$.
This implies that the topological vertex formalism with $O5$-plane discussed in \cite{Kim:2017jqn} is consistent with
the technique to compute Seiberg-Witten curve from the 5-brane web with $O5$-plane in \cite{Hayashi:2017btw} and thus, they justify each other.


\section{Conclusion and Discussion}\label{sec:concl}
In this paper, we first consider Seiberg-Witten curve obtained by toric-like dot diagram which is the dual graph of a class of $(p, q)$ 5-brane web diagram with $O5$-plane depicted in Figure \ref{Fig:SpN-Nf-web}. Its special case for $(N, N_f) = (1, 0)$ appears in the classification list of phase diagrams aiming at defining discrete theta angles and two different phase diagrams are connected by generalized flop transitions. We discuss the  boundary conditions satisfied by Seiberg-Witten curve \eqref{eq:sec2SWcurve} and the prepotential in terms of the period integrals \eqref{eq:sec2SW1}. Secondly, we study 5d $\mathcal{N}=1$ Nekrasov partition functions for $Sp(N)$ gauge theories with $N_f (\leq 2N + 3)$ flavors based on new method of ``topological vertex formalism with $O5$-plane'' proposed in \cite{Kim:2017jqn}. Inspired by work of Nekrasov-Okounkov \cite{Nekrasov:2003rj}, we rewrite the Nekrasov partition function in terms of profile functions for random partition diagrams. After taking thermodynamic limit, we can obtain the saddle point equation for profile functions. By introducing Resolvent, we can reproduce the Seiberg-Witten curve \eqref{eq:SWrederiv} and derive the boundary conditions \eqref{eq:BC1}, \eqref{eq:BC2} and the prepotential in terms of the cycle integrals \eqref{eq:intABMzRz}.
It means that the thermodynamic limit of Nekrasov partition function for 5-brane web with $O5$-plane
agrees with the prepotential in terms of
the Seiberg-Witten curve obtained from $(p, q)$ 5-brane with $O5$-plane. This gives further evidence for mirror symmetry conjecture which relates Nekrasov partition function with Seiberg-Witten curve in the case with the orientifold plane. Especially, based on two different Seiberg-Witten curves from 5-brane web diagram with and without $O5$-plane, we verify the agreement for the prepotentials for 5d $\mathcal{N}=1$ pure $Sp(1)=SU(2)$ gauge theory with discrete theta angle 0.

In this paper, we have restricted our study to the case with $N_f (\le 2N+3)$ flavors for simplicity. It would be straightforward to generalize to the case with $2N+4 \le N_f \le 2N+6$. It would be also possible to generalize to the $OSp$ linear quiver gauge theory.
For future perspective, it is interesting to 
generalize our work to non-Lagrangian field theories constructed from 5-brane web with $O5$-plane.
On the other hand, it is also fascinating to consider similar stories in other limits like Nekrasov-Natashivili limit \cite{Nekrasov:2009rc}, other types of orientifold planes appearing in different dimensions. 
It would be also interesting to study from the viewpoint of holomorphic anomaly equation \cite{Bershadsky:1993ta, Bershadsky:1993cx} or blow-up formula \cite{Nakajima:2003pg}, whose generalizations and applications have been studied in \cite{Nakajima:2005fg, Gottsche:2006bm, Nakajima:2009qjc, Gottsche:2010ig, Keller:2012da, Gu:2017ccq, Huang:2017mis, Gu:2018gmy, Kim:2019uqw, Gu:2019dan, Gu:2019pqj, Gu:2020fem, Kim:2020hhh}.
All in all, we believe that the hidden mathematical structures corresponding to orientifold 5-brane are also charming!

\acknowledgments
This project started around January 2019 which is the first one in the authors' serial explorations about the theory and applications of orientifold 5-brane in mathematics and physics.
We would like to thank Sung-Soo Kim for the early stage of the collaboration and comments for the draft.
We thank Hirotaka Hayashi, Kimyeong Lee, Yongchao Lu, Yuji Sugimoto, Xing-Yue Wei and Xinyu Zhang for useful discussion and comments. 
We would like to thank all the teachers and friends we met before.
Especially, the first author would like to thank Bohui Chen, An-min Li, Guosong Zhao for their constant support
and also thank the 3rd Pan-Pacific International Conference on Topology and Applications (3rd PPICTA), Sichuan Normal University and Southwest Geometry Conference for the invitations to present preliminary results of this paper.
Some ideas and discussions have been benefited from the authors' visits to Sichuan University, Peking University, Beijing Normal University, Sun Yat-sen University, IAS of Zhejiang University, SISSA, Korea Institute for Advanced Study, Khazar University, Mathematical Sciences Research Institute of Berkeley, University of Utah, Oklahoma State University and New Zealand Mathematics Research Institute.
Parts of the key computations have been done in Daci Temple, Yanjiyou Coffee bar, Xipuchuntian Tea house in Chengdu. Xiaobin Li is supported by NSFC grant No. 11501470, No. 11426187, No. 11791240561 and partially supported by NSFC grant No. 11671328. Futoshi Yagi is supported by the NSFC grant No. 11950410490, Fundamental Research Funds for the Central Universities A0920502051904-48 and Start-up research grant A1920502051907-2-046, and
in part by Recruiting Foreign Experts Program No. T2018050 granted by SAFEA.

\appendix


\section{Explicit expression for the Seiberg-Witten curve}\label{app:SW}
In this section, we write down the Seiberg-Witten curve more explicitly by rewriting some of the coefficients in terms of the parameters introduced in section \ref{sec:SW}.
As summarized in section \ref{sec:summary}, the Seiberg-Witten curve is given in the following form:
\begin{align}\label{eq:appSW}
t^2 + P(w) t + Q(w) = 0
\end{align}
with
\begin{align}
&P(w) = \sum_{n=1}^{N+2} C_{1,n} ( w^n + w^{-n} ) + C_{1,0},
\cr
&Q(w) = C \prod_{i=1}^{N_f} (w-e^{ - \beta m_i}) (w^{-1} - e^{ - \beta m_i}).
\end{align}

As mentioned around the end of section \ref{sec:SWcurve}, we have a degree of freedom to rescale $t$ because the Seiberg-Witten 1-form
$
\lambda_{SW} = \log w \, d (\log t)
$
is invariant under the rescaling of $t$. Suppose we redefine $t$ and $C_{1,n}$ by the rescaling $t \to C^{\frac{1}{2}} t$ and $C_{1,n} \to C^{\frac{1}{2}} C_{1,n}$, respectively. After dividing both hand sides by $C$ in \eqref{eq:appSW}, we have the same Seiberg-Witten curve but now with
\begin{align}\label{eq:C1}
C =1.
\end{align}
Under this convention, the condition \eqref{eq:q} indicates
\begin{align}
C_{1,N+2} 
= e^{ \beta (m_0 - \frac{1}{2} \sum_{i=1}^{N_f} m_i )}
\end{align}
Here, the ambiguity of the overall $\pm$ sign has been fixed again by using the redefinition $t \to -t$, which can be done without changing \eqref{eq:C1}. 

From the constraints \eqref{eq:Doubleroot}, we can rewrite $C_{1,1}$ and $C_{1,0}$ included in the expression above in terms of other coefficients as follows. If $N$ is even,
\begin{align}\label{eq:C10C11even}
C_{1,1} &= \pm e^{- \frac{1}{2} \beta \sum_{i=1}^{N_f} m_i} \chi_{c,s} - \sum_{k=1}^{\frac{N}{2}} C_{1,2k+1},
\cr
C_{1,0} &= \mp 2 e^{- \frac{1}{2} \beta \sum_{i=1}^{N_f} m_i} \chi_{s,c}
- 2e^{ \beta (m_0 - \frac{1}{2} \sum_{i=1}^{N_f} m_i )}  - 2 \sum_{k=1}^{\frac{N}{2}} C_{1,2k},
\end{align}
while if $N$ is odd,
\begin{align}\label{eq:C10C11odd}
C_{1,1} &= \pm e^{- \frac{1}{2} \beta \sum_{i=1}^{N_f} m_i} \chi_{c,s}
- e^{\beta (m_0 - \frac{1}{2} \sum_{i=1}^{N_f} m_i) }
- \sum_{k=1}^{\frac{N-1}{2}} C_{1,2k+1}
\cr
C_{1,0} &= \mp 2 e^{- \frac{1}{2} \beta \sum_{i=1}^{N_f} m_i} \chi_{s,c}
 - 2 \sum_{k=1}^{\frac{N+1}{2}} C_{1,2k},
\end{align}
Here, we have defined
\begin{align}
&\chi_s := \frac{1}{2}
\left(
\prod_{i=1}^{N_f} (e^{ + \frac{1}{2}\beta m_i} + e^{ - \frac{1}{2}\beta m_i} )
+ \prod_{i=1}^{N_f} (e^{ + \frac{1}{2}\beta m_i} - e^{ - \frac{1}{2}\beta m_i} )
\right),
\cr
&\chi_c := \frac{1}{2}
\left(
\prod_{i=1}^{N_f} (e^{ + \frac{1}{2}\beta m_i} + e^{ - \frac{1}{2}\beta m_i} )
- \prod_{i=1}^{N_f} (e^{ + \frac{1}{2}\beta m_i} - e^{ - \frac{1}{2}\beta m_i} )
\right).
\end{align}
Especially, they are defined as $\chi_s = 1$, $\chi_c = 0$ for $N_f=0$.

The expressions in \eqref{eq:C10C11even} and \eqref{eq:C10C11odd}  have two types of ambiguity in the first term: the choice of the sign $\pm$ and the choice of $\chi_s, \chi_c$. The first ambiguity is absorbed by the redefinition of $m_0 \to m_0 + \pi i / \beta$ together with $C_{1,n} \to - C_{1,n}$ $(2 \le n \le N+1)$ and $t \to -t$. Since this ambiguity does not give any significant difference, we choose $+$ for $C_{1,1}$, which means $-$ for $C_{1,0}$, just to fix the convention. The second ambiguity is also absorbed by the redefinition $m_1 \to - m_1$ if $N_f \ge 1$. However, if $N_f=0$, the second ambiguity gives the essential difference, which corresponds to the different discrete theta angle of the gauge theory \cite{Hayashi:2017btw}. The difference of the discrete theta angle can be distinguished from the 5-brane web diagram. Figure \ref{Fig:SpN-Nf-web} corresponds to the discrete theta angle 0 for $N$ odd while the discrete theta angle $\pi$ for $N$ even. This diagram is reproduced by the tropical limit of the Seiberg-Witten curve if we choose $\chi_s$ for $C_{1,1}$, which means $\chi_c$ for $C_{1,0}$. Although this choice is not essential for $N_f \ge 1$, it would be natural to use this choice also for $N_f \ge 1$  in this paper because Figure \ref{Fig:SpN-Nf-web} can be reproduced in the tropical limit with this choice if $m_i > 0$ $(i=1,2,\cdots, N_f)$ and $-m_0$ is chosen to be large enough. For the theory with $N_f=0$ and with the other discrete theta angle, which is $\pi$ for $N$ odd and $0$ for $N$ even, we should choose the opposite choice, which is $\chi_c$ for $C_{1,1}$ and $\chi_s$ for $C_{1,0}$.

Note that $C_{1,n}$ $(2 \le n \le N+1)$ are left undetermined. These $N$ parameters are interpreted as Coulomb branch parameters. In order to make the expression simpler, we introduce the notation
\begin{align}
C_{1,n} 
= e^{ \beta (m_0 - \frac{1}{2} \sum_{i=1}^{N_f} m_i )}  U_n.
\qquad (2 \le n \le N+1)
\end{align}

In summary, the Seiberg-Witten curve for 5d $\mathcal{N}=1$ $Sp(N)$ gauge theory with $N_f (\le 2N+3)$ flavors is given in the form
\begin{align}\label{eq:SWexplicit}
t^2 + \left [ e^{ \beta (m_0 - \frac{1}{2} \sum_{i=1}^{N_f} m_i )}  \,\, p(w) \right] t + \prod_{i=1}^{N_f} (w-e^{ - \beta m_i}) (w^{-1} - e^{ - \beta m_i}) = 0,
\end{align}
where $p(w)$ is given as follows depending on $N$, $N_f$ and the discrete theta angle:
\begin{itemize}
\item For $N$ even and $1 \le N_f \le 2N+3$,
\begin{align}
p(w) = & (w^{N+2}+w^{-N-2})
+ \sum_{n=2}^{N+1} U_{n} (w^{n}+w^{-n})
\cr
& + \left( e^{ - \beta m_0} \chi_{s} - \sum_{k=1}^{\frac{N}{2}} U_{2k+1} \right) (w+w^{-1})
 - 2 \left( e^{ - \beta m_0} \chi_{c} + 1 + \sum_{k=1}^{\frac{N}{2}} U_{2k} \right).
 \cr
\end{align}
\item For $N$ even, $N_f = 0$, and discrete theta angle 0
\begin{align}
p(w) = & (w^{N+2}+w^{-N-2})
+ \sum_{n=2}^{N+1} U_{n} (w^{n}+w^{-n})
\cr
&  - \sum_{k=1}^{\frac{N}{2}} U_{2k+1} (w+w^{-1})
 - 2 \left( e^{ - \beta m_0} + 1 + \sum_{k=1}^{\frac{N}{2}} U_{2k} \right).
\end{align}
\item For $N$ even, $N_f = 0$, and discrete theta angle $\pi$
\begin{align}
p(w) = & (w^{N+2}+w^{-N-2})
+ \sum_{n=2}^{N+1} U_{n} (w^{n}+w^{-n})
\cr
& + \left( e^{ - \beta m_0}  - \sum_{k=1}^{\frac{N}{2}} U_{2k+1} \right) (w+w^{-1})
 - 2 \left( 1 + \sum_{k=1}^{\frac{N}{2}} U_{2k} \right).
\end{align}
\item For $N$ odd and $1 \le N_f \le 2N+3$
\begin{align}
p(w) = &(w^{N+2}+w^{-N-2})
+ \sum_{n=2}^{N+1} U_n (w^{n}+w^{-n})
\cr
& + \left( e^{-\beta m_0} \chi_{s} - 1
- \sum_{k=1}^{\frac{N-1}{2}} U_{2k+1} \right) (w+w^{-1})
- 2 \left( e^{-\beta m_0}  \chi_{c}  + \sum_{k=1}^{\frac{N+1}{2}} U_{2k} \right).
\cr
\end{align}
\item For $N$ odd, $N_f = 0$, and discrete theta angle $0$
\begin{align}\label{eq:SWNf0theta0}
p(w) = &(w^{N+2}+w^{-N-2})
+ \sum_{n=2}^{N+1} U_n (w^{n}+w^{-n})
\cr
& + \left( e^{-\beta m_0} - 1
- \sum_{k=1}^{\frac{N-1}{2}} U_{2k+1} \right) (w+w^{-1})
- 2 \sum_{k=1}^{\frac{N+1}{2}} U_{2k} .
\end{align}
\item For $N$ odd, $N_f = 0$, and discrete theta angle $\pi$
\begin{align}
p(w) = &(w^{N+2}+w^{-N-2})
+ \sum_{n=2}^{N+1} U_n (w^{n}+w^{-n})
\cr
& - \left( 1
+ \sum_{k=1}^{\frac{N-1}{2}} U_{2k+1} \right) (w+w^{-1})
- 2 \left( e^{-\beta m_0}  + \sum_{k=1}^{\frac{N+1}{2}} U_{2k} \right).
\end{align}
\end{itemize}

The expression for Seiberg-Witten curve obtained above with $N=1$ agrees with the Seiberg-Witten curve computed in \cite{Hayashi:2017btw} up to the convention change $m_1 \to - m_1$ for $N_f \ge 1$. 



\section{Comparison of the Seiberg-Witten curves with and without $O5$-plane}\label{App:SWwwoO5}

In this Appendix, we check the agreement about the prepotentials for 5d $\mathcal{N}=1$ pure $Sp(1)=SU(2)$ gauge theory with discrete theta angle 0, obtained from the two different Seiberg-Witten curves based on the 5-brane web diagrams with and without $O5$-plane.

\subsection{The Seiberg-Witten curve for pure $SU(2)$ gauge theory without $O5$-plane}\label{app:SWSU2}

\begin{figure}
\centering
\begin{minipage}{7cm}
\centering
\includegraphics[width=3cm]{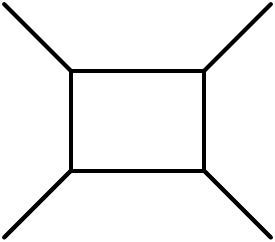}
\caption{5-brane web for 5d pure $SU(2)$ gauge theory with discrete theta angle 0.}
\label{fig:5branepureSU2}
\end{minipage}
\begin{minipage}{7cm}
\centering
\includegraphics[width=3cm]{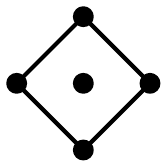}
\caption{Dual graph of the 5-brane web for 5d pure $SU(2)$ gauge theory with discrete theta angle 0.}
\label{fig:5branepureSU2dual}
\end{minipage}
\end{figure}

The Seiberg-Witten curve for the 5d $\mathcal{N}=1$ pure $SU(2)$ gauge theory with discrete theta angle 0 has been studied in various literatures including \cite{Nekrasov:1996cz, Lawrence:1997jr, Brandhuber:1997ua, Aharony:1997bh}.
In order for this Appendix to be self-contained, we review this Seiberg-Witten curve based on the 5-brane web diagrams without $O5$-plane in Figure \ref{fig:5branepureSU2} and the prepotential obtained from it.
Analogous to \eqref{eq:SW-general}, the Seiberg-Witten curve is given by the polynomial
\begin{align}\label{eq:SW-general2}
\sum_{(m,n) \in \mathbb{Z}^2} C_{m,n} t^m w^n = 0,
\end{align}
where $(m,n)$ represents the integer coordinates for the dots and the sum is over all the dots included in the dual graph in Figure \ref{fig:5branepureSU2dual}. The central dot corresponds to $C_{1,0}$.

By using the rescaling of $t$ as well as the multiplication of a constant to \eqref{eq:SW-general2}, we can fix $C_{2,0}=C_{0,0}=1$. The Seiberg-Witten curve is given in the form
\begin{align}\label{eq:SWPtilde}
t^2 + \tilde{P}(w;U) t + 1 =0.
\end{align}
Since we do not have $O5$-plane, the invariance under $w \to w^{-1}$ is not imposed a priori. However, we can choose the convention to satisfy $\tilde{P}(w)=\tilde{P}(w^{-1})$ by using the rescaling of $w$ in this case.
Denoting the two solutions of \eqref{eq:SWPtilde} for $t$ as
\begin{align}\label{eq:tt}
\tilde{t}_{\pm}(w;U) := \frac{1}{2} \left( - \tilde{P}(w;U) \pm \sqrt{\tilde{P}(w;U){}^2 - 4} \right),
\end{align}
the instanton factor $q=e^{-\beta m_0}$ is introduced as their ratio at small $w$
\begin{align}
\frac{\tilde{t}_{+}(w;U)}{\tilde{t}_{-}(w;U)}
= \frac{1} {\tilde{P}(w;U){}^2} \left( 1 + \mathcal{O}(w) \right) = q w^{-2} + \mathcal{O}(w^{-1})
\quad \text{as} \quad w \to 0
\end{align}
analogous to \eqref{eq:q}. Thus, we obtain
\begin{align}
\tilde{P}(w;U) = q^{-\frac{1}{2}} \left( w - U + w^{-1} \right),
\end{align}
where $U$ is the Coulomb branch parameter.

The Seiberg-Witten 1-form $\lambda_{SW}$ is given by
\begin{align}\label{eq:SW1intp}
\lambda_{SW}
&= - \frac{1}{2 \pi i \beta} (\log w) d (\log t)
\cr
&
= \frac{1}{2 \pi i \beta} (\log t) d (\log w) -  \frac{1}{2 \pi i \beta} d (\log t \log w).
\end{align}
Substituting $t=\tilde{t}_{\pm}(w)$ into the first line in \eqref{eq:SW1intp}, and denoting them as $\lambda_{SW}^{\pm}$, we find
\begin{align}
\lambda_{SW}^{\pm}
=\pm \frac{\log w}{2 \pi i \beta} \frac{\frac{\partial \tilde{P}(w;U)}{\partial w} }{\sqrt{ \tilde{P}(w;U)^2 - 4} } dw
= \pm \frac{\log w}{2 \pi i \beta} \frac{w-w^{-1}}{\sqrt{  \prod_{I=1}^2  (w- e^{- \beta \alpha^{+}_{I}} )(w- e^{- \beta \alpha^{-}_{I}} )  } } dw.
\end{align}
Here, we have defined $e^{- \beta \alpha^{\pm}_{I}}$ ($I=1,2$) to be the solutions of the $\tilde{P}^2(w;U)-4=0$, which are explicitly written as
\begin{align}\label{eq:rootPt}
&e^{- \beta \alpha^{\pm}_{1}}=  \frac{1}{2} (U \mp 2q^{\frac{1}{2}}) - \frac{1}{2} \sqrt{U^2 \mp 4q^{\frac{1}{2}} U + 4q -4},
\cr
&e^{- \beta \alpha^{\pm}_{2}}=  \frac{1}{2} (U \pm  2q^{\frac{1}{2}}) + \frac{1}{2} \sqrt{U^2 \pm 4q^{\frac{1}{2}} U + 4q -4}.
\end{align}
The convention has been fixed by imposing
$e^{- \beta \alpha^{-}_{1}} < e^{- \beta \alpha^{+}_{1}} \ll e^{- \beta \alpha^{-}_{2}} < e^{- \beta \alpha^{+}_{2}}$ in the parameter region $q U^2 \ll 1$, $U \gg 1$ with $q$ and $U$ being real positive, at which region the length of the cuts are short and the two cuts are far from each other.

We define $A$-cycle to be the contour going around the branch cut between $e^{- \beta \alpha^{-}_{1}}$ and $e^{- \beta \alpha^{+}_{1}}$ counterclockwise
in the complex plane, on which $\lambda^{-}_{SW}(w)$ is defined. Then, $B$-cycle is defined in such a way that $A \cdot B = 1$ is satisfied.
With this convention, the prepotential is given as
\begin{align}\label{eq:pureSU2SWsol}
a(U) &= \oint_A \lambda_{SW} = - \frac{1}{2 \pi i \beta} \int_{A} \frac{ (\log w) (w-w^{-1})}{\sqrt{  \prod_{I=1}^2  (w- e^{- \beta \alpha^{+}_{I}})(w-e^{- \beta \alpha^{-}_{I}})  } }  dw,
\cr
\frac{\partial F}{\partial a}(U) &= 2 \pi i \beta \oint_B \lambda_{SW} = - 2  \int^{\alpha_2^-}_{\alpha_1^+} \frac{ (\log w) (w-w^{-1})}{\sqrt{  \prod_{I=1}^2 (w- e^{- \beta \alpha^{+}_{I}})(w-e^{- \beta \alpha^{-}_{I}})} }  dw.
\end{align}

We first evaluate them for special value of $U$ for each. The expression for $a(U)$ simplifies when $|U| \gg 1$
since $A$-cycle shrinks in this limit as
\begin{align}
 e^{- \beta \alpha^{+}_{1}} = U^{-1} + \mathcal{O} (U^{-2}),
\qquad
 e^{- \beta \alpha^{+}_{2}}  =  U + \mathcal{O} (1)
\qquad
\text{as} \quad U \to \infty.
\end{align}
Changing the integration variable as $w = U^{-1} \tilde{w}$, we find that
$A$-cycle is now the contour going around the simple pole at $\tilde{w}=1$ counterclockwise
and its integral is given by the residue as
\begin{align}\label{eq:specialA}
a(U) & = - \frac{1}{2 \pi i \beta} \oint_{A} \frac{\log (U^{-1} \tilde{w}) (- \tilde{w}) }{1-\tilde{w}} d\tilde{w} + \mathcal{O}(U^{-1})
= \frac{1}{\beta} \log U + \mathcal{O}(U^{-1}) \qquad \text{as} \quad U \to \infty.
\end{align}

The expression for $\frac{\partial F}{\partial a}(U)$ simplifies when $U = 2 + 2q^{\frac{1}{2}}$ since $B$-cycle shrinks in this limit as
\begin{align}
e^{- \beta \alpha^{+}_{1}} = e^{- \beta \alpha^{-}_{2}} = 1.
\end{align}
The integrand in \eqref{eq:pureSU2SWsol} is finite around $w=-1$ while the integral region is vanishing. Therefore, we find that the B-cycle integral vanishes in this case:
\begin{align}\label{eq:specialB}
\frac{\partial F}{\partial a}(U= 2 + 2q^{\frac{1}{2}}) & = 0.
\end{align}

As for generic value of $U$, it is not straightforward to compute \eqref{eq:pureSU2SWsol} directly. So, we consider their partial derivatives in terms of $U$
\begin{align}\label{eq:dadUomega}
\frac{\partial a}{\partial U} = \oint_A \Omega ,
\qquad
\frac{\partial}{\partial U} \left( \frac{\partial F}{\partial a} \right) = 2 \pi i \beta \oint_B \Omega,
\end{align}
where $\Omega$ is given as
\begin{align}\label{eq:holo}
\Omega := \frac{\partial \lambda_{SW}}{\partial U}
= \frac{1}{2 \pi i \beta} \frac{\partial}{\partial U} \left( \log t \right) d(\log w) + (\text{total derivative})
\end{align}
due to the second line in \eqref{eq:SW1intp}.
Since the total derivative term does not contribute to either of the cycle integrals, we abbreviate it in the following.

Substituting $t=\tilde{t}^{\pm}(w)$ in \eqref{eq:tt} into \eqref{eq:holo} and denoting them as $\Omega^{\pm}$, respectively, we find
\begin{align}\label{eq:dUdt}
\Omega^{\pm}
=& \mp \frac{1}{2 \pi i \beta}  \frac{\frac{\partial \tilde{P}(w;U)}{\partial U}}{\sqrt{\tilde{P}(w;U){}^2 - 4} } \frac{dw}{w},
\end{align}
Then, by using $\alpha_I^{\pm}$ given in \eqref{eq:rootPt}, we can rewrite \eqref{eq:dUdt} as
\begin{align}
\Omega^{\pm} = \mp  \frac{1}{2 \pi i \beta} \frac{dw}{\sqrt{  \prod_{I=1}^2 (w- e^{- \beta \alpha^{+}_{I}})(w-e^{- \beta \alpha^{-}_{I}})} }
\end{align}
From this expression, we can identify $\Omega$ as the holomorphic 1-form on the Seiberg-Witten curve.

Then, \eqref{eq:dadUomega} is given as
\begin{align}
\frac{\partial a}{\partial U}
&=
- 2 \lim_{\epsilon \to 0+} \int^{{e^{- \beta \alpha^{+}_{1}}} + i \epsilon}_{ {e^{- \beta \alpha^{-}_{1}}} + i \epsilon} \Omega^-
\cr
&=
- \frac{1}{\pi i \beta } \lim_{\epsilon \to 0+} \int^{ e^{- \beta \alpha^{+}_{1}} + i \epsilon}_{ e^{- \beta \alpha^{-}_{1}} + i \epsilon} \frac{dw}{\sqrt{  \prod_{I=1}^2 (w- e^{- \beta \alpha^{+}_{I}})(w-e^{- \beta \alpha^{-}_{I}})} },
\cr
\frac{\partial }{\partial U} \left( \frac{\partial F}{\partial a} \right)
&= 2 \pi i \beta \left(
\int^{ e^{- \beta \alpha^{-}_{2}} }_{ e^{- \beta \alpha^{+}_{1}} } \Omega^+ + \int^{ e^{- \beta \alpha^{+}_{1}} }_{ e^{- \beta \alpha^{-}_{2}} } \Omega^-
\right)
\cr
& = - 2 \int^{ e^{- \beta \alpha^{-}_{2}} }_{ e^{- \beta \alpha^{+}_{1}} }  \frac{dw}{\sqrt{  \prod_{I=1}^2 (w- e^{- \beta \alpha^{+}_{I}})(w-e^{- \beta \alpha^{-}_{I}})} }.
\end{align}
They are given in terms of the complete elliptic integral of the first kind
\begin{align}
K(k) = \int^{1}_0 \frac{dt}{\sqrt{(1-t^2)(1-k^2 t^2)}}
\end{align}
by using the formula
\begin{align}\label{eq:formulaK}
\int^{x_2}_{x_1} \frac{dx}{\sqrt{\prod_{i=1}^4(x-x_i)}  }
= \displaystyle \frac{2K\left( \sqrt{ \frac{(x_1-x_2)(x_3-x_4) }{(x_2-x_3)(x_4-x_1)} } \right)}{\sqrt{(x_2-x_3)(x_4-x_1)}}.
\end{align}
We find \eqref{eq:dadUomega} to be explicitly given as
\begin{align}\label{eq:Uderivresult}
\frac{\partial a}{\partial U}
&= - \frac{2}{\pi i \beta \delta(U)} K \left( \frac{\tilde{\delta}(U)}{\delta(U)} \right),
\cr
\frac{\partial }{\partial U} \left( \frac{\partial F}{\partial a} \right)
&= - \frac{4}{\tilde{\delta}(U)}
K \left( \frac{\delta(U)}{\tilde{\delta}(U)} \right),
\end{align}
where
\begin{align}\label{eq:defkkinv}
\delta(U)^2 &:= - \sqrt{U^4 - 8(q+1) U^2 + 16(q-1)^2},
\cr
\tilde{\delta}(U)^2 &: = \frac{1}{2}U^2 - 2(q+1) + \frac{1}{2}\delta(U)^2
\end{align}

%


From the computations above combined with the results in \eqref{eq:specialA} and \eqref{eq:specialB}, the Seiberg-Witten solution is given in the form
\begin{align}\label{eq:SU2SWfinal}
a (U)
&= \lim_{U_0 \to \infty}
\left( \frac{1}{\beta} \log U_0  - \frac{2}{\pi i \beta} \int^U_{U_0} \frac{1}{\delta(U')} K \left( \frac{\tilde{\delta}(U')}{\delta(U')} \right) dU' \right),
\cr
\frac{\partial F}{\partial a} (U)
&= - 4 \int^U_{2+2q^{\frac{1}{2}}} \frac{1}{{\delta}(U')} K \left( \frac{\delta(U')}{\tilde{\delta}(U')} \right) dU'.
\end{align}

\subsection{The Seiberg-Witten curve for pure $Sp(1)$ gauge theory with $O5$-plane}

\begin{figure}
\centering
\begin{minipage}{8cm}
\centering
\includegraphics[width=6cm]{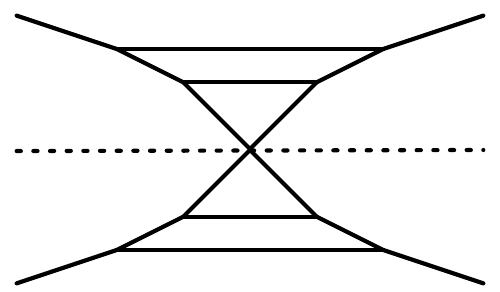}
\caption{5-brane web for 5d pure $Sp(1)$ gauge theory with discrete theta angle 0.}
\label{fig:5branepureSp1}
\end{minipage}
\begin{minipage}{6cm}
\centering
\includegraphics[width=1.8cm]{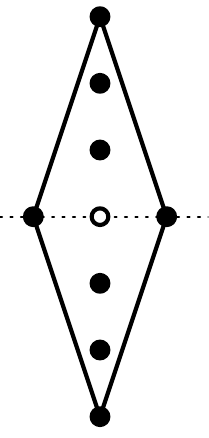}
\caption{Dual graph of the 5-brane web for 5d pure $SU(2)$ gauge theory with discrete theta angle 0.}
\label{fig:5branepureSp1dual}
\end{minipage}
\end{figure}

We consider the 5d pure $Sp(1)$ gauge theory with the discrete theta angle 0 obtained from the 5-brane web diagram with $O5$-plane.
The corresponding 5-brane web diagram and its dual graph is given in Figure \ref{fig:5branepureSp1} and Figure \ref{fig:5branepureSp1dual}, respectively.

The Seiberg-Witten curve for this theory is obtained by substituting $N=1$ and $N_f=0$ to \eqref{eq:SWexplicit} with \eqref{eq:SWNf0theta0}.
The Seiberg-Witten curve is written as
\begin{align}
t^2 + P(w; U) t + 1 =0
\end{align}
with
\begin{align}
P(w;U) = q^{-1}  \left[ (w^3 + w^{-3}) - U (w^2 + w^{-2}) + (q-1) (w+w^{-1}) + 2U \right].
\end{align}
The Seiberg-Witten 1-form \eqref{eq:SW1cancelled} reduces to
\begin{align}\label{eq:SW1pureSp1}
\lambda_{SW}^{\pm} (w)
=&
\mp \frac{\log w}{4 \pi i \beta} \frac{\tilde{\Delta}(w) }{\sqrt{\Delta(w)} } dw
\end{align}
in this case, since $Q(w)=1$.
Here, $\tilde{\Delta}(w)$ and $\Delta(w)$ are introduced as in \eqref{eq:PQK} and \eqref{eq:def-deltatilde}, which are
\begin{align}\label{eq:PPpDDp}
2 \frac{\partial P(w;U)}{\partial w}  &= q^{-1} w^{-2} (w-1)(w+1) \tilde{\Delta}(w),
\cr
P(w;U)^2 - 4 &= q^{-2} w^{-2} (w-1)^2(w+1)^2 \Delta (w).
\end{align}
in this case. They are explicitly given by
\begin{align}\label{eq:DSp1fact}
\tilde{\Delta}(w) = &6(w^2+w^{-2}) - 4U(w+w^{-1}) + 2(2+q),
\cr
\Delta(w) =& \Bigl( (w^2+w^{-2}) - (U+2) (w +w^{-1})+ (2+q+2U) \Bigr)
\cr
& \qquad \times \Bigl(  (w^2+w^{-2}) - (U-2) (w +w^{-1})+ (2+q-2U) \Bigr) .
\end{align}

As discussed in \eqref{eq:Deltafactor}, we write
\begin{align}\label{eq:DeltaSp1roots}
\Delta = w^{-4} \prod_{I=1}^2 (w- e^{- \beta \alpha^{-}_{I}} )(w-e^{- \beta \alpha^{+}_{I}})(w-e^{\beta \alpha^{-}_{I}})(w-e^{\beta \alpha^{+}_{I}})
\end{align}
where $e^{- \beta \alpha^{\pm}_{I}} $ ($I=1,2$) are explicitly given by
\begin{align}\label{eq:alphax}
&e^{- \beta \alpha^{\pm}_{I}}= \frac{x_{I}^{\pm} - \sqrt{(x_I^{\pm})^2-4}}{2}
\qquad (I=1,2)
\end{align}
with
\begin{align}
x_1^{\pm} :=& \frac{1}{2} \left( (U \pm 2) + \sqrt{ (U\mp 2)^2 -4q} \right),\cr
x_2^{\pm} :=& \frac{1}{2} \left( (U \mp 2) - \sqrt{ (U \pm 2)^2 -4q} \right).
\end{align}
Defining $A_I$-cycle and $B_I$-cycle as depicted in Figure \ref{Fig:ABM}, the prepotential is given as the following Seiberg-Witten solution
\begin{align}\label{eq:Sp1ABint}
a(U) &= \oint_{A_1} \lambda_{SW}
= \frac{1}{4 \pi i \beta} \oint_{A_1} \frac{(\log w) \tilde{\Delta}(w)}{\sqrt{ \Delta }} \frac{dw}{w}, \qquad
\cr
\frac{\partial F}{\partial a}(U) &= 2 \pi i \beta \oint_{B_1} \lambda_{SW}
= \int_{\alpha_1^+}^{\alpha_2^-}
\frac{(\log w) \tilde{\Delta}(w)}{\sqrt{ \Delta }} \frac{dw}{w}.
\end{align}
where $A_1$ goes around the branch cut between $\alpha_1^-$ and $\alpha_1^+$.

Parallel to Appendix \ref{app:SWSU2}, we first evaluate them for special value of $U$ for each. The expression for $a(U)$ simplifies when $|U| \gg 1$
since $A_1$-cycle shrinks in this limit as
\begin{align}
&e^{- \beta \alpha^{\pm}_{1}} = U^{-1} + \mathcal{O} (U^{-3}),
\qquad
e^{- \beta \alpha^{\pm}_{2}}  =  \mp 1 + \mathcal{O} (U^{-\frac{1}{2}})
\qquad
\text{as} \quad U \to \infty.
\end{align}
Changing the integration variable as $w = U^{-1} \tilde{w}$, we find that
$A_1$-cycle is now the contour going around the simple pole at $\tilde{w}=1$ counterclockwise
and its integral is given by the residue as
\begin{align}\label{eq:specialASp1}
a(U) & = \frac{1}{4 \pi i \beta} \oint_{A_1} \frac{\log (U^{-1} \tilde{w}) (6 \tilde{w}^{-2} - 4 \tilde{w}^{-1} )}{1 - \tilde{w}} d\tilde{w} + \mathcal{O}(U^{-1})
\cr
& = \frac{1}{\beta} \log U + \mathcal{O}(U^{-1}) \qquad \text{as} \quad U \to \infty.
\end{align}
The expression for $\frac{\partial F}{\partial a}(U)$ simplifies when $U = 2 + 2q^{\frac{1}{2}}$ since $B_1$-cycle shrinks in this limit as
\begin{align}
&e^{- \beta \alpha^{+}_{1}} = e^{- \beta \alpha^{-}_{2}} = \frac{1}{2} \left( 2+ q^{\frac{1}{2}} - (4q^{\frac{1}{2}} + q)^{\frac{1}{2}} \right).
\end{align}
The integrand in \eqref{eq:Sp1ABint} is finite around $w=e^{- \beta \alpha^{+}_{1}} = e^{- \beta \alpha^{-}_{2}} $ while the integral region is vanishing. Therefore, we find that the B-cycle integral vanishes in this case:
\begin{align}\label{eq:specialBSp1}
\frac{\partial F}{\partial a}(U= 2 + 2q^{\frac{1}{2}}) & = 0.
\end{align}

In the following, we consider the partial derivative of in terms of $U$ parallel to Appendix \ref{app:SWSU2}.
By repeating the analogous discussion from \eqref{eq:dadUomega} to \eqref{eq:dUdt}, we obtain
\begin{align}\label{eq:dadUomegaforSp1}
\frac{\partial a}{\partial U} = \oint_A \Omega ,
\qquad
\frac{\partial}{\partial U} \left( \frac{\partial F}{\partial a} \right) = 2 \pi i \beta \oint_B \Omega,
\end{align}
with
\begin{align}\label{eq:Sp1holo1}
\Omega^{\pm}  = \frac{\partial}{\partial U} \lambda_{SW}^{\pm}
= \mp \frac{1}{2 \pi i \beta} \frac{ \frac{\partial P(w;U)}{\partial U}}{\sqrt{P(w;U){}^2 - 4} } \frac{dw}{w}.
\end{align}
Since
\begin{align}
\frac{\partial P(w;U)}{\partial U} = - q^{-1} w^{-2} (w-1)^2 (w+1)^2,
\end{align}
together with \eqref{eq:PPpDDp},
we find that the holomorphic 1-form is given as
\begin{align}
\Omega^{\pm}\label{eq:Sp1holo2}
= \pm \frac{1}{2\pi i \beta} \frac{w-w^{-1}}{\sqrt{ \Delta(w) } } \frac{dw}{w}.
\end{align}

In order to further simplify the computation, we introduce the coordinate transformation
\begin{align}
x:= w+w^{-1}.
\end{align}
Since
\begin{align}
x_I^{\pm} = e^{- \beta \alpha^{\pm}_{I}} + e^{\beta \alpha^{\pm}_{I}}
\qquad (I=1,2)
\end{align}
is satisfied by definition \eqref{eq:alphax}, $\Delta(w)$ in \eqref{eq:DeltaSp1roots} is rewritten as
\begin{align}
\Delta(x) = \prod_{I=1}^2 (x-x_I^+)(x-x_I^-).
\end{align}
With this convention, the holomorphic 1-form \eqref{eq:Sp1holo2} simplifies as
\begin{align}
\Omega^{\pm}
=& \pm \frac{1}{2 \pi i \beta} \frac{1}{\sqrt{ \Delta (x)} } dx.
\end{align}
Also, the $A_1$-cycle, which goes around the branch cut between $e^{- \beta \alpha^{-}_{1}} $ and $e^{- \beta \alpha^{+}_{1}} $
maps to the contour which goes around the branch cut between $x_1^-$ and $x_1^+$.
Thus, we find
\begin{align}
\frac{\partial a}{\partial U}
&=
- \frac{1}{\pi i \beta } \lim_{\epsilon \to 0_+}
\int^{x_1^+ + i \epsilon}_{x_1^- + i \epsilon} \frac{dx}{\sqrt{\prod_{I=1}^2  (x-x_I^+ )(x-x_I^- ) } }
= - \frac{2}{\pi i \beta \delta(U)} K \left( \frac{\tilde{\delta}(U)}{\delta(U)} \right),
\cr
\frac{\partial }{\partial U} \left( \frac{\partial F}{\partial a} \right)
&=
- 2 \int^{x_2^-}_{x_1^+}  \frac{dx}{\sqrt{\prod_{I=1}^2  (x-x_I^+ )(w-x_I^- ) } }
= - \frac{4}{ \tilde{\delta}(U)}
K \left( \frac{\delta(U)}{\tilde{\delta}(U)} \right).
\end{align}
where $\delta(U)$ and $\tilde{\delta}(U)$ are identical  to \eqref{eq:defkkinv}.

Combined with the result in \eqref{eq:specialASp1} and \eqref{eq:specialBSp1},
\begin{align}
a (U)
&= \lim_{U_0 \to \infty}
\left( \frac{1}{\beta} \log U_0  - \frac{2}{\pi i \beta} \int^U_{U_0} \frac{1}{\delta(U')} K \left( \frac{\tilde{\delta}(U')}{\delta(U')} \right) dU' \right),
\cr
\frac{\partial F}{\partial a} (U)
&= - 4 \int^U_{2+2q^{\frac{1}{2}}} \frac{1}{{\delta}(U')} K \left( \frac{\delta(U')}{\tilde{\delta}(U')} \right) dU'.
\end{align}
This agrees with the result \eqref{eq:SU2SWfinal}.
Therefore, we find that the Seiberg-Witten curves from the 5-brane web diagram with and without $O5$-plane are equivalent to each other.


\section{Parametrization}\label{app:param}

In this Appendix, we discuss the parametrization of the 5-brane web diagram for 5d $Sp(N)$ gauge theory with $N_f$ flavors.
In Appendix \ref{app:Kahler} we discuss the derivation of the parametrization \eqref{eq:Kahlerparameter} and reproduce \eqref{eq:AN1} directly from 5-brane web diagram.
In Appendix \ref{app:Sduality}, we focus on the parametrization of pure $Sp(1)$ gauge theory and discuss the transformation of the parameters induced by S-duality.

\subsection{Derivation on the K\"ahler parameters}\label{app:Kahler}

As mentioned in \eqref{eq:Q-length}, the K\"ahler parameters of the Calabi-Yau 3-folds can be obtained from the corresponding 5-brane web diagram.
It is expected to be the case also for 5-brane web diagrams with $O5$-plane.

\begin{figure}
\centering
\includegraphics[width=14cm]{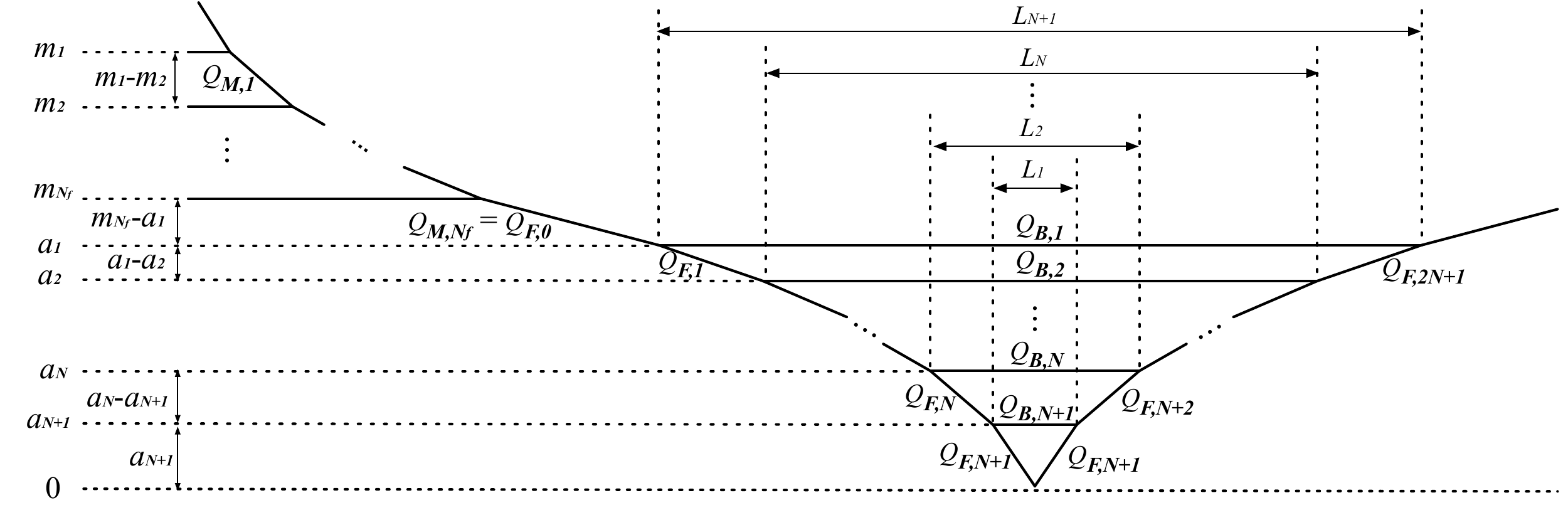}
\caption{Setup of parameters}
\label{fig:setupparam}
\end{figure}

As in Figure \ref{fig:setupparam}, we denote the height of the internal D5-branes to be $a_I$ ($I=1,2,\cdots$) while the height of the external D5-branes to be $m_i$ ($i=1,2,\cdots$).
From this convention, it is straightforward to find the distances between two adjacent D5-branes, from which we find
\begin{align}
&Q_{M,i} = e^{ - \beta (m_i - m_{i+1})} \quad (i=1,2,\cdots, N_f-1),
\cr
&Q_{M,N_f} = Q_{F,0} = e^{ - \beta (m_{N_f} - a_{1})}
\cr
& Q_{F, I} = Q_{F, 2N+2-I} =e^{ - \beta (a_{I} - a_{I+1})} \quad (I=1,2,\cdots, N),
\cr
& Q_{F, N+1}  =e^{ - \beta a_{N+1}}
\end{align}
as given in \eqref{eq:Kahlerparameter}.

\begin{figure}
\centering
\includegraphics[width=10cm]{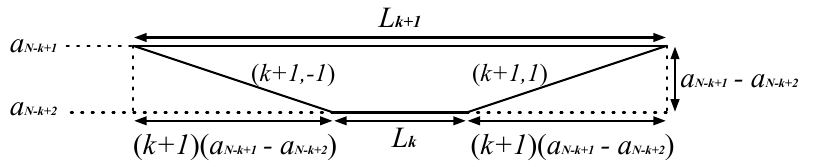}
\caption{A face that appears in the 5-brane web diagram for 5d $Sp(N)$ gauge theory with $N_f$ flavors.}
\label{fig:face}
\end{figure}

Suppose we denote the lengths of the $k$-th internal D5-brane from the bottom as $L_{k}$ ($k=1,2,\cdots, N+1$), it is related to the remaining K\"ahler parameters $Q_{B,I}$ as
\begin{align}
Q_{B,I} = e^{- \beta L_{N+2-I}} \qquad (I=1,2,\cdots, N+1).
\label{eq:QBI}
\end{align}
In order to find $L_k$, we concentrate on one face as depicted in Figure \ref{fig:face}
This leads to the following relation
\begin{align}
L_{k+1} = L_k + 2(k+1) (a_{N-k+1} - a_{N-k+2}).
\end{align}
This is valid also for $k=0$ if we define $L_{0} = a_{N+2} = 0$. Thus, we use this relation recursively to obtain
\begin{align}\label{eq:Lk}
L_k = 2 k a_{N+2-k} - 2 \sum_{j=1}^{k-1} a_{N+2-j}.
\end{align}
Together with \eqref{eq:QBI}, we reproduce
\begin{align}
Q_{B,I} = \exp \left[ - \beta \left( 2 (N+2-I) a_{I} - 2 \sum_{J=I+1}^{N+1} a_{J} \right) \right] \qquad (I=1,2,\cdots, N+1)
\end{align}
given in \eqref{eq:Kahlerparameter}.

\begin{figure}
\centering
\begin{minipage}{8cm}
\includegraphics[width=7.9cm]{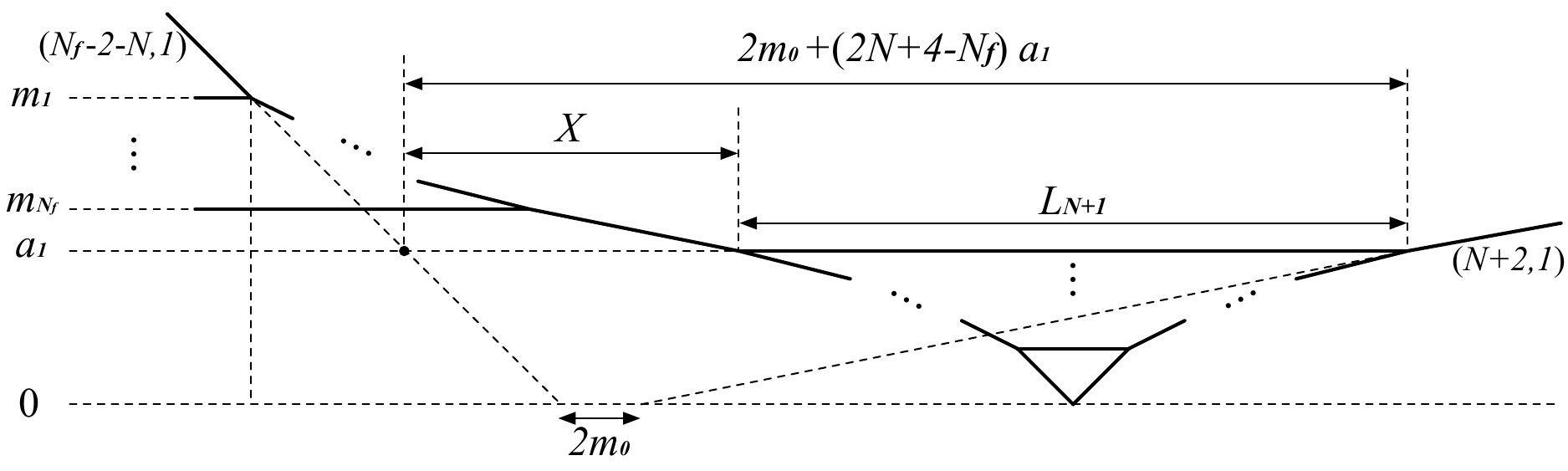}
\end{minipage}
\begin{minipage}{6cm}
\includegraphics[width=5.9cm]{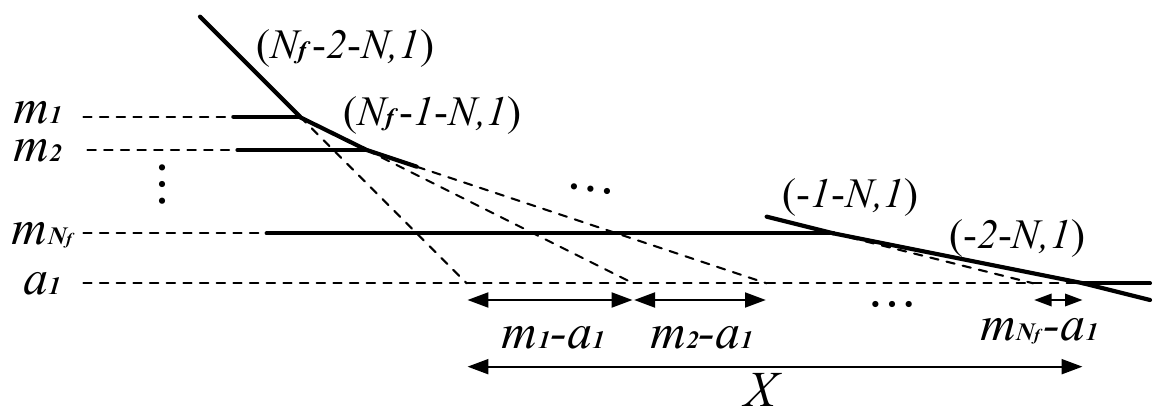}
\end{minipage}
\caption{Left: The computation of $L_{N+1}$. Right: The computation for the length $X$ on the left figure.}
\label{fig:L1}
\end{figure}

In the following, we relate $m_0$ with other parameters. As in Figure \ref{fig:L1},
we identify the distance between the two points which are obtained by extrapolating the external $(N_f-2-N,1)$ 5-brane and $(N+2,1)$ 5-brane to the $O5$-plane as $2m_0$.
In this convention, if we extrapolate them to the height of $a_1$ instead of to the $O5$-plane, the distance between the two points  $2m_0 + (2N+4-N_f)a_1$.
As can be read off from the right of Figure \ref{fig:L1}, $L_{N+1}$ is obtained by subtracting $X = \sum_{i=1}^{N_f} (m_i - a_1)$ from this distance.
Therefore, $L_{N+1}$ can be rewritten in terms of this $m_0$ as
\begin{align}
L_{N+1} = 2m_0 + (2N+4) a_1 - \sum_{i=1}^{N_f} m_i.
\end{align}
Compare it with \eqref{eq:Lk}, we find the relation
\begin{align}
\sum_{I=1}^{N+1} a_I + m_0 - \frac{1}{2} \sum_{i=1}^{N_f} m_j = 0,
\end{align}
as mentioned at \eqref{eq:AN1A}.
By using this relation, $Q_{B,I}$, $Q_{F,N}$, $Q_{F,N+1}$ and $Q_{F,N+2}$ can be also rewritten as
\begin{align}\label{eq:paramm0}
Q_{B,I} & = \exp \left[ - \beta \left(  2m_0 - \sum_{i=1}^{N_f} m_i +2(N+3-I) a_I + 2\sum_{J=1}^{I-1} a_J \right) \right]  \quad (I=1,2,\cdots, N),
\cr
Q_{B,N+1} & = Q_{F,N+1} ^2 = \exp \left[ -  \beta \left(  - 2m_0 + \sum_{i=1}^{N_f} m_i - 2\sum_{J=1}^{N} a_J \right) \right],
\cr
Q_{F,N}& =Q_{F,N+2} = \exp \left[ - \beta \left(  m_0 - \frac{1}{2} \sum_{i=1}^{N_f} m_i +2 a_N + \sum_{J=1}^{N-1} a_J \right) \right] .
\end{align}

\subsection{Transformation induced by S-duality for pure $Sp(1)$ gauge theory}\label{app:Sduality}
\begin{figure}
\centering
\begin{minipage}{7.5cm}
\centering
\includegraphics[width=7cm]{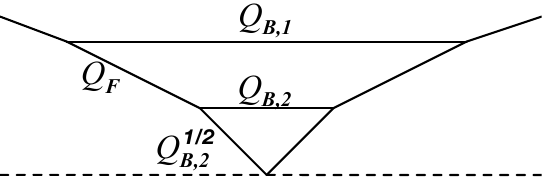}
\caption{Parametrization for the 5-brane web with $O5$-plane for the pure $Sp(1)$ gauge theory.}
\label{fig:sp1paramset}
\end{minipage}
\begin{minipage}{6cm}
\centering
\includegraphics[width=4cm]{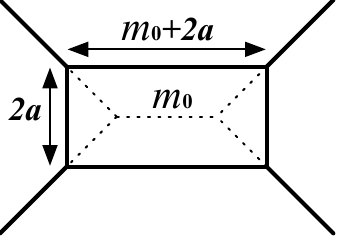}
\caption{Parametrization for the 5-brane web without $O5$-plane for the pure $SU(2)$ gauge theory.}
\label{fig:P1xP1}
\end{minipage}
\end{figure}

In this paper, we find the partition function for 5d $Sp(N)$ gauge theory with $N_f$ flavors.
For the special case where $N=1$ and $N_f=0$, the partition function for the pure $Sp(1)$ gauge theory is already computed with the same method in \cite{Kim:2017jqn}.
However, the parametrization in this paper looks different from the one used in this literature.
On the one hand, we use the parametrization \eqref{eq:paramm0}, which reduces to
\begin{align}\label{eq:conv1}
Q_F = e^{- \beta (m_0 + 2 a)}, \quad
Q_{B,1} = e^{- \beta (2 m_0 + 6a)}, \quad
Q_{B,2} = e^{- \beta (- 2 m_0 - 2a)},
\end{align}
where $Q_F$, $Q_{B,1}$ and $Q_{B,2}$ are depicted in Figure \ref{fig:sp1paramset}.
On the other hand, in \cite{Kim:2017jqn}, they are parametrized as
\begin{align}\label{eq:conv2}
Q_F = e^{- 2 \beta a}, \quad
Q_{B,1} = e^{- \beta (m_0 + 6a)}, \quad
Q_{B,2} = e^{- \beta (m_0 - 2a)}.
\end{align}

Here, we discuss that they are actually related by the S-duality.
In order to understand this point, we discuss the 5-brane web diagram without $O5$-plane for the pure $SU(2)$ gauge theory.
The parametrization for this diagram is given in Figure \ref{fig:P1xP1}.
As discussed in \cite{Aharony:1997bh, Bao:2011rc}, the S-duality, which exchange the D5-brane and the NS5-brane, leads to the parameter exchange
\begin{align}
m_0 + 2a \leftrightarrow 2a,
\end{align}
or equivalently,
\begin{align}\label{eq:Strans}
m_0 \to - m_0 , \quad
a \to a + \frac{1}{2} m_0.
\end{align}
In the corresponding local $\mathbb{P}^1 \times \mathbb{P}^1$ geometry, this corresponds to the exchange of the base $\mathbb{P}^1$ and the fiber $\mathbb{P}^1$ discussed in \cite{Katz:1997eq}. Taking into account the topological vertex formalism \cite{Aganagic:2003db}, it is straightforward to see that the topological string partition function is invariant under this transformation \cite{Bao:2011rc, Mitev:2014jza}.

The partition functions obtained from the 5-brane web diagrams with and without $O5$-plane should agree with each other.
Thus, the partition function for 5d $Sp(N)$ gauge theory with $N_f$ flavors with $N=1$ and $N_f=0$ should be also invariant under the transformation \eqref{eq:Strans}.
This transformation relates the convention  \eqref{eq:conv1} in this paper with the convention  \eqref{eq:conv2} in \cite{Kim:2017jqn} and thus, we find the consistency.


\section{IMS prepotential from the tropical limit}\label{app:IMS}

The Seiberg-Witten curve is constructed in such a way to reproduce the 5-brane web diagram in the tropical limit. In this Appendix, we comment on the Seiberg-Witten solution from this point of view. On the one hand, we can compute the prepotential in the tropical limit from the 5-brane web diagram. In the tropical limit, the $A_I$-cycle integrals reduce to the distance between the $O5$-plane and the corresponding color D5-brane. Analogously,  the $M_i$-cycle integrals reduce to the distance between the $O5$-plane and the corresponding flavor D5-brane.
The $B_I$-cycles reduce to the boundary of the faces in the 5-brane web and the $B_I$-cycle integrals reduce to the area of the corresponding face, which is given by
\begin{align}
(\text{Area})
&= \frac{1}{2} (L_{N-I+2} + L_{N-I+1}) (a_I - a_{I+1})
\cr
&=(N-I+2)a_I{}^2 - (N-I)a_{I+1}{}^2 - 2 a_I a_{I+1} - 2 (a_I - a_{I+1}) \sum_{J=I+2}^{N+1} a_J,
\end{align}
where $L_k$ is given in \eqref{eq:Lk}.

On the other hand, since the tropical limit is interpreted as the decompactification limit of the 5d gauge theory on $\mathbb{R}^4 \times S^1$, the prepotential in this limit is known to be given by IMS prepotential \cite{Intriligator:1997pq}
\begin{align}
F= & m_0 \sum_{I} a_I{}^2
+ \frac{1}{6} \left(  \sum_{s = \pm 1} \sum_{I<J} \left| a_I + s a_J \right|^3 + 8 \sum_{I} a_I{}^3 \right)
- \frac{1}{12} \sum_{i,I} \sum_{s=\pm 1} \left| a_I + s m_i \right|^3
\end{align}
which simplifies as
\begin{align}
F= &\frac{1}{2} \left( 2 m_0 - \sum_{i=1}^{N_f} m_i \right) \sum_{I=1}^N a_I{}^2
+ \frac{1}{3} \sum_{I=1}^N (N-I+4)a_I{}^3
+ \sum_{1 \le I<J \le N}  a_I a_J{}^2
\end{align}
if we assume the parameter region
$m_1 > m_2 > \cdots >m_{N_f} > a_1 > a_2 > \cdots  > a_{N+1}>0$.
From this prepotential, we can compute its partial derivative in terms of $a_I$. We can check explicitly that these two computations agree taking into account the relation \eqref{eq:AB-integral} and \eqref{eq:AN1}.


\section{Proof of the key identity}\label{app:proofkey}
In this appendix, we give the proof of the key identity \eqref{eq:keyid} used in this paper, which is given by
\begin{align}\label{eq:keyidApp}
R_{ \boldsymbol{\lambda}   \boldsymbol{\mu} } \bigl( e^{ -\beta (a-b)} \bigr)
= \exp \left [
\frac{1}{4} \int^{\infty}_{-\infty}\int^{\infty}_{-\infty}
f_{ \boldsymbol{\lambda} }''(x-a) f_{ \boldsymbol{\mu} ^T}''(y-b) \gamma_{\hbar} (x-y) dx dy
\right] \quad  (a\neq b).
\end{align}
The profile function $f_{ \boldsymbol{\lambda} }(x|\hbar)$ is denoted as $f_{ \boldsymbol{\lambda} }(x)$ for short in this Appendix.

The idea of the proof is to show that both sides of the  key identity admit the same expressions. On the one hand, the l.h.s. of the identity \eqref{eq:keyidApp} is
\begin{align*}
R_{ \boldsymbol{\lambda}  \boldsymbol{\mu} } (Q)
:= & \prod_{i=1}^{\infty} \prod_{j=1}^{\infty} \left(1 - Q g^{i+j - \lambda_i - \mu_j - 1} \right)
\cr
 = &  \left[\prod_{i=1}^{\lambda_1^T} \prod_{j=1}^{\mu_1^T} (1 - Q g^{i + j - \lambda_i - \mu_j - 1})\right] \times \left[\prod_{i=1}^{\lambda_1^T} \prod_{j=\mu_1^T}^{\infty} (1 - Q g^{i + j - \lambda_i - \mu_j - 1})\right]
\cr
 \times & \left[\prod_{i=\lambda_1^T}^{\infty} \prod_{j=1}^{\mu_1^T} (1 - Q g^{i + j - \lambda_i - \mu_j - 1})\right] \times \left[\prod_{i=\lambda_1^T}^{\infty} \prod_{j=\mu_1^T}^{\infty} (1 - Q g^{i + j - \lambda_i - \mu_j - 1})\right],
\end{align*}
where we put $Q=e^{ -\beta (a-b)}$.
Since $\lambda_i = 0$ for $i > \lambda_1^T$ and $\mu_j = 0$ for $j > \mu_1^T$, $R_{\boldsymbol{\lambda}  \boldsymbol{\mu}}(Q)$ can be reduced to the following form
\begin{align}\label{eq:Rdiv}
R_{ \boldsymbol{\lambda}  \boldsymbol{\mu} } (Q)
 = &  \left[\prod_{i=1}^{\lambda_1^T} \prod_{j=1}^{\mu_1^T} (1 - Q g^{i + j - \lambda_i - \mu_j - 1})\right] \times \left[\prod_{i=1}^{\lambda_1^T} \prod_{j=\mu_1^T + 1}^{\infty} (1 - Q g^{i + j - \lambda_i - 1})\right]
\cr
 \times & \left[\prod_{i=\lambda_1^T + 1}^{\infty} \prod_{j=1}^{\mu_1^T} (1 - Q g^{i + j - \mu_j - 1})\right] \times \left[\prod_{i=\lambda_1^T + 1}^{\infty} \prod_{j=\mu_1^T + 1}^{\infty} (1 - Q g^{i + j - 1})\right].
\end{align}
Furthermore, the item in the second bracket in \eqref{eq:Rdiv} can be written as
\begin{align}\label{eq:II}
& \prod_{i=1}^{\lambda_1^T} \prod_{j=\mu_1^T + 1}^{\infty} (1 - Q g^{i + j - \lambda_i - 1})
\cr
 =& \left[\prod_{i=1}^{\lambda_1^T} \prod_{j=1}^{\lambda_i} (1 - Q g^{i + j - \lambda_i - 1})\right] \times \left[\prod_{i=1}^{\lambda_1^T} \prod_{j=\lambda_i + 1}^{\infty} (1 - Q g^{i + j - \lambda_i - 1})\right] \times \left[\prod_{i=1}^{\lambda_1^T} \prod_{j= 1}^{\mu_1^T } (1 - Q g^{i + j - \lambda_i - 1})\right]^{-1}
\cr
=& \left[\prod_{i=1}^{\lambda_1^T} \prod_{j_1 = 1}^{\lambda_i} (1 - Q g^{i - j_1})\right] \times \left[\prod_{i=1}^{\lambda_1^T} \prod_{j_2 = 1}^{\infty} (1 - Q g^{i + j_2 - 1})\right] \times \left[\prod_{i=1}^{\lambda_1^T} \prod_{j= 1}^{\mu_1^T } (1 - Q g^{i + j - \lambda_i - 1})\right]^{-1}.
\end{align}
At the last equality, we have introduced a new label $j_1, j_2$, which is related to the original label $j$ as
$j_1= \lambda_i - j + 1 $ and $j_2= j -\lambda_i$.
Similarly, by exchanging $\mu \leftrightarrow \lambda$ and $i \leftrightarrow j$ in the computation above,
we find that the item in the third bracket in \eqref{eq:Rdiv} can be written as
\begin{align}\label{eq:III}
& \prod_{i=\lambda_1^T +1}^{\infty} \prod_{j= 1}^{\mu_1^T} (1 - Q g^{i + j - \mu_j - 1})
\cr
=& \left[\prod_{i_1 =1}^{\mu_j} \prod_{j=1}^{\mu_1^T} (1 - Q g^{- i_1 + j})\right] \times \left[\prod_{i_2 = 1}^{\infty} \prod_{j= 1}^{\mu_1^T} (1 - Q g^{i_2 + j - 1})\right] \times \left[\prod_{i=1}^{\lambda_1^T} \prod_{j= 1}^{\mu_1^T } (1 - Q g^{i + j - \mu_j - 1})\right]^{-1}.
\cr
\end{align}
Combining \eqref{eq:II} and \eqref{eq:III}, $R_{\boldsymbol{\lambda}  \boldsymbol{\mu}}(Q)$ has the following expression:
\begin{align*}
R_{ \boldsymbol{\lambda}  \boldsymbol{\mu} } (Q)
= & \prod_{i=1}^{\infty} \prod_{j=1}^{\infty} \left(1 - Q g^{i+j-1} \right) \times
\left[\prod_{i=1}^{\lambda_1^T} \prod_{j=1}^{\lambda_i} (1 - Q g^{i - j })\right] \times \left[\prod_{i=1}^{\mu_j} \prod_{j=1}^{\mu_1^T} (1 - Q g^{-i + j})\right]
\cr
 \times & \prod_{i=1}^{\lambda_1^T} \prod_{j=1}^{\mu_1^T} \frac{(1 - Q g^{i + j - 1})(1 - Q g^{i + j - \lambda_i - \mu_j - 1})}{(1 - Q g^{i + j - \lambda_i - 1})(1 - Q g^{i + j - \mu_j - 1})}.
\end{align*}
Here, note that, when $Q=e^{- \beta (a-b)} (a \neq b)$ and $g=e^{-\beta h}$, we have
\begin{align*}
 \prod_{i=1}^{\infty} \prod_{j=1}^{\infty} \left(1 - Q g^{i+j-1} \right)
&=  \exp \left[ \sum_{i=1}^{\infty} \sum_{j=1}^{\infty} \log \left(1 - e^{- \beta (a-b)} e^{-\beta h(i+j - 1)} \right) \right]
\cr
&= \exp \left[
- \sum_{i=1}^{\infty} \sum_{j=1}^{\infty}  \sum_{n=1}^{\infty}  \frac{1}{n} e^{-\beta n(a-b)} e^{-\beta n (i-1) \hbar} e^{- \beta n j \hbar}
 \right]
 \cr
&= \exp \left[
\sum_{n=1}^{\infty} \frac{1}{n}
\frac{e^{-\beta n(a-b)}}{(1-e^{-\beta n \hbar} )(1-e^{\beta n \hbar})}
 \right]
 \cr
&
=
\exp \left[
\gamma_{\hbar} (a-b)
 \right].
\end{align*}
Using this identity, $R_{ \boldsymbol{\lambda}  \boldsymbol{\mu} }(e^{- \beta (a-b)})$ has the final expression
\begin{align}\label{eq:Rfinal}
R_{ \boldsymbol{\lambda}  \boldsymbol{\mu} }(e^{-\beta(a-b)})
&= \exp (\gamma_{\hbar}(a-b))
\cr
&\times
\prod_{i=1}^{\lambda_1^T} \prod_{n=1}^{\lambda_i}  \left( 1-e^{-\beta (a - b + h( i - n ) ) } \right) \times \prod_{j=1}^{\mu_1^T} \prod_{m=1}^{\mu_j}
\left(1-e^{-\beta (a - b + h(j - m)) } \right)
\cr
&
\times \prod_{i=1}^{\lambda_1^T} \prod_{j=1}^{\mu_1^T}
\frac{
\left(1-e^{-\beta (a - b + h( i + j  - \lambda_i - \mu_j - 1)  ) } \right)
\left( 1-e^{-\beta (a - b + h( i + j - 1 )  ) } \right)
}{
\left(1-e^{-\beta (a - b + \hbar ( i + j  - \mu_j  - 1) ) } \right)
\left(1-e^{-\beta (a - b + \hbar ( i + j - \lambda_i -  1) ) } \right)
},
\end{align}
where we used $Q=e^{- \beta (a-b)} (a \neq b)$ and $g=e^{-\beta h}$.

On the other hand, the r.h.s. of the key identity \eqref{eq:keyidApp}  is
\begin{align}\label{eq:rhs}
(\text{r.h.s.}) = &\exp \left [
\frac{1}{4} \int^{\infty}_{-\infty}\int^{\infty}_{-\infty}
f_{ \boldsymbol{\lambda} }''(x-a) f_{  \boldsymbol{\mu} ^T}''(y-b) \gamma_{\hbar} (x-y) dx dy
\right]
\cr
=&  \exp \left [
\frac{1}{4} \int^{\infty}_{-\infty} \biggl( \int^{\infty}_{-\infty}
f_{\boldsymbol{\lambda}}''(x-a)  \gamma_{\hbar} (x-y) dx \biggr) f_{ \boldsymbol{\mu} ^T}''(y-b) dy
\right].
\end{align}
We first concentrate on the $x$-integral part of this expression.
By using the expression for the second derivative of profile function
\begin{align*}
f_{\boldsymbol{\lambda}}'' (x)
&= 2 \delta (x)
+ 2 \sum_{i=1}^{\infty} \biggl[
\delta (x - \hbar (  i - 1 - \lambda_i)  )
- \delta ( x - \hbar(  i  -  \lambda_i ) )
- \delta ( x - \hbar (i - 1) )
+ \delta ( x - \hbar i )
\biggr],
\end{align*}
which is given in the first line in \eqref{eq:profile-sec},
we find
\begin{align}\label{eq:xint}
& \int^{\infty}_{-\infty}
f_{\boldsymbol{\lambda}}''(x-a) \gamma_{\hbar} (x-y)  dx
\cr
& = 2 \gamma_{\hbar} (a-y)
+ 2 \sum_{i=1}^{\infty} \biggl[
\gamma_{\hbar} (a + \hbar (  i - 1 - \lambda_i)-y)
-  \gamma_{\hbar} ( a + \hbar(  i  -  \lambda_i )-y)
\cr
& \qquad\qquad\qquad\qquad\qquad\qquad\qquad
- \gamma_{\hbar} ( a + \hbar (i - 1)-y)
+  \gamma_{\hbar} (a + \hbar i-y)
\biggr].
\end{align}
By using the following recursive relation
\begin{align}\label{eq:recursion-r}
\gamma_{\hbar}(x-h)-\gamma_{\hbar}(x)-\gamma_{\hbar}(x+(k-1)h)+\gamma_{\hbar}(x+kh) = \sum_{l=0}^{k-1} \log(1- e^{-\beta(x+lh)})
\end{align}
with the identification $x:= a + \hbar (  i - \lambda_i) - y, \quad k:= \lambda_i$,
the items in the bracket in \eqref{eq:xint} can be reduced to
\begin{align*}
&\gamma_{\hbar} (a + \hbar (  i - 1 - \lambda_i ) -  y )  - \gamma_{\hbar}(a + \hbar (  i - \lambda_i) - y)
- \gamma_{\hbar} (a + \hbar (  i - 1) - y  )  +  \gamma_{\hbar} (a + i \hbar - y )
\cr
=  &\sum_{\ell=0}^{\lambda_i-1} \log (1-e^{-\beta (a + \hbar ( \ell + i - \lambda_i) - y) })
\cr
= & \sum_{n=1}^{\lambda_i} \log (1-e^{-\beta (a + \hbar ( i - n ) - y) }) \quad ( \text{set} \quad n= \lambda_i - l ).
\end{align*}
Therefore, the $x$-integral part of the r.h.s. of the  key identity is
\begin{align*}
& \int^{\infty}_{-\infty}
f_{\boldsymbol{\lambda}}''(x-a) \gamma_{\hbar} (x-y) dx
= 2 \gamma_{\hbar} (a-y)
+ 2 \sum_{i=1}^{\infty} \sum_{n=1}^{\lambda_i} \log (1-e^{-\beta (a + \hbar ( i - n ) - y) }),
\end{align*}
which reduces the item in the bracket in \eqref{eq:rhs} as
\begin{align}\label{eq:y-int}
&
\frac{1}{4} \int^{\infty}_{-\infty}\int^{\infty}_{-\infty}
f_{\boldsymbol{\lambda}}''(x-a) f_{\boldsymbol{\mu}^T}''(y-b) \gamma_{\hbar} (x-y) dx dy
\cr
&=
\frac{1}{2} \int^{\infty}_{-\infty} \gamma_{\hbar} (a-y) f_{\boldsymbol{\mu}^T}''(y-b) dy
\cr
& \qquad \qquad
+ \frac{1}{2} \sum_{i=1}^{\infty} \sum_{n=1}^{\lambda_i}
\int^{\infty}_{-\infty} \log (1-e^{-\beta (a + \hbar ( i - n ) - y) }) f_{\boldsymbol{\mu}^T}''(y-b) dy.
\end{align}
Next, we go on to the $y$-integral part of the r.h.s.~of the  key identity.
Here, we use the expression for $f_{\boldsymbol{\mu}^T}''(y-b)$ given in the second line in \eqref{eq:profile-sec}. That is,
\begin{align}\label{eq:profile-second2}
f_{\boldsymbol{\mu}^T}''(y-b)
= &
2 \delta (y-b)
+ 2 \sum_{j=1}^{\infty} \Bigl[
\delta (y-b + \hbar ( j - 1 - \mu_j )  )
\cr
& \quad - \delta (y-b + \hbar ( j - \mu_j ) )
- \delta ( y-b + \hbar (j - 1))
+ \delta ( y-b + \hbar  j )
\Bigr].
\end{align}
The first term in \eqref{eq:y-int}, which we denote (I), is computed as
\begin{align}\label{eq:FirstTerm}
(\text{I}) := & \frac{1}{2} \int^{\infty}_{-\infty} \gamma_{\hbar} (a-y) f_{\boldsymbol{\mu}^T}''(y-b) dy
\cr
= & \gamma_{\hbar} (a-b)
+ \sum_{j=1}^{\infty}
\biggl[
\gamma_{\hbar} (a-b + \hbar ( j - 1 - \mu_j ))
- \gamma_{\hbar} (a-b + \hbar ( j - \mu_j ) )
\cr
& \qquad \qquad \qquad \qquad \qquad \qquad
- \gamma_{\hbar} (a-b + \hbar (j - 1))
+ \gamma_{\hbar} (a-b + \hbar  j )
\biggr].
\end{align}
By using the same recursive relation \eqref{eq:recursion-r} with the identification $x:= a - b + \hbar ( j - \mu_j), \,\, k:=\mu_j$, we find
\begin{align*}
&\gamma_{\hbar} (a - b + \hbar (  j - 1 - \mu_j )  )  - \gamma_{\hbar}(a - b + \hbar (  j - \mu_j) )
- \gamma_{\hbar} (a - b + \hbar ( j - 1)  )  +  \gamma_{\hbar} (a - b + j \hbar  )
\cr
=  &\sum_{\ell=0}^{\mu_j-1} \log (1-e^{-\beta (a - b + \hbar ( \ell + j - \mu_j) ) })
\cr
=&  \sum_{n=1}^{\mu_j} \log (1-e^{-\beta (a - b + \hbar (  j - n ) ) }) \quad (\text{set} \quad \ell=\mu_j - m).
\end{align*}
Then, we obtaine
\begin{align*}
(\text{I}) 
= \gamma_{\hbar} (a-b)
+ \sum_{j=1}^{\infty}  \sum_{m=1}^{\mu_j} \log (1-e^{-\beta (a - b + \hbar (  j - m ) ) }).
\end{align*}
Next, we compute the second term in \eqref{eq:y-int}, which we denote (II). By using \eqref{eq:profile-second2}, we find
\begin{align*}
(\text{II}) := & \frac{1}{2} \sum_{i=1}^{\infty} \sum_{n=1}^{\lambda_i} \int^{\infty}_{-\infty} \log (1-e^{-\beta (a + \hbar ( i - n ) - y) }) f_{\boldsymbol{\mu}^T}''(y-b) dy
\cr
= &\sum_{i=1}^{\infty} \sum_{n=1}^{\lambda_i} \log (1-e^{-\beta (a - b + \hbar ( i - n ) ) })
\cr
& +  \sum_{i=1}^{\infty} \sum_{n=1}^{\lambda_i} \sum_{j=1}^{\infty}
\biggl[
\log (1-e^{-\beta (a - b + \hbar ( i + j - \mu_j - n - 1)  ) })
- \log (1-e^{-\beta (a - b + \hbar ( i + j  - \mu_j - n ) ) })
\cr
&
\qquad -  \log (1-e^{-\beta (a - b + \hbar ( i + j - n - 1) ) })
+ \log (1-e^{-\beta (a - b + \hbar ( i + j - n )  ) })
\biggr]
\end{align*}
Combining (I) and (II), we obtain
\begin{align*}
& \exp \left [
\frac{1}{4} \int^{\infty}_{-\infty}\int^{\infty}_{-\infty}
f_{\boldsymbol{\lambda}}''(x-a) f_{\boldsymbol{\mu}^T}''(y-b) \gamma_{\hbar} (x-y) dx dy
\right]
\cr
=  &\exp \left[\gamma_{\hbar} (a-b)  \right]
\times \prod_{j=1}^{\infty}  \prod_{m=1}^{\mu_j}
\left(1-e^{-\beta (a - b + \hbar (  j - m ) ) } \right)
\times \prod_{i=1}^{\infty} \prod_{n=1}^{\lambda_i}  \left( 1-e^{-\beta (a - b + \hbar ( i - n ) ) } \right)
\cr
&
\times \prod_{i=1}^{\infty}\prod_{n=1}^{\lambda_i}  \prod_{j=1}^{\infty}
\frac{
\left(1-e^{-\beta (a - b + \hbar ( i + j - \mu_j - n - 1)  ) } \right)
\left( 1-e^{-\beta (a - b + \hbar ( i + j - n )  ) } \right)
}{
\left(1-e^{-\beta (a - b + \hbar ( i + j  - \mu_j - n ) ) } \right)
\left(1-e^{-\beta (a - b + \hbar ( i + j - n - 1) ) } \right)
}
\cr
=  &\exp \left[\gamma_{\hbar} (a-b)  \right]
\times \prod_{j=1}^{\mu_1^T}  \prod_{m=1}^{\mu_j}
\left(1-e^{-\beta (a - b + \hbar ( j - m ) ) } \right)
\times \prod_{i=1}^{\lambda_1^T} \prod_{n=1}^{\lambda_i} \left( 1-e^{-\beta (a - b + \hbar ( i - n ) ) } \right)
\cr
&
\times \prod_{i=1}^{\lambda_1^T}\prod_{n=1}^{\lambda_i}  \prod_{j=1}^{\mu_1^T}
\frac{
\left(1-e^{-\beta (a - b + \hbar ( i + j - \mu_j - n - 1)  ) } \right)
\left( 1-e^{-\beta (a - b + \hbar ( i + j - n )  ) } \right)
}{
\left(1-e^{-\beta (a - b + \hbar ( i + j  - \mu_j - n ) ) } \right)
\left(1-e^{-\beta (a - b + \hbar ( i + j - n - 1) ) } \right)
}.
\end{align*}
Because the product over $n$ will produce simple expression
\begin{align*}
& \prod_{n=1}^{\lambda_i}
\frac{
\left(1-e^{-\beta (a - b + \hbar ( i + j - \mu_j - n - 1)  ) } \right)
\left( 1-e^{-\beta (a - b + \hbar ( i + j - n )  ) } \right)
}{\left(1-e^{-\beta (a - b + \hbar ( i + j  - \mu_j  - n) ) } \right)
\left(1-e^{-\beta (a - b + \hbar ( i + j - n - 1) ) } \right)
}
\cr
& =
\frac{\left(1-e^{-\beta (a - b + \hbar ( i + j - \mu_j - \lambda_i - 1)  ) } \right)
\left( 1-e^{-\beta (a - b + \hbar ( i + j - 1 )  ) } \right)
}{\left(1-e^{-\beta (a - b + \hbar ( i + j  - \mu_j  - 1) ) } \right)
\left(1-e^{-\beta (a - b + \hbar ( i + j - \lambda_i -  1) ) } \right)
},
\end{align*}
the r.h.s. of the  key identity is
\begin{align}\label{eq:rhsfinal}
& \exp \left [
\frac{1}{4} \int^{\infty}_{-\infty}\int^{\infty}_{-\infty}
f_{\boldsymbol{\lambda}}''(x-a) f_{\boldsymbol{\mu}^T}''(y-b) \gamma_{\hbar} (x-y) dx dy
\right]
\cr
=  &\exp \left[ \gamma_{\hbar} (a-b)  \right]
\times \prod_{j=1}^{\mu_1^T}  \prod_{m=1}^{\mu_j}
\left(1-e^{-\beta (a - b + \hbar (  j - m ) ) } \right)
\times \prod_{i=1}^{\lambda_1^T} \prod_{n=1}^{\lambda_i}  \left( 1-e^{-\beta (a - b + \hbar ( i - n ) ) } \right)
\cr
&
\times \prod_{i=1}^{\lambda_1^T} \prod_{j=1}^{\mu_1^T}
\frac{
\left(1-e^{-\beta (a - b + \hbar ( i + j  - \lambda_i - \mu_j - 1)  ) } \right)
\left( 1-e^{-\beta (a - b + \hbar ( i + j - 1 )  ) } \right)
}{
\left(1-e^{-\beta (a - b + \hbar ( i + j  - \mu_j  - 1) ) } \right)
\left(1-e^{-\beta (a - b + \hbar ( i + j - \lambda_i -  1) ) } \right)
}.
\end{align}

By comparing \eqref{eq:Rfinal} and \eqref{eq:rhsfinal}, we arrive at the conclusion \eqref{eq:keyidApp}.

\section{Derivation of \eqref{eq:Xis0} from \eqref{eq:CX=0}} \label{app:X0}
In this appendix, we derive \eqref{eq:Xis0} from \eqref{eq:CX=0}. That is, we show that if
$
\sum_{\boldsymbol{\lambda} } C_{ \boldsymbol{\lambda} \boldsymbol{\mu} \boldsymbol{\nu}} X_{\boldsymbol{\lambda}} = 0
$
is satisfied for arbitrary $\boldsymbol{\mu}$ and $\boldsymbol{\nu}$, where $X_{\boldsymbol{\lambda}} $ is an arbitrary factor which depends on $\boldsymbol{\lambda}$ 
but does not depend on $\boldsymbol{\mu}$ and $\boldsymbol{\nu}$,
then 
$
X_{\boldsymbol{\lambda}}
$
vanishes for all Young diagram $\boldsymbol{\lambda}$.
It turns out that this statement holds even if we relax the condition by choosing $\boldsymbol{\nu} = \varnothing$. 
That is, if 
\begin{align}\label{eq:CX0nu0}
\sum_{\boldsymbol{\lambda} } C_{ \boldsymbol{\mu} \varnothing \boldsymbol{\lambda} } X_{\boldsymbol{\lambda}} = 0.
\end{align}
is satisfied for arbitrary $\boldsymbol{\mu}$, then $X_{\boldsymbol{\lambda}} $ vanishes, 
where we use the cyclic symmetry of the topological vertex.

By multiplying the Schur functions $s_{\boldsymbol{\mu}^T} (x)$ to the left hand side of \eqref{eq:CX0nu0}
and by summing over all the Young diagrams for $\boldsymbol{\mu}$, we obtain 
\begin{align}\label{eq:sC0}
\sum_{\boldsymbol{\mu}} s_{\boldsymbol{\mu}^T} (x) 
\sum_{\boldsymbol{\lambda} } C_{\boldsymbol{\mu} \varnothing \boldsymbol{\lambda} } X_{\boldsymbol{\lambda}}
= 
\sum_{\boldsymbol{\lambda} }
\left[ \prod_{i=1}^{\infty} \prod_{j=1}^{\infty} \left( 1 + x_i g^{j-\frac{1}{2} - \lambda_j} \right) \right]
\tilde{X}_{\boldsymbol{\lambda}} 
\end{align} 
with
\begin{align}
\tilde{X}_{\boldsymbol{\lambda}} := g^{\frac{|| \boldsymbol{\lambda}||^2}{2}} \tilde{Z}_{\boldsymbol{\lambda}}(g)  X_{\boldsymbol{\lambda}}, 
\end{align}
where we use the formula
\begin{align}\label{eq:xgX0}
\sum_{\boldsymbol{Y} } s_{\boldsymbol{Y^T}  } (x) s_{\boldsymbol{Y} } (y) 
= \prod_{i=1}^{\infty} \prod_{j=1}^{\infty}(1 + x_i y_j).
\end{align}
Therefore, \eqref{eq:CX0nu0} indicates that the right hand side of \eqref{eq:sC0} vanishes for arbitrary $x_i$.

Suppose a Young diagram $\boldsymbol{Y}$ is given. For given $\boldsymbol{Y}$, we choose
\begin{align}
x_i = - g^{ i - \frac{1}{2} - Y^T_i}.
\end{align}
Then, \eqref{eq:CX0nu0} and \eqref{eq:sC0} leads to
\begin{align}\label{eq:RXt0}
\sum_{\boldsymbol{\lambda} } R_{\boldsymbol{Y}^T \boldsymbol{\lambda}} (1) \tilde{X}_{\boldsymbol{\lambda}} = 0,
\end{align} 
where $R_{\boldsymbol{Y}^T \boldsymbol{\lambda}} (Q) $ is defined in \eqref{eq:defofR}. 
Here, we use the identity
\begin{align}
R_{\boldsymbol{Y}^T \boldsymbol{\lambda} } (Q) = 
R_{\varnothing \varnothing} (Q) N_{\boldsymbol{Y} \boldsymbol{\lambda} }  (Q), 
\end{align}
with the unrefined Nekrasov factor
\begin{align}
N_{\boldsymbol{Y} \boldsymbol{\lambda} }  (Q)
= \prod_{(i,j) \in \boldsymbol{Y}} (1-Q g^{Y_i + \lambda^T_j - i - j +1})
\prod_{(i,j) \in \boldsymbol{\mu}} (1-Q g^{ - Y^T_j - \lambda_i + i + j - 1}).
\end{align} 
It is known \cite{Cheng:2018wll} that $N_{\boldsymbol{Y} \boldsymbol{\lambda}}  (1) \neq 0$ if and only if $\boldsymbol{Y} = \boldsymbol{\lambda}$.
Therefore, \eqref{eq:RXt0} leads to $\tilde{X}_{\boldsymbol{Y}} =0$ and thus,  $X_{\boldsymbol{Y}} =0$.
Since this argument holds for any given Young diagram ${\boldsymbol{Y}}$, we have derived that 
\begin{align}
X_{\boldsymbol{\lambda}} =0
\end{align}
for all $\boldsymbol{\lambda}$.

\bibliographystyle{JHEP}
\bibliography{ref}
\end{document}